\documentclass[12pt]{article}
\usepackage{amsmath}
\usepackage{amssymb}
\usepackage{latexsym}
\usepackage{epsfig}

\newcommand{\be}{\begin{equation}}
\newcommand{\ee}{\end{equation}}

\def\ket#1{\left| #1 \right\rangle}

\newcommand{\sectiono}[1]{\section{#1}\setcounter{equation}{0}}

\newcommand{\ra}{\rangle}
\newcommand{\la}{\langle}

\newcommand{\Hf}{ {\cal V}^{ (f) } }
\newcommand{\Hg}{ {\cal V}^{(g)} }
\newcommand{\Hfp}{ {\cal V}^{(f)  +} }
\newcommand{\Hgp}{ {\cal V}^{(g) +} }
\newcommand{\Hfm}{ {\cal V}^{(f)  -} }

\begin{document}

\begin{titlepage}
\rightline{\today}
\rightline{\tt hep-th/0611110}
\rightline{\tt DESY-06-199} \rightline{\tt MIT-CTP-3778}
\rightline{\tt YITP-SB-06-44}
\begin{center}
\vskip 1.0cm
{\Large \bf {Analytic Solutions for Tachyon Condensation}}\\
\vskip 0.4cm
{\Large \bf {with General Projectors}}
\vskip 1.0cm
{\large {Yuji Okawa${}^1$, Leonardo Rastelli${}^2$,
and Barton Zwiebach${}^3$}}
\vskip 1.0cm
{\it {${}^1$ DESY Theory Group}}\\
{\it {Notkestrasse 85}}\\
{\it {22607 Hamburg, Germany}}\\
yuji.okawa@desy.de
\vskip 0.5cm
{\it {${}^2$ C.N. Yang Institute for Theoretical Physics}}\\
{\it {Stony Brook University}}\\
{\it {Stony Brook, NY 11794, USA}}\\
leonardo.rastelli@stonybrook.edu
\vskip 0.5cm
{\it {${}^3$ Center for Theoretical Physics}}\\
{\it {Massachusetts Institute of Technology}}\\
{\it {Cambridge, MA 02139, USA}}\\
zwiebach@lns.mit.edu
\vskip 1.0cm
{\bf Abstract}
\end{center}

\noindent
The tachyon vacuum solution of Schnabl is
based on the wedge states,
which close under the star product
and interpolate between the identity state
and the sliver projector.
We use reparameterizations to solve the long-standing problem
of finding an analogous family of states for arbitrary projectors
and to construct analytic solutions based on them.
The solutions simplify for special projectors
and allow explicit calculations in the level expansion.
We test the solutions in detail
for a one-parameter family of special projectors
that includes the sliver and the butterfly.
Reparameterizations further allow a one-parameter deformation
of the solution for a given projector,
and in a certain limit the solution
takes the form of an operator insertion on the projector.
We discuss implications of our work for vacuum string field theory.

\medskip

\end{titlepage}

\newpage

\baselineskip=18pt

\tableofcontents

\sectiono{Introduction}

The first analytic solution
of open string field theory (OSFT) \cite{Witten:1985cc},
corresponding to condensation of the open string tachyon,
was recently constructed by Schnabl \cite{Schnabl:2005gv}
and further studied in \cite{Okawa:2006vm, Fuchs:2006hw,
Rastelli:2006ap,  Ellwood:2006ba, Fuchs:2006an, Fuji:2006me}.
The starting point of \cite{Schnabl:2005gv} is
a clever gauge-fixing condition,
which makes the infinite system of equations of motion
amenable to a recursive analysis.
Schnabl's gauge choice
for the open string field $\Psi$  is
\be
{\cal B}_0 \Psi = 0 \, ,
\ee 
where ${\cal B}_0$ is the antighost
zero mode in the conformal frame
$z = f_{\cal S}(\xi) = \frac{2}{\pi} \arctan \xi$
of the sliver\footnote{
For convenience,  
 we have rescaled the traditional 
conformal frame of the sliver  
by a factor of $2/\pi$.  
This does not change  
the sliver state because of the $SL(2,R)$ invariance of the vacuum,
nor does it affect 
the definition of~${\cal B}_0$.}: 
\begin{equation}
\begin{split}
{\cal B}_0 & \equiv \oint \frac{dz}{2 \pi i} \, z \, b(z)
= \oint \frac{d\xi}{2 \pi i} \,
\frac{f_{\cal S}(\xi)}{f_{\cal S}'(\xi)} \, b(\xi) \\
& = \oint \frac{d\xi}{2 \pi i} \, (1+ \xi^2) \,
\arctan \xi \,\, b(\xi)
= b_0 + \frac{2}{3} \, b_2 - \frac{2}{15} \, b_4 +  \dots\, \, .
\end{split}
\end{equation}
The sliver state $W_\infty$ is
a nontrivial 
projector of the  open string star algebra,
{\it i.e.}, a string field 
different from the identity
that squares to itself \cite{Rastelli:2000iu, Kostelecky:2000hz,
Rastelli:2001jb, Rastelli:2001vb}.
The wedge states $W_\alpha$ with $\alpha \geq 0$
are a family of states
which interpolate
between the sliver $W_\infty$
and the identity $W_0 \equiv {\cal I }$,
and they obey the abelian relation:
\be
W_\alpha * W_\beta = W_{\alpha + \beta} \,.
\ee 
Schnabl's solution is constructed
in terms of a state $\psi_\alpha$, with $\alpha \geq 0$, 
which is the wedge state $W_{\alpha+1}$
with suitable operator insertions.
One  defines the derivative state  
\be
\psi'_\alpha \equiv {d\psi_\alpha\over d\alpha} \,,\quad   \alpha \geq 0, 
\ee
and then Schnabl's solution can  be written as follows:\footnote{
We use the conventions of \cite{Okawa:2006vm},
and the solution differs from that in \cite{Schnabl:2005gv}
by an overall sign.
See the beginning of section 2 of \cite{Okawa:2006vm}
for more details.}  
\begin{equation}
\Psi 
= \lim_{N \to \infty} \Bigl[ \, - \psi_N+
\sum_{n=0}^N \psi'_n \, \Bigr] \,.
\label{solPsi}
\end{equation}
A simple description of the states $\psi_\alpha$ was
presented 
in \cite{Okawa:2006vm} using the CFT formulation of OSFT \cite{LC}. 

While the sliver was historically
the first example of a projector,
it was soon realized that infinitely many projectors
exist \cite{Gaiotto:2002kf}.
Let us restrict 
attention to the subset of string fields known as surface states.
A surface state  is  specified  by a local
coordinate map $z=f(\xi)$ 
from the canonical 
half-disk $\mathbb{D}^+ \equiv \{ \xi \; |\; \Im \xi \geq 0\, , |\xi| \leq 1 \}$
to a region in the upper-half plane (UHP)
$\mathbb{H} \equiv \{ z \; | \; \Im z \geq 0 \}$.
The surface  state $| f \rangle $
is defined by
its inner product
\be
\label{def_overlap_surf}
\langle \, \phi, f \, \rangle 
= \langle \, f \circ \phi(0) \, \rangle_\mathbb{H}
\ee
with any state $\phi$ in the Fock space.
The condition that $| f \rangle $ is a 
projector is
$f(i) = \infty$ \cite{Gaiotto:2002kf}, namely,
the local coordinate curve goes to the
boundary of $\mathbb{H}$
at the open string midpoint $\xi=i$.
(Throughout this paper, we will restrict our considerations
to ``single-split'' projectors, {\it i.e.}, surface states
whose coordinate curve goes to infinity {\it only}
at the open string midpoint.)
The associated open string functional $\Psi_f [ X(\sigma)]$
is {\it split},
namely,
it is the product of a functional of the left
half of the string times a functional of the right half of the string,
$\Psi_f [ X] = \Psi^L_f (X_L) \Psi^R_f(X_R)$. In the half-string formalism
of OSFT \cite{Rastelli:2001rj, Gross:2001rk}, where string fields are regarded as operators acting on the space
of half-string functionals, surface state projectors are interpreted 
as rank-one projectors~\cite{Gaiotto:2002kf}. From this viewpoint,
all surface state projectors should be equivalent.
This is the intuition provided by finite dimensional
vector spaces, where all rank-one projectors
are related by similarity transformations.

These observations raise the natural question of whether Schnabl's solution,
based on the sliver projector, can be generalized
 to solutions based on a generic surface state projector. In this
 paper we find that this is indeed the case.
We also find, however, that the solution technically simplifies
for the subclass of  {\it special} projectors~\cite{Rastelli:2006ap},
 which includes the sliver as its canonical representative.
While we give a  geometric description of the solution
associated with
a general projector,
with the technology currently available 
we are able to evaluate its explicit Fock space expansion
only when the projector is special.

It is useful at this point
 to recall some  facts about special projectors \cite{Rastelli:2006ap}.
The crucial algebraic property of a special projector
 is that  the zero mode ${\cal L}_0$
of the energy-momentum tensor 
 in the frame of the projector\footnote{The definition of a special projector
 further requires the conformal frame $f(\xi)$ 
 to obey certain regularity conditions~\cite{Rastelli:2006ap} 
 which guarantee that the operator $L^+= {1\over s}({\cal L}_0 + {\cal L}_0^\star)$ has
 a non-anomalous left/right decomposition.}, 
 \be
 {\cal L}_0 \equiv \oint \frac{dz}{2 \pi i}  \,z T(z)  = 
   \oint \frac{d\xi}{2 \pi i}   \frac{f(\xi)}{f'(\xi)}  T(\xi) \, ,
 \ee
 and its BPZ conjugate 
${\cal L}_0^\star$ obey 
 \be \label{Lalgebra}
[ {\cal L}_0 , {\cal L}_0^\star] = s ({\cal L}_0 + {\cal L}_0^\star) \, , \quad s > 0 \,.
 \ee
 The sliver is a special projector with $s=1$
and the butterfly is a special projector with $s=2$.
The sliver and the butterfly fit into an infinite
``hypergeometric'' collection of special
projectors --- one projector
$P_\infty^{(s)}$ for each real
$s \geq 1$ --- which 
was briefly described in~\cite{Rastelli:2006ap}. 
 We believe that the hypergeometric collection contains 
 all the single-split special projectors.
  For special projectors, we shall use the notations
 \be \label{Snotations}
 L \equiv \frac{{\cal L}_0}{s} \, , \quad L^\star  \equiv \frac{{\cal L}_0 ^\star} {s} \, .
  \ee
In terms of $L$ and $L^\star$ the algebra (\ref{Lalgebra}) takes the canonical
form
\be \label{canalg}
[L, L^\star] = L+ L^\star \,.
\ee
For any special projector $P_\infty$ 
a family of states
$P_\alpha$ with $\alpha \geq 0$
analogous to the wedge state family of the sliver
is described by the following simple expression:
   \be
  P_\alpha =  e^{-\frac{\alpha}{2} L^+ }  {\cal I } \, ,\quad L^+ \equiv L + L^* \, .
  \ee
The states in the family interpolate
between the projector $P_\infty$
and the identity $P_0 \equiv {\cal I}$, and
they 
obey the relation: 
\be
P_\alpha * P_\beta = P_{\alpha + \beta} \, .
\label{abelian-relation}
\ee
  
We are now in the position to outline our strategy.
 Our starting point is the fact that
{\it all} single-split, twist-invariant projectors
can be related to one another by  a reparameterization of the 
 open string coordinate.  
  Reparameterizations are generated
 infinitesimally by the star-algebra 
derivations $K_n = L_n - (-)^n L_{-n}$ 
 and are familiar gauge symmetries of OSFT \cite{Witten:1986qs, Qiu:1987dv}.  
 Given  a generic  projector $P_\infty$, there exists a finite reparameterization
 that relates it to the sliver, formally
 implemented by an operator $e^H$,
 with $H$ a linear combination of $K_n$'s:
  \be
  P_\infty = e^H \, W_\infty \, , \quad H = \sum_{n=1}^\infty  a_n K_n \, .
  \ee
 Acting with $e^H$ on  the solution $\Psi_{W_\infty}$ associated with the sliver $W_\infty$, we find
the solution $\Psi_{P_\infty}$ associated with $P_\infty$:
\be
\Psi_{P_\infty}  \equiv e^H \Psi_{W_\infty} \,.
\ee
By construction,
 $\Psi_{P_\infty}$ is gauge equivalent to $ \Psi_{W_\infty}$.
 The idea of using  reparameterizations as a solution-generating technique was already noted in~\cite{Okawa:2006vm}. 
 The  solution $\Psi_{P_\infty}$  will take the  form (\ref{solPsi}),  
 with the replacement
 of all the elements associated with the sliver by the corresponding
 elements associated with $P_\infty$. In particular,  we can define an
abelian family 
of states 
interpolating between the identity and a generic projector $P_\infty$
simply by taking $P_\alpha \equiv e^H W_\alpha$.
If we write
\be
W_\alpha =   e^{-\frac{\alpha}{2} L^+_{\cal S} } \, {\cal I } \, ,
\ee
where the  subscript ${\cal S}$   
in $ L^+_{\cal S}$ denotes that it is an  operator related to the sliver,
we have
\be
P_\alpha \equiv e^H\, W_\alpha =  e^H \,e^{-\frac{\alpha}{2} L_{\cal S}^+ } \, e^{-H} \, e^{H} \, {\cal I } 
\equiv e^{-\frac{\alpha}{2} L^+ } \, {\cal I } \, .
\ee
Note
that the identity is annihilated by $H$
and  we have {\it defined}
\be
\label{Ldef}
L^+ \equiv e^H \, L_{\cal S}^+ \, e^{-H} \,.
\ee
Similarly
we take
\be \label{Lgeneral}
L \equiv e^H \, L_{\cal S} \, e^{-H} \, ,\quad L^\star \equiv e^H\,  L_{\cal S}^\star\, e^{-H}\,.
\ee
The operators $L$ and 
 $L^\star$ are BPZ conjugates
of each other since $H^\star = - H$, and they
obey the canonical algebra (\ref{canalg}).
If the projector $P_\infty$ is special, 
the definition (\ref{Lgeneral})
turns out to coincide with (\ref{Snotations}),
but for general projectors
the operator $L$ is not proportional to 
${\cal L}_0$.

It is in practice prohibitively difficult to determine the operator
$H$.  The construction, while motivated by the above considerations, must
be realized differently. 
 The main result of this paper  is to give  a geometric description of the
reparameterization 
procedure and a concrete implementation using the CFT
  language of OSFT. 
In particular we provide a geometric description
for the family of interpolating states $P_\alpha$ associated
with an arbitrary projector
that makes the abelian relation (\ref{abelian-relation}) obvious.

The description simplifies further for the case of
a special projector. It should be emphasized that the geometrical 
construction of the family of states has been a long-standing 
question --- there have been
several attempts 
for the butterfly.
In this paper we find out that the answer is quite simple
if one uses the conformal frame of the projector itself.

It is remarkable that projectors play a central role
in the construction
of the analytic  tachyon solution.
Projectors have been intensively studied
in the context of vacuum string field theory
(VSFT) \cite{vsft, Gaiotto:2001ji}.
In its simplest incarnation, VSFT is
the conjecture that the OSFT action expanded
around the  tachyon vacuum has a
kinetic operator ${\cal Q}$ of the form \cite{Gaiotto:2001ji}:
 \be \label{QVSFT}
 {\cal Q} = \frac{c(i) - c(-i)}{2 i} \,.
 \ee
Taking a matter/ghost factorized ansatz for classical solutions, $\Psi = \Psi_g \otimes \Psi_m$,
the VSFT equations of motion reduce to projector equations for the 
matter part $\Psi_m$. 
VSFT correctly describes
the classical dynamics of D-branes
\cite{Rastelli:2001jb, Rastelli:2001vb, Okawa:2002pd},
but it is somewhat singular.
For example, 
the overall constant in front of the VSFT action
must be taken to be formally
infinite.  It is believed that VSFT arises from OSFT, expanded around the
tachyon vacuum, by a singular field redefinition.
Moreover,  the operator (\ref{QVSFT}) is expected to be the leading
term of a more complicated kinetic operator that involves
the matter energy-momentum 
tensor as well, as discussed in more detail in \cite{Drukker:2005hr}. 
One specific example of  such a field redefinition
given in \cite{Gaiotto:2001ji} was
the reparameterization that maps wedge states to one another,
which in a singular limit formally maps 
all wedge states to the sliver. Interestingly, this
reparameterization emerges naturally in the context of this paper.
Indeed, it turns out that for each projector $P_\infty$
there is a reparameterization that leaves
the projector invariant but
maps  the states in the interpolating
family to one another. 
It takes $P_\alpha$ to $P_{e^{2\beta}\alpha}$,
where $\beta$ is an arbitrary real number.
If we implement this reparameterization on the 
sliver-based solution and 
take the large $\beta$ limit, all wedge states approach the sliver
and the solution takes the form of an 
operator insertion on the sliver. A closely related
approach in constructing a solution in a series expansion
was proposed some time ago
in~\cite{Okawa:2003zc} and investigated further 
in~\cite{Yang:2004xz}. It would be 
interesting 
to find a systematic way to derive the kinetic operator
of VSFT starting from a 
suitably reparameterized version of the tachyon vacuum solution.

\medskip

We begin in 
section 2 with a general introduction to reparameterizations.
After reviewing basic definitions and algebraic properties,
we explain why any two regular 
twist-invariant surface states 
can be related by  a reparameterization. 
The geometrical reason is simple. 
A surface state can be 
defined by
what we call the  {\it reduced} surface:
it is 
the  surface $\mathbb{H}$ for the inner product
in (\ref{def_overlap_surf})
minus the local coordinate patch.
In this picture the open string
is a parameterized 
boundary curve created by removing the patch. The two string
endpoints and the string midpoint define three special points on the boundary
of the reduced surface.
Given two surface states, 
the Riemann mapping theorem ensures
that there is a conformal map between the reduced surfaces
that maps the two endpoints and the midpoint of
one string into those of the other.
This map defines a relationship between
the parameterizations of the two open strings;
this is the induced reparameterization.  When the surface state
is a projector, 
the reduced surface is split in two at the point
where the open string midpoint reaches
the boundary of the full 
surface.
When we map the reduced surfaces of two projectors
to each other,
each of the split surfaces of one reduced surface
is mapped to a split surface of the other reduced surface.
Since each split surface has only two special points
(a string midpoint and a string endpoint),
the conformal map has a one-parameter ambiguity.\footnote{The maps of the two split surfaces are related by a symmetry constraint, so there are no two independent parameters.}

In section 3 we use the above insights 
to give the geometric construction of
the abelian family $P_\alpha$ associated with a generic projector.
In fact, once we choose a map 
$R$ that relates the sliver to the chosen projector,
the surface states $P_\alpha$ are
obtained from the wedge states by a reparameterization
naturally induced by $R$.
This construction represents the surface states $P_\alpha$
using the conformal frame of the projector:
the local coordinate patch is  that of the projector but
the surface only covers part of the UHP.
The geometric description of the surface states $P_\alpha$
simplifies in this conformal frame --- a fact
that was missed in the earlier attempts
to describe them.  
In \S 3.2 
we specialize to special projectors, for which
we find remarkable simplification.
The reparameterization map that relates 
the sliver to the special
projector in the hypergeometric collection
with the parameter $s$
can be chosen to be
simply $R(z) = z^s$,
where $z$ is the coordinate in the UHP. For any  fixed $s$, 
the regions of the UHP
needed to represent states $P_\alpha$ with  different values of $\alpha$ 
are related to one another by  rescaling.   
This is  related to  the fact
that for special projectors the operator $L$
defined in  (\ref{Lgeneral})  is   
proportional to ${\cal L}_0$,
which is the dilation operator in the conformal frame of the projector. 

In section 4 we begin by discussing the algebraic framework of the
tachyon vacuum solution.
We then present our main result,
the CFT construction of the solution
using reparameterizations.
We also present a detailed analysis
of various operator insertions in the CFT description
and derive useful formulas.
In section 5 we use the operator formalism to derive
an expression for the solutions
associated with special projectors.
The solution 
is written
as a sequence of normal-ordered operators acting on the vacuum
and can be readily expanded in level.
Our expression has two parameters, $s \in [1, \infty)$
labeling the special projectors
and $\beta\in (-\infty, \infty)$
labeling  
 the reparameterizations of the solution
that leave the projector invariant.

In section 6 we give the level expansion
of the solutions for special projectors
up to level four.
We first set $\beta=0$ and examine
the dependence of the energy on $s$
to level zero, two, and four.
We find that as the level is increased
the energy density
approaches the expected value that cancels the D-brane tension.
The solutions constructed by our method
can be written in terms of
even-moded total Virasoro operators
and even-moded antighost operators
in addition to the modes of $c$ ghost.
This structure imposes additional constraints,
and thus the solutions belong
to a resticted sector of the universal subspace of the CFT. 
We then examine
the most accurate expression for the solution in the Siegel gauge
computed in~\cite{Gaiotto:2002wy}
and find evidence that
it does not belong to the restricted universal subspace
at level four.
We thus conclude that
the solution in the Siegel gauge cannot be
obtained by our construction.
In \S 6.4
we examine the solution for
a fixed value of $s$ 
and in the limit as $\beta$ becomes large. 
The leading term in the solution
takes the form of an insertion
of the  $c$ ghost in $P_\beta$
multiplied by $e^{2\beta/s}$
and by a finite, calculable coefficient.
We offer some concluding remarks in section 7.

\sectiono{Reparameterizations}

In this section we describe some general facts
about reparameterizations.
The first three subsections are for
a review of well-known material. 
In \S 2.1 we define the notion of
midpoint-preserving
reparameterization $\varphi$ 
of the open string coordinate,  $t \to t'=\varphi(t)$, with $t= e^{i\sigma}$.   
Corresponding
to $\varphi$ there is an operator $U_\varphi$ acting on the space of string fields
that obeys a number of algebraic properties, as explained in \S 2.2.
The transformation
$\Psi \to U_\varphi \Psi$ is
a gauge transformation of OSFT with a vanishing inhomogeneous term,
as we review in \S 2.3.
Finally, in \S 2.4 we explain the key idea:
any two regular  twist-invariant surface states can be related
to one another by a unique reparameterization. For surface states that 
correspond to single-split projectors, 
an interesting and useful ambiguity arises.

In the rest of the paper we shall use these facts to find solutions
of OSFT corresponding to a general projector, starting from Schnabl's solution
corresponding to the sliver. By construction, all these solutions will be
gauge equivalent.

\subsection{Definitions}
 
 Let us  start by recalling  the  definition
 of midpoint-preserving reparameterizations (henceforth,
 simply reparameterizations) \cite{Witten:1986qs}.  A reparameterization
 of the open string coordinate is a map 
 $\sigma \to  \sigma' = \rho(\sigma)$
(with  $\sigma\, , \sigma' \in [0 \, , \, \pi]$)  that obeys
\be \label{r}
 \rho(\pi - \sigma) = \pi - \rho(\sigma)\,.
 \ee
Note that this is a much stronger condition on $\rho$
than 
just fixing the midpoint
$\sigma = \pi/2$: it implies that points at equal
parameter
distance from the midpoint remain at equal
parameter distance after the map. 
We will use the coordinate $t \equiv \exp(i \sigma)$
defined on the unit semicircle in the upper half plane. 
It follows from (\ref{r}) that a map $t \to t' = \varphi(t)$ (with $|t | = | t'| =1$, $\Re t \geq 0 \, ,\Re t' \geq 0$)
is a reparameterization if
 \be \label{phi}
\varphi\left(-\frac{1}{t}\right) = -\frac{1}{\varphi(t)} \,.
\ee
For an infinitesimal reparameterization we write the general ansatz 
\be
\label{ansatz_rep}
\varphi(t) =  t + \epsilon \, v(t) + O(\epsilon^2)\,,
\ee
where $\epsilon$ is an infinitesimal real parameter and $v(t)$ is
a complex vector. We deduce from (\ref{phi}) that the  vector field
 $v(t)$ must be BPZ odd: 
 \be
 \label{c_bpz}
 v\Bigl(-{1\over t}\Bigr) = {1\over t^2} \,  v(t)  \,.
 \ee
Hence $v(t)$ 
 is a linear combination of the BPZ odd vector
fields $v_{K_n}$  
corresponding to the familiar derivations
$K_n = L_n - (-1)^n L_{-n}$:
\be \label{vK}
v (t)=  \sum_{n=1}^\infty a_n v_{K_n} = 
\sum_{n=1}^\infty a_n \, \left (t^{n+1} - (-1)^n t^{-n+1} \right) \,.
\ee
By definition,  reparameterizations preserve
the unit norm of $|t|$. Using (\ref{ansatz_rep}) this condition gives 
\be
\label{unit_norm_cond}
t \, v(t)^* + t ^* v(t) = 0\,,
\ee
which implies that  the coefficients
$a_n$ in (\ref{vK}) satisfy
\be \label{realityan}
a_n =( -)^n a_n^* \, .
\ee
We see that $a_n$ must be real for  $n$ even and imaginary for $n$ odd.

A finite reparameterization $\varphi (t)$ can be obtained by exponentiation of a vector $v(t)$ of the form (\ref{vK}): 
\be
\exp (v(t) \partial_t ) \; t  =  \varphi(t) \,.
\ee
Indeed, the 
condition (\ref{c_bpz}) implies that $\varphi(t)$ satisfies (\ref{phi}).
Moreover, (\ref{unit_norm_cond}) implies that $\varphi(t)$ has unit
norm. 
In general $\varphi(t)$ is
defined only
on the unit semicircle with $|t | = 1$
and cannot be
extended to a holomorphic function inside the local coordinate half-disk 
$\mathbb{D}^+$.   
If $v$ is a finite
linear combination of $v_{K_n}$ vectors, $\varphi(t)$ can be extended to a finite annulus in the upper-half plane $\mathbb{H}$
containing the unit semicircle.

\subsection{The operator $U_\varphi$}

We now consider the operator $U_\varphi$ that implements a finite reparameterization.
The operator is defined to act on
any operator $\mathcal{O} (t)$ in the CFT
as
\be
\label{rep_op}
U_\varphi \, \mathcal{O} (t)  U_\varphi^{-1} =  \varphi \circ \mathcal{O} (t) \,.
\ee
This is the same relation one has for operators that realize the conformal
maps used for surface states, the difference being that here the action on $\mathcal{O}$ is  only defined for $|t|=1$ and  typically does not extend to 
the origin. We write\footnote{We use the symbol $H$
rather than 
$K$ since we 
reserve the latter for the operator introduced 
in \cite{ Rastelli:2006ap}:   $K \equiv \widetilde L^+ \equiv L^+_R - L^+_L$.}
\be
U_\varphi =  e^{-H}\,  , \quad H = \sum_{n=1}^\infty a_n K_n \,,   \quad a_n = (-1)^n a_n^* \,.
\ee
We can verify that the reality condition on the coefficients $a_n$ guarantees 
that $U_\varphi$ preserves the reality condition of the string field. In OSFT the string field
$\Psi$  obeys the reality condition: 
 \be
 \Psi = {\rm hc}^{-1} \circ {\rm bpz} (\Psi) \, .
 \ee
BPZ conjugation (bpz) and hermitian conjugation (hc)
act on Virasoro generators as follows: 
\be
 {\rm bpz} (L_n) = (-1)^n L_{-n} \, , \qquad  {\rm hc}(L_n) = L_{-n} \,.
\ee
For any operator $\mathcal{O}$ we let $\mathcal{O}^\star$ denote its BPZ 
conjugate. Recalling that BPZ conjugation is a linear transformation
while hermitian conjugation is an {\it anti}-linear transformation, we easily
check that reparameterizations preserve the reality
of the string field:
\be
\begin{split}
{\rm hc}^{-1} \circ {\rm bpz}\left( U_\varphi |\Psi\rangle \right) &  = 
{\rm hc}^{-1} \left(  \langle {\rm bpz} (\Psi) |   e^{\sum_{n}  a_n K_n} \right)\\
&= e^{-\sum_n (-1)^n a_n^*  K_n} |{\rm hc}^{-1} \circ {\rm bpz} (\Psi) \rangle \\
& =  e^{\sum_n  a_n K_n} |\Psi \rangle  = U_\varphi |\Psi\rangle \,.
 \end{split}
\ee

 The operator $U_\varphi$ obeys the following 
formal properties: 
\begin{eqnarray}
U_\varphi^\star &  =  & U_\varphi^{-1}\,, \label{U1}  \\[0.5ex]
\left[ Q_B,  U_\varphi \right] & = &  0\,, \label{U2}\\[0.5ex]
U_\varphi\, {\cal I} &  =  & U_\varphi^\star \, {\cal I} = {\cal I} \,, \label{U3} \\[0.5ex]
U_\varphi \Psi_1  * U_\varphi \Psi_2 &  = & U_\varphi  (  \Psi_1 * \Psi_2  ) \,,
 \qquad \forall \; \Psi_1\, , \Psi_2 \,. \label{U4}  
\end{eqnarray}
These identities are the exponentiated version of the
following familiar
properties of $H= \sum_{n=1}^\infty a_n K_n$:
\begin{eqnarray}
H^\star &  =  & - H  \,, \label{K1} \\ 
\left[Q_B, H\right] & = &\phantom{-} 0\,, \label{K2}\\
H \, {\cal I} &  =  &\phantom{-} 0\,, \label{K3} \\
H \Psi_1  * \Psi_2  + \Psi_1 * H \Psi_2 &  = &  H (  \Psi_1 * \Psi_2  )  \,,  \qquad \forall \; \Psi_1\, , \Psi_2\,.
\label{K4}
\end{eqnarray}
The properties (\ref{U1})--(\ref{U4}) can also be
understood from the viewpoint of OSFT
without reference to the operator $H$.
For example,
since points at equal parameter distance from the midpoint
remain at equal parameter
distance after reparameterizations,
(\ref{U4}) follows at once from the picture of the star product
as gluing of half
open string functionals. Similarly, (\ref{U3}) follows,
 at least formally, from the understanding of the identity string field as the functional
that identifies the left and the the right halves of the open string.
In \cite{Rastelli:2006ap} it was found that the 
property (\ref{K3}) may fail to hold for certain singular BPZ odd operators
$H$. 
The finite reparameterizations 
that we explicitly consider in this paper
appear to be perfectly smooth, and we believe that they obey all the formal
properties
(\ref{U1})--(\ref{U4}).  
 Following the discussion of \cite{Rastelli:2006ap},
we note that a regular $H$ should admit a  left/right decomposition
$H = H_L + H_R$ that is non-anomalous:
\be
\label{nonanom}
[\, H_L \,,  H_R] = 0 \,, \quad H_L (A * B) =  (H_L A)* B\,, \quad
H_R (A* B) =  A*  (H_R B) 
\ee
for general string fields $A$ and $B$.

\subsection{Reparameterizations as gauge symmetries}

Reparameterizations are well-known 
gauge symmetries of OSFT. 
(See, for example, \cite{Qiu:1987dv} for an early general discussion.)
Infinitesimal gauge transformations take the familiar form
\be 
\label{gaugetr}
\delta_\Lambda \, \Psi = Q_B \Lambda + \Psi * \Lambda - \Lambda * \Psi \, ,
\ee
where, in the classical theory,  $\Psi$ carries  ghost number one  and the gauge parameter
$\Lambda$ carries ghost number zero.
Choose now $\Lambda = H_R {\cal I} = -H_L {\cal I}$.
The inhomogeneous  term in (\ref{gaugetr}) vanishes since
$[Q_B, H_R] = 0$ and $Q_B {\cal I} = 0$. Using (\ref{nonanom}) we have
\be  
\delta_{H_R {\cal I} } \Psi =  \Psi *( H_R {\cal I}) + (H_L{\cal I}) * \Psi =H_R( \Psi * {\cal I}) + H_L ({\cal I} * \Psi) =(H_R + H_L) \Psi = H \Psi \,.
\ee
This shows that the infinitesimal reparameterization generated by $H$ can be viewed as an infinitesimal
gauge transformation with gauge parameter $H_R {\cal I}$. Exponentiating this relation,
we claim that  
\be \label{finitegauge}
U_\varphi \Psi \equiv  e^H \Psi = {\cal U}_\varphi^{-1}  * \Psi * {\cal U}_\varphi \, ,
\ee
where the string fields ${\cal U}_\varphi$ and $ {\cal U}_\varphi^{-1}$
are defined by 
\begin{eqnarray}
{\cal U}_\varphi & \equiv & \exp_* (H_R {\cal I} ) \equiv {\cal I} + H_R {\cal I} + \frac{1}{2} H_R {\cal I} * H_R {\cal I} + \dots \frac{1}{n!}( H_R {\cal I})^n + \dots \,, \\ 
{\cal U}_\varphi ^{-1}& \equiv & \exp_* (-H_R {\cal I} ) \equiv {\cal I} - H_R {\cal I} + \frac{1}{2} H_R {\cal I} * H_R {\cal I} + \dots \frac{(-1)^n}{n!}( H_R {\cal I})^n + \dots \,,
\end{eqnarray}
and they obey
\be
{\cal U}_\varphi^{-1} * {\cal U}_\varphi = {\cal U}_\varphi* {\cal U}_\varphi ^{-1} = {\cal I} \,.
\ee
It is straighforward to check that for arbitrary string field $A$,
\be
 \exp_* (H_L {\cal I} ) * A =  e^{H_L} A \,, \qquad
 \hbox{and}  \qquad  A *   \exp_* (H_R {\cal I} )  =  e^{H_R} A \,.
\ee
These identities, together with  $[H_L , H_R]=0$,
can be used to show that the equality in (\ref{finitegauge}) holds.
The right-hand side of (\ref{finitegauge}) has the structure of a finite  gauge-transformation in OSFT:
\be
\Psi~ \to~  {\cal U}_\varphi^{-1}  * \Psi * {\cal U}_\varphi  + {\cal U}_\varphi^{-1}  * Q_B {\cal U}_\varphi  \, ,
\ee
where  the inhomogeneous term ${\cal U}_\varphi^{-1}  * Q_B {\cal U}_\varphi $ is identically zero.

Since reparameterizations are gauge symmetries, it is
clear that they map a classical 
solution of OSFT to other gauge-equivalent
classical solutions. If $\Psi$ is a solution
then  $U_\varphi \Psi$ is also a solution, as is verified
 using the formal properties (\ref{U2}) and (\ref{U4}):
\be
Q_B \Psi + \Psi * \Psi = 0~ \longrightarrow ~U_\varphi (Q_B \Psi + \Psi * \Psi) = 0 ~\longrightarrow~
Q_B U_\varphi  \Psi + U_\varphi \Psi * U_\varphi \Psi = 0 \,.  
\ee
It is also clear that $\Psi$ and $U_\varphi$ have the same vacuum energy.
Indeed, using (\ref{U1}) and (\ref{U2}),
\be \label{Uquad}  
\langle U_\varphi \Psi, Q_B U_\varphi  \Psi \rangle =  \langle {\rm bpz }(U_\varphi \Psi) |\, Q_B  U_\varphi |\Psi \rangle = 
\langle {\rm bpz} (\Psi) |\, U^{-1}_\varphi   U_\varphi Q_B |  \Psi \rangle  
 = \langle \Psi, Q_B \Psi \rangle\,. 
\ee
Furthermore, from (\ref{U1}) and (\ref{U4}),
\be \label{Ucub}
\langle U_\varphi \Psi, U_\varphi \Psi * U_\varphi \Psi \rangle \equiv
\langle {\rm bpz} (U_\varphi \Psi) |  U_\varphi \Psi * U_\varphi  \Psi \rangle =
\langle {\rm bpz} (\Psi) | U_\varphi^{-1}   U_\varphi | \Psi *   \Psi \rangle = \langle \Psi , \Psi * \Psi \rangle\,.
\ee
The two equations
(\ref{Uquad}) and (\ref{Ucub}) guarantee that if  the equations of motion
for $\Psi$ are obeyed when contracted with $\Psi$ itself, the same is true
for $U_\varphi \Psi$.

\subsection{Reparameterizations of surface states}

We now explain how reparameterizations can be used
to relate surface states.
Consider a twist-invariant surface states $| f  \ra$,
specified as usual by a local coordinate
map $z=f(\xi)$  from the canonical half-disk
$\mathbb{D}^+$ to
a region in 
the upper half plane $\mathbb{H}$. 
(Both $\mathbb{D}^+$ and $\mathbb{H}$ are defined
above (\ref{def_overlap_surf}).)
We   
denote by $\Hf$ the {\it reduced} surface corresponding to the
surface state $| f  \ra$.  The reduced surface is defined as the complement of the local
coordinate half-disk in $\mathbb{H}$:
\be
\Hf \equiv \mathbb{H}/f(\mathbb{D}^+)\,.
\ee
The reduced surface $\Hf$ has two types of boundary.  The first  type is the boundary where
open string boundary conditions apply; it is the part of the
boundary of $\mathbb{H}$
which belongs to $\Hf$.
The second type  is provided by the coordinate curve $C_f$ which
represents the open string:
\be
\label{c_f_curve}
{\cal C}_f \equiv \{ f(t) \in \mathbb{H}\,, \, |t|=1\, ,\, \Im (t) \geq 0\}\,.
\ee
Let us assume for the time being that
the local coordinate curve
does not go to infinity anywhere.  Then  $\Hf $
has the topology of a disk. The  twist invariance
$f(-\xi) = - f(\xi)$, together with the standard conjugation
symmetry $(f(\xi))^* = f(\xi^*)$,  
implies that $f(\xi) = - (f(-\xi^*))^*$ so  $\Hf$ is invariant
under a reflection about the imaginary $z$ axis. We now
claim that given any two such surface states
$| f \ra$ and $|g \ra$, there exists a reparameterization
$\varphi$ (depending of course on $f$ and $g$)
that relates them: 
\be \label{fgrep}
|g \ra = U_\varphi   | f \ra\, .
\ee
This is shown as follows.
By the Riemann mapping theorem, there exists a holomorphic map
$z'=\widehat R(z)$ relating the reduced surfaces $\Hf$ and  
$\Hg$:
\be
\Hg =\widehat R (\Hf)\,.
\ee
We construct the map using the symmetry of the problem:
first we uniquely map the region
to the right of the imaginary axis of $\Hf$
to that of $\Hg$
by requiring that $f(1)$, $f(i)$,
and infinity are mapped to
$g(1)$, $g(i)$, and infinity, respectively. 
We then extend the map to the left
of the imaginary line using Schwarz's reflection principle, which
applied here gives $\widehat R(z) = - (\widehat R (-z^*))^*$.
The map $\widehat R$ so constructed takes the local coordinate
curve  ${ C}_f $ to the local coordinate curve  ${C}_g$
(defined by ({\ref{c_f_curve}) 
with $f$ replaced by $g$): 
\be
{ C}_g = \widehat R ( { C}_f )  \, .
\ee
A reparameterization $ t' = \varphi (t)$
of the two coordinate curves is defined implicitly
by the relation
\be \label{Rf}
\widehat R ( f(t) ) \equiv g (  \varphi(t) ) \,.
\ee
It  follows from the above construction that $\varphi$ is a
reparameterization. Indeed one readily verifies that
\be
\begin{split}
\widehat R \Bigl( f\Bigl(-{1\over t}\Bigr)\Bigr) & = \widehat R (f(-t^*)) =
\widehat R ( - (f(t))^*) = -(\widehat R(f(t)))^*\\[1.0ex]
&= - (g(\varphi (t)))^* = g(- (\varphi (t))^*) = g \Bigl( -{1\over \varphi (t)}\Bigr)\,,
\end{split}
\ee
which establishes that (\ref{phi}) holds.

We 
now give a formal argument that explains why (\ref{fgrep}) holds.
The surface state $\langle f |$ is defined by its overlap
with a generic state $ |\Psi \rangle$.
Without loss of generality, we can restrict to states $| \Psi \rangle = |  X_b \rangle $ which are eigenstates of the position operator\footnote{For notational simplicity, we are focussing on the
matter part of the CFT.}
$\hat X(t)$,
\be
\hat X(t) | X_b \rangle = X_b(t)  | X_b \rangle \,.
\ee
The overlap $\la f | X_b \ra $
is computed
by  the path-integral over $\Hf$,
where we impose open string boundary conditions on the portion
of the boundary with  $\Im z = 0$ and the boundary conditions $X(f(t)) =
X_b(t)$  on the coordinate curve ${\cal C}_f$. Schematically,
\be
\langle f |  X_b \rangle =   \int_{z \in \Hf } \,  [d   X(z)] \, e^{-S_{BCFT}[ X]} \; \quad {\rm with} \; \; X(f(t)) \equiv X_b(t)  \; {\rm on} \; \;{\cal C}_f   \,.
\ee
Applying the reparameterization $z \to z'=\widehat R(z)$, we see that $\langle f | X_b \rangle$
is equivalently computed by the path-integral over $\Hg$,
provided we keep track of how the boundary conditions are mapped,
\be \label{pathVg}
\la f |  X_b \ra = \int_{z' \in \Hg } \,  [d   X(z')] \, e^{-S_{BCFT}[ X]} \; \quad {\rm with} \;  \;X(g(t')) \equiv X_b(\varphi^{-1}(t'))  \; {\rm on} \; \;{\cal C}_g   \,.
\ee
The path-integral in (\ref{pathVg})
can now be interpreted as computing the overlap of the surface state $\langle g|$
with the position eigenstate $|X_b \circ \varphi^{-1} \rangle$. Thus
\be \label{fundamental}
\la f |  X_b \ra = \la g | X_b \circ \varphi^{-1} \ra \,.
\ee
To proceed, we note that the reparameterization $U_\varphi$ that gives
\be
U_\varphi \hat X(t) U_\varphi^{-1}=
\hat X(\varphi(t))\,,
\ee
will also give
\be  \label{willalso}
U_\varphi |X_b \rangle = |X_b \circ \varphi^{-1} \rangle\,.
\ee
Indeed
\be
\hat X(t) \, U_\varphi |X_b \rangle = U_\varphi U_\varphi^{-1} \,   \hat X(t)  \, U_\varphi |X_b \rangle =
U_\varphi  \hat X(\varphi^{-1}(t) )\, |X_b \rangle = X_b(\varphi^{-1}(t)) \, U_\varphi |X_b \rangle \, ,
\ee
confirming that $U_\varphi |X_b \rangle$ is the $\hat X(t)$ eigenstate of eigenvalue
$X_b \circ \varphi^{-1} (t)$, as stated in (\ref{willalso}).
Back in (\ref{fundamental}), we see that
\be
\la f | X_b \ra =  \la g |   \, U_\varphi  | X_b  \ra \, \qquad \forall \;  |X_b\rangle \, ,
\ee
which implies
\be
\la g | = \la f | \, U_\varphi^{-1} = \la f | \, U^\star_\varphi \, \, .
\ee
This is the BPZ conjugate of the claimed relation (\ref{fgrep}).

\smallskip

So far we have assumed that the coordinate curves ${ C}_f$
and ${ C}_g$ do not reach infinity. It is vital for us to consider projectors,
for which the coordinate curve {\it does} reach infinity at the open
string midpoint: $f(i) = \infty$. If we assume that
the midpoint is the  only point for which $f(t)$ is
infinite  then
the reduced surface $\Hf$ splits into  
two disks $\Hfm$  and $\Hfp$, with  $\Re z < 0 $ and $\Re z > 0$,
respectively,
joined at the point
at infinity.  The claim (\ref{fgrep})
still holds in this case: any two such
twist-invariant
projectors $|f \ra$ and $| g \ra$ can be related by  a reparameterization~$\varphi$.
We define the map $\widehat R$ for $\Hfp$ and, as before, we
extend it to $\Hfm$.
Again, the map $\widehat R:  \Hfp \to \Hgp$,
is guaranteed to exist by the Riemann mapping theorem, but this time it is
not unique. While before $f(i)$ and $\infty$ provided two different points
whose maps could be constrained,
now they are the same one.
We partially fix the $SL(2,R)$ 
symmetry  by requiring
that  $f(1)$ and $f(i) = \infty$ are mapped
to $g(1)$ and $g(i) = \infty$, respectively. There
is one degree of freedom left unfixed, so there exists
a  one parameter family of analytic maps from $ \Hfp$ to $\Hgp$.
This redundancy will play
an important role in the following.

Finally, we note that  we can never hope to relate
regular surface states to projectors using  reparameterizations,
since  the topologies of the reduced surfaces $\Hf$  
are different in the two classes.

\sectiono{Abelian families for general projectors}

The basic building block of Schnabl's solution is the state
$\psi_\alpha$, which is constructed from the wedge state $W_{\alpha+1}$
by adding suitable operator insertions. 
In this section we generalize
the wedge states $W_\alpha$, associated with 
the sliver $W_\infty$,
to states $P_\alpha$ associated with 
a generic twist-invariant projector $P_\infty$.
In the next section we shall deal with the operator insertions
 and construct the analog of the state $\psi_\alpha$ for a generic projector.
 
 As we have explained in \S 2.4, given  a projector
 $P_\infty$, there exists a reparameterization $\varphi$ that relates it
to the sliver: 
 \be
 W_\infty = U_\varphi  P_\infty \,.
 \ee
(There is in fact a one-parameter family of such
reparameterizations. 
For now 
we simply choose one of them.) We define $P_\alpha$  by 
\be
P_\alpha \equiv U_\varphi^{-1} W_\alpha = U_{\varphi^{-1}} W_\alpha \,.
\ee
It follows from (\ref{U3}) that $P_0 = {\cal I}$
and from (\ref{U4}) that the states $P_\alpha$ obey
the same abelian relation as $W_\alpha$:
\be
P_\alpha * P_\beta = U_{\varphi^{-1}} W_\alpha * U_{\varphi^{-1}} W_\alpha = U_{\varphi^{-1}}
 (W_\alpha * W_\beta)
= U_{\varphi^{-1} } (W_{\alpha + \beta}) = P_{\alpha + \beta}\,.
\ee
In \S 3.1 we give a geometric construction of  $P_\alpha$ 
by determining the shape of
the associated one-punctured disk ${\cal P}_\alpha$ 
 in the presentation where the local
coordinate patch  
 is that of the projector~$P_\infty$.
In \S 3.2 we focus on special projectors, for which the construction
simplifies considerably and the reparameterization to the sliver can be given
in closed form. For a special projector the corresponding abelian family obeys 
 a remarkable geometric property: 
 the surfaces ${\cal P}_\alpha$ with different values of $\alpha$ are related
 to one another by overall conformal scaling.

\subsection{Abelian families by reparameterizations}

Given a single-split, twist-invariant projector $|f \rangle$, we wish to find a
reparameterization that relates it to the sliver.
In the notations
of \S 2.4, we write the sliver 
  as $|W_\infty \rangle \equiv |g \rangle$
with $z'=g(\xi) = \frac{2}{\pi} \arctan(\xi)$ and look for a one-parameter family
of conformal maps $\widehat R_\beta: \Hf \to \Hg$. 
From now on we shall drop the superscript
in ${\cal V}^{(f)} \to{\cal V}$,
and we 
rename the sliver's coordinate $z' \to  z_{\cal S}$ 
and the sliver's region ${\cal V}^{(g)} \to {\cal U}$.

To describe the conformal maps $\widehat R_\beta (z)$  we 
need to define a set of curves and regions in the conformal plane.
We denote by $C_0^+$ and $C_0^-$
 the right and left parts, respectively, 
of the coordinate curve $C_0$ of the projector $|f \rangle$.
It is convenient to extend  $C_0^+$ and $C_0^-$ by complex conjugation to curves
on the full plane, making the extended curves invariant under
complex conjugation. For twist invariance of the projectors, 
the curve $C_0^-$ is determined
by $C_0^+$:  $z \in C_0^-$ if
$-z \in C_0^+$.
The curve
$C_0^-$ is the mirror image of $C_0^+$ across the imaginary
axis. (See Figure~\ref{rz01f}.)

Let $\mathcal{V}^+$ 
denote the region of the $z$-plane to the right of $C_0^+$ and
let $\mathcal{V}^-$ 
denote the region of the $z$-plane to the left of $C_0^-$. Since the
coordinate curves reach the point at infinity, both $\mathcal{V}^+$ and $\mathcal{V}^-$
are conformally equivalent to the UHP,
with the role of the real axis 
in the UHP played by the curves $C_0^\pm$.  The union
of $\mathcal{V}^+$ and $\mathcal{V}^-$ is $\mathcal{V}$,  the surface of the
projector minus its coordinate disk.
Let us define analogous regions $\mathcal{U}^\pm$
for the sliver as follows:
\be
\mathcal{U}^+ = \Bigl\{ z_{\cal S}  \Bigl| ~ 
\Re (z_{\cal S})
\geq {1\over 2}~ \Bigr\}\,,
\quad  \mathcal{U}^- = \Bigl\{ z_{\cal S}   
\Bigl| ~ 
\Re (z_{\cal S})
\leq -{1\over 2}~ \Bigr\}\,.
\ee
It is also useful to define vertical
lines $V_\alpha^\pm$ in the sliver frame:
\be
V_\alpha^\pm 
= \Bigl\{ ~ z_{\cal S}~  \Bigl|  ~
\Re (z_{\cal S})
= \pm{1\over 2} (1+\alpha) \Bigr\}\,.
\ee
The boundaries of $\mathcal{U}^\pm$ are $V^\pm$.  
Both $\mathcal{U}^+$ and $\mathcal{U}^-$
are conformally equivalent to the UHP,
with the role of the real axis 
in the UHP played by the lines $V_0^+$
and~$V_0^-$. (See Figure~\ref{rz01f}.)

\begin{figure}
\centerline{\hbox{\epsfig{figure=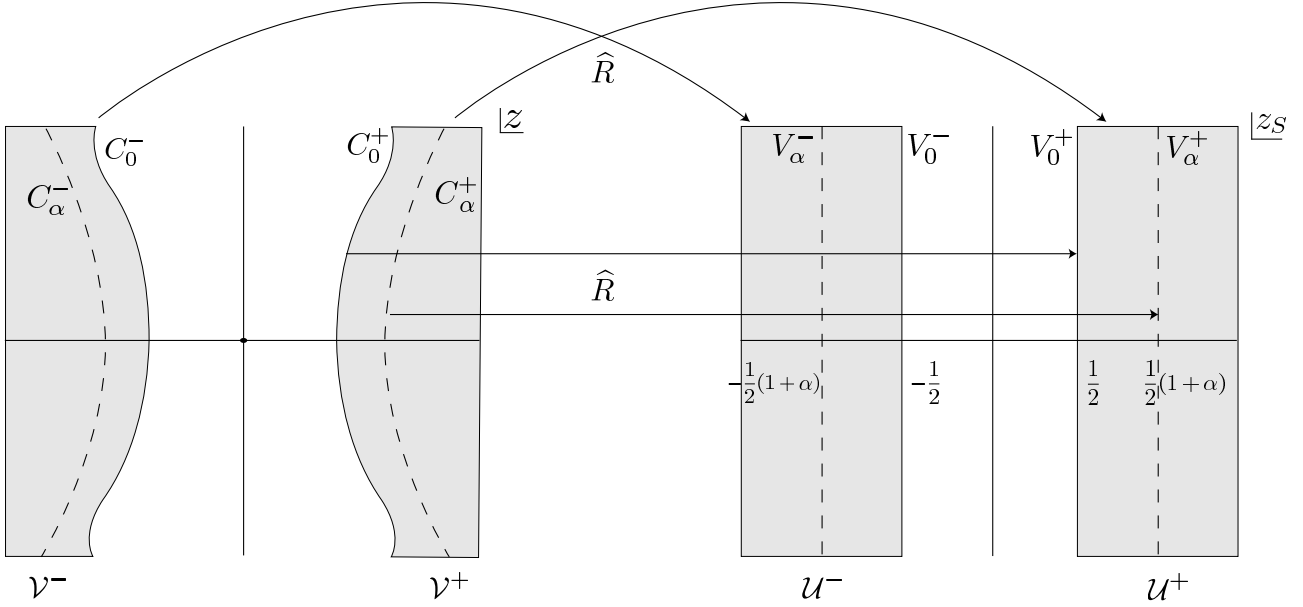, height=8.0cm}}}
\caption{Left: Coordinate curves $C_0^\pm$ of the projector and 
(shaded) regions
$\mathcal{V}^\pm$ to the left and right of the coordinate disk.
Right:  Coordinate curves $V_0^\pm$ for the sliver and (shaded) regions
$\mathcal{U}^\pm$ to the left and right of the coordinate disk.
The map $\widehat{R}$
relates the reduced surfaces of the two projectors.
It takes $\mathcal{V}^\pm$ to $\mathcal{U}^\pm$
and defines the reparameterization that relates the two projectors.
}
\label{rz01f}
\end{figure}

We are interested in the map  
\be
\label{right_to_right}
 R: \mathcal{V}^+~ \to ~\mathcal{U}^+  \,,\qquad   z_{\cal S} = R (z)  \,.
\ee
The map must exist since both regions are conformal to the UHP.
Of course, the map will take
the boundary $C_0^+$ to the boundary $V_0^+$.
We impose two additional
conditions: 
\begin{enumerate}
\item  The intersection of $C_0^+$
with the real axis 
is mapped to $z_{\cal S} = 1/2$.

\item The point at infinity on $C_0^+$ is mapped to the point at infinity on 
$V_0^+$. 
\end{enumerate}
The map $R$ commutes with
the operation of complex conjugation:
$R(z^*) = (R(z))^*$. Thus
 the portion of the real
axis contained in $\mathcal{V}^+$
is mapped to the portion of the real 
axis contained in~$\mathcal{U}^+$.
We can then define the map $\widehat R$ that maps the whole
${\cal V}$ to the whole ${\cal U}$ as follows: 
\be
\label{iothj} 
\widehat R (z) = 
\begin{cases} 
\phantom{-} R(z)   \quad &\hbox{if} ~ z\in
\mathcal{V}_0^+\,, 
\\[1.0ex]
\; \;  -R(-z)   \quad &\hbox{if} ~ z\in \mathcal{V}_0^-\,. 
\end{cases} 
\ee
It is easy 
to check that $\widehat R$ is an odd function: 
\be
{\widehat R}(-z)= - 
{\widehat R} (z)\,.
\ee
 The map $ \widehat R$ describes a reparameterization between the projector and the
sliver.  Indeed, letting $f(\xi)$ denote the coordinate function of the
projector and $f_{\cal S} (\xi_{\cal S})$ denote the coordinate function of the sliver,
we have the relation $\xi_{\cal S} = f^{-1} \circ \widehat  R \circ f(\xi)$. 
As befits  a reparameterization, it satisfies the condition in (\ref{phi}). 

As we have already remarked,   the reparameterization $\widehat R(z)$ is not unique: we only specified
two out of the three conditions needed
to determine a map $\mathbb{H} \to \mathbb{H}$ uniquely.
The remaining ambiguity is that of post-composition with the self maps
of $\mathcal{U}^+$ that leave the points $z_{\cal S}= 1/2$ and $z_{\cal S}=\infty$ invariant.  
Given a function $R_0(z)$ that realizes
the map in (\ref{right_to_right}) with the conditions
listed above,  we can generate
a one-parameter 
family $R_\beta (z)$ of maps that satisfy the same conditions
as follows:
\be
\label{deform_R}
R_\beta (z)  \equiv e^{-2\beta}\, \Bigl(R_0(z)-{1\over 2} \Bigr)  + {1\over 2}  \,,
\ee
with $-\infty <\beta < \infty$ an arbitrary real constant. 
It is clear that the map is a scaling about $z= 1/2$
with scale factor~$e^{-2\beta}$.
With $R_\beta$ replacing $R$ in (\ref{iothj})
we obtain 
a family $ \widehat R_\beta$ of reparameterizations.
We will later use
this ambiguity to produce, for any fixed projector,
a family of solutions parameterized by~$\beta$.

\medskip

Let us
continue our analysis, assuming that a 
choice of $\widehat R$ has been made
for the projector under consideration.  
Since the function $ \widehat R(z)$ is invertible
we can define the curves $C_\alpha^\pm$ as the image under the inverse function 
$ \widehat R^{-1}$
of the vertical lines~$V_\alpha^\pm$:
\be
C_\alpha^\pm
 \equiv \widehat R^{-1} (V_\alpha^\pm)\,.
\ee
It follows from $\widehat R(C_\alpha^\pm) = V_\alpha^\pm$ that
\be
\label{inverse_map_real}
\Re(\widehat R(z) ) = {1\over 2} (1+ \alpha) \,, \qquad  z \in C_\alpha^+\,.
\ee
The various lines $V_\alpha^\pm$ and $C_\alpha^\pm$ are shown in Fig~\ref{rz01f}.

We now proceed to 
the key step in the construction: we 
introduce a family  $P_\alpha$ of states
associated with the projector that is related
by a reparameterization 
to the wedge states.
Consider first the surface $\mathcal{W}_\alpha$
for the wedge state $W_\alpha$ given by 
\be
\label{wedge_from_sliver_new}
\hbox{Wedge state surface}~ \mathcal{W}_\alpha:\quad 
-{1\over 2}(1+\alpha) \leq \Re (z_{\cal S}) \leq
{1\over 2}(1+\alpha)\,. 
\ee
This surface is shown on the right side of Figure~\ref{rz02f}.
We write
\be
\label{wedge_nottion}
\mathcal{W}_\alpha =  (V_\alpha^- \,, V_\alpha^+)\,, 
\ee
where $(C, C')$ denotes the region between the curves $C$ and $C'$.
The coordinate disk for $\mathcal{W}_\alpha$ is $(V_0^-, V_0^+)$. 
Using $z_{\cal S}^{\pm}$
for coordinates on $V_\alpha^{\pm}$,
the identification for the surface is described as follows:
\be
\label{z_ident_sliv_new} 
z_{\cal S}^+ - z_{\cal S}^- = 1+\alpha\,.
\ee  

\begin{figure}
\centerline{\hbox{\epsfig{figure=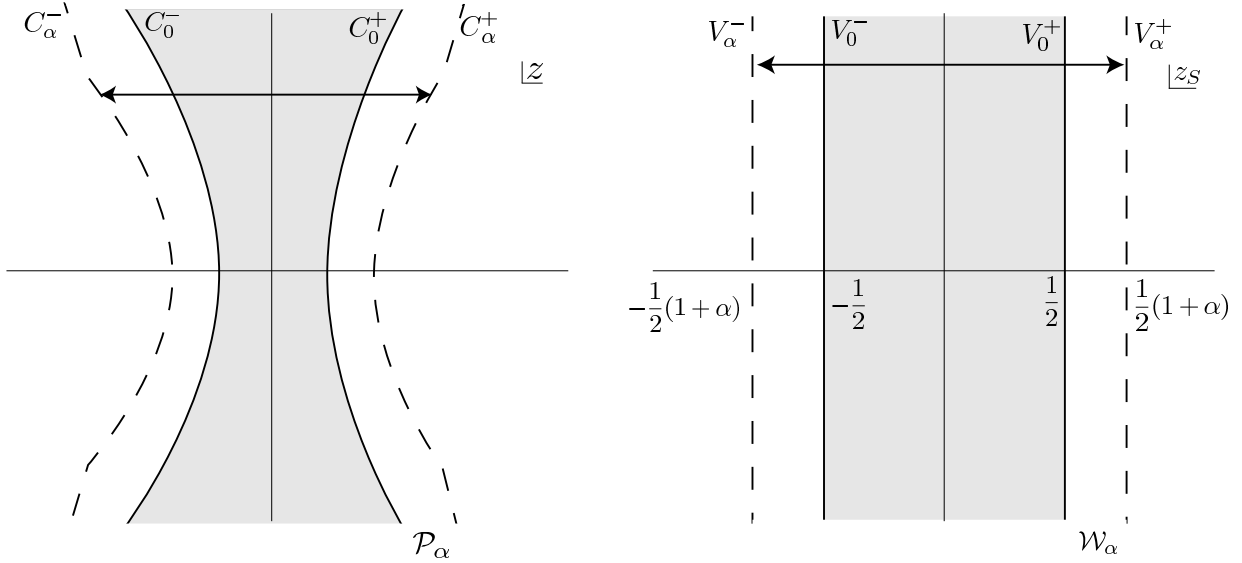, height=7.5cm}}}
\caption{Left: The surface $\mathcal{P}_\alpha$ with its coordinate
disk shaded.  Right:  The wedge surface $\mathcal{W}_\alpha$ with its
coordinate disk shaded.}
\label{rz02f}
\end{figure}

\noindent
Now define the surface 
\be
\label{def_q_alpha}
\mathcal{P}_\alpha \equiv (C_\alpha^-,C_\alpha^+) = \bigl(~ {\widehat R}^{-1}(V_\alpha^-)\,,
\, {\widehat R}^{-1}(V_\alpha^+)\,  \bigr)  ,
\ee
with the identification inherited from that of
the vertical lines
in (\ref{z_ident_sliv_new}).  The surface $\mathcal{P}_\alpha$ is shown
on the left side of Figure~\ref{rz02f}. 
The coordinate disk in $\mathcal{P}_\alpha$ is the region
($C_0^-, C_0^+$), or $\mathcal{P}_0$ without the identification.
It follows that the complement of the coordinate
disk in $\mathcal{W}_\alpha$ is mapped by ${\widehat R}^{-1}$ to the 
complement of the coordinate disk in $\mathcal{P}_\alpha$.  We have thus related
the states $P_\alpha$ and $W_\alpha$ by a reparameterization.   
Using $z^\pm $ for coordinates
on $C_\alpha^\pm$, the identification (\ref{z_ident_sliv_new})
becomes 
\be
\widehat R( z^+) - \widehat R(z^-)  =  1+ \alpha\,. 
\ee
Using (\ref{iothj}) this gives
\be
\label{indent_after_rep}
R( z^+) +  R(-z^-)  =  1+ \alpha\,. 
\ee
A few comments are in order. Since $\mathcal{P}_0$ is the coordinate disk of 
the projector with its boundaries identified, this is simply another
surface for the identity state.  Moreover, the limit of
$\mathcal{P}_\alpha$ as $\alpha \to \infty$ is expected to be
the surface for 
the projector itself.
In fact,
the curves to be identified are going to infinity, 
and the identification becomes immaterial
because infinity is a single point in the UHP.
We thus obtain the UHP
with the coordinate patch  
of the projector --- this is the surface for the projector.

\medskip

In order to describe 
star products of wedge states it is convenient 
to use an alternative presentation of the region 
(\ref{wedge_from_sliver_new}). We use the transition function
(\ref{z_ident_sliv_new}) to move
the region $(V_\alpha^- \,, V_0^-)$
to the right of $V_\alpha^+$.
Since the image of $z_{\cal S}^-= -1/2$ is  
$z_{\cal S}^+ = (1+ 2\alpha)/2$, 
we have 
\be
\label{wedge_from_sliver_99}
 \mathcal{W}_\alpha =  (V_0^-, V_{2\alpha}^+)\,,
\ee
with the identification in (\ref{z_ident_sliv_new}) still
operational. (See Figure~\ref{rz03f}.)
Similarly, the surface $\mathcal{P}_\alpha$
can also be represented as 
\be
\mathcal{P}_\alpha = (C_0^-, C_{2\alpha}^+)\,,
\ee
with the identification in (\ref{indent_after_rep}) still
operational. (See Figure~\ref{rz03f}.)

\begin{figure}
\centerline{\hbox{\epsfig{figure=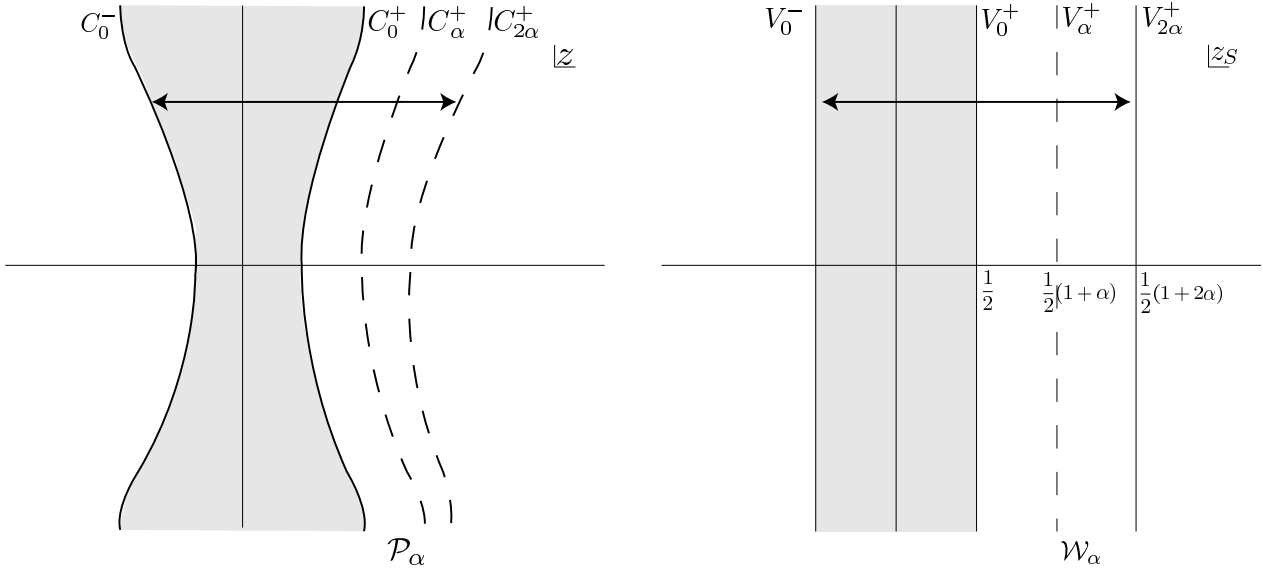, height=7.5cm}}}
\caption{Left: The surface $\mathcal{P}_\alpha$ presented
as the region between $C_0^-$ and $C_{2\alpha}^+$.  
Right:  The wedge surface $\mathcal{W}_\alpha$ presented
as the region between $V_0^-$ and $V_{2\alpha}^+$.}
\label{rz03f}
\end{figure}

The gluing for the star product of wedge states
is performed simply by translation with a real parameter
in the sliver frame.
Using the representation (\ref{wedge_from_sliver_99}),
the two vertical lines 
to be glued are always in $\mathcal{U}^+$.
This induces
the identification between two $C^+$ curves in~$\mathcal{V}^+$
for the star product of the states $P_\alpha$.
If the curve $C_\alpha^+$ 
described with a coordinate $z_<$
is to be glued to $C_{\alpha+\gamma}^+$ with a coordinate $z_>$,  
then $z_<$ and $z_>$ are related by
\begin{equation}
\label{same_side_gluing}
R(z_>) - R(z_<) = \frac{\gamma}{2} \,.
\end{equation}
The right-hand side is 
the real translation parameter that relates the
curves $R(C_\alpha^+)$ and $R(C_{\alpha+ \gamma}^+)$.

\begin{figure}
\centerline{\hbox{\epsfig{figure=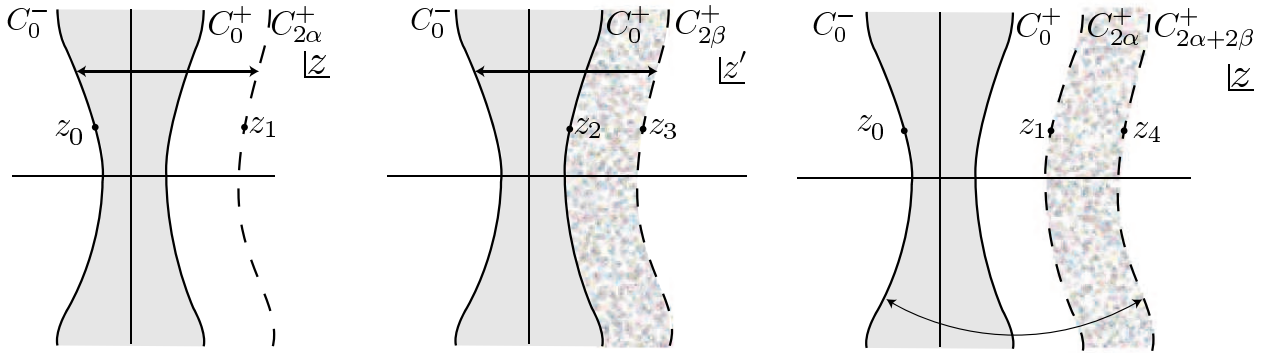, height=6.7cm}}}
\caption{Left: $\mathcal{P}_\alpha$ presented
as the region between $C_0^-$ and $C_{2\alpha}^+$.  
Middle: $\mathcal{P}_\beta$ presented
as the region between $C_0^-$ and $C_{2\beta}^+$.
Right: The surface $\mathcal{P}_{\alpha + \beta}$ obtained by gluing
the complement of the coordinate disk in $\mathcal{P}_\beta$ to
$\mathcal{P}_\alpha$.}
\label{rz04f}
\end{figure}

We now demonstrate the abelian relation
$P_\alpha * P_\beta = P_{\alpha + \beta}$ geometrically.
We present
$\mathcal{P}_\alpha$ as the region $(C_0^-,C_{2\alpha}^+)$
and $\mathcal{P}_\beta$  as the region $(C_0^-, C_{2\beta}^+)$, 
as shown in Figure~\ref{rz04f}. 
The surface for $P_\alpha * P_\beta$
is obtained by mapping the region $(C_0^+, C_{2\beta}^+)$
in $\mathcal{P}_\beta$
to the immediate right of
$C_{2\alpha}^+ \in \mathcal{P}_\alpha$
and by gluing together
$C_{2\alpha}^+\in  \mathcal{P}_\alpha$
and $C_0^+ \in \mathcal{P}_\beta$.
Using coordinates
$z\in \mathcal{P}_\alpha$ and $z' \in \mathcal{P}_\beta$, 
the gluing identification
that follows from (\ref{same_side_gluing}) is
\be
R(z) - R(z') = \alpha\,.
\ee
When  $z' \in C_{2\beta}^+$, we have 
\be
\Re (R(z)) =  \Re (R(z')) + \alpha = {1\over 2} (1+ 2\beta) + \alpha  
= {1\over 2} (1+ 2(\alpha+\beta))\,,
\ee
where we made use of (\ref{inverse_map_real}).  It thus follows that, 
after gluing, the image of $C_{2\beta}^+$
in the $z$-plane is the curve~$C_{2\alpha + 2\beta}^+$. The composite surface
is the region $(C_0^-, C_{2\alpha + 2\beta}^+)$ shown
on the right side 
of Figure~\ref{rz04f}.  To fully confirm 
that this is simply $\mathcal{P}_{\alpha+ \beta}$ we must
examine 
the identification between $C_0^-$ and $C_{2(\alpha+ \beta)}^+$.
Let $z_0\in C_0^-$ and $z_1\in C_{2\alpha}^+$
denote two points identified in $\mathcal{P}_\alpha$
(see Figure~\ref{rz04f}):
\be
\label{is_this_necessary}
R(z_1) + R(-z_0) = 1+ \alpha \,.
\ee
Let $z_2\in C_0^+ \in \mathcal{P}_\beta$ denote the point identified with
$z_1$ by the following relation: 
\be
R(z_1) - R(z_2) = \alpha\,.
\ee
Let $z_3\in C_{2\beta}^+ \in \mathcal{P}_\beta$ be the point
associated with $z_2$ on account of   
having the same imaginary value after mapping by $R$:
\be
R(z_3) - R(z_2) = \beta \,.
\ee
Finally, let $z_4 \in C_{2(\alpha+ \beta)}^+$ in the $z$-plane denote
the point glued to $z_3$:
\be
R(z_4) - R(z_3) = \alpha \,.
\ee
The relation between $z_4$ and $z_0$ is the identification derived from
the gluing procedure.  To find this relation we note that  the last three 
equations imply that $R(z_1) = R(z_4) - \beta$.
Together with (\ref{is_this_necessary}) we obtain
$R(z_4) + R(-z_0) = 1+ \alpha +\beta$,
which is the expected gluing relation on $\mathcal{P}_{\alpha+ \beta}$. This 
completes the verification that  $P_\alpha * P_\beta = P_{\alpha+ \beta}$.

\subsection{Abelian families for special projectors}

For single-split special projectors,
the maps $R(z)$ that relate them to the sliver
are explicitly given by
\be
\label{rs_f_sp}
R(z) = z^s \,,
\ee
where $s$ is the parameter appearing in the algebra $[\mathcal{L}_0, 
\mathcal{L}_0^\star] = s ( \mathcal{L}_0+ 
\mathcal{L}_0^\star)$ of the special projector. We will explain (\ref{rs_f_sp}) in \S 5.1.
 The full map 
from the complement of the coordinate disk 
in the projector to 
the complement of the coordinate disk of the sliver 
given by (\ref{iothj}) is
\be
\label{iothj_butt}
\widehat R (z) = 
\begin{cases} 
\phantom{-} z^s  \quad &\hbox{if} ~ z\in \mathcal{V}^+\,, \\[1.0ex]
-(-z)^s    \quad &\hbox{if} ~ z\in \mathcal{V}^- \,.
\end{cases} 
\ee
It follows from (\ref{rs_f_sp}) that the coordinate curve $C_0^+$ is
the $s$-th root of the sliver line $V_0^+$.
Similarly $C_\alpha^+$ is the
$s$-th root of $V_\alpha^+$.
The  
surface $\mathcal{P}_\alpha$
associated with a special projector with parameter $s$ is
shown in~Figure~\ref{orzf5}.

\begin{figure}
\centerline{\hbox{\epsfig{figure=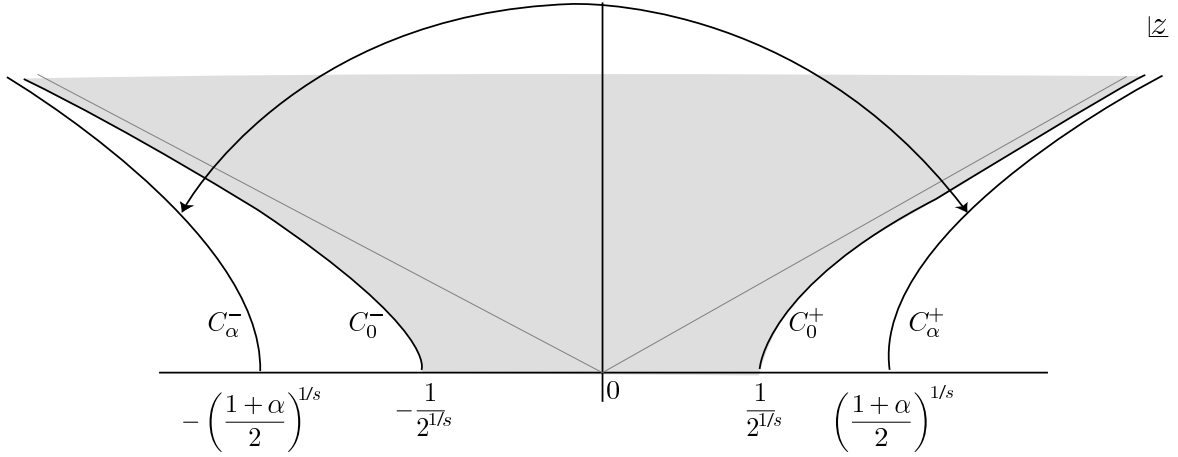, height=6.0cm}}}
\caption{The surface 
$\mathcal{P}_\alpha$ for an arbitrary special projector with 
parameter $s$. The curves $C_\alpha^-$ and $C_\alpha^+$
are identified via the
relation~$({z^+})^s + (-{z^-})^s  =  1+ \alpha$. 
The local coordinate patch is the region between
$C_0^-$ and $C_0^+$.}
\label{orzf5}
\end{figure}

Another key feature 
of special projectors is that we can
write the  map  from $\mathcal{P}_\alpha$
to $\mathbb{H}$ in terms of
 the map $z= f(\xi)$ that defines
the projector. Recall that $f(\xi)$ maps
the upper-half disk of $\xi$ to the region $(C_0^-, C_0^+)$
--- this is $\mathcal{P}_0$
without the identification 
The map $f(\xi)$ is  known explicitly for special projectors, as we shall review in \S 5.1.

 The first step in constructing the map from $\mathcal{P}_\alpha$
to $\mathbb{H}$ is
relating the curves $C_\alpha^+$ to the curve $C_0^+$.  From the relation 
(\ref{inverse_map_real})  we have 
\be
\label{inverap_real}
\Re (z^s)  = {1\over 2} (1+ \alpha)
\quad \hbox{for} 
\quad  z \in C_\alpha^+\,\quad
\hbox{and}\quad 
\Re (z^s ) = {1\over 2}
\quad \hbox{for} 
\quad  z \in C_0^+\,.
\ee
It follows that $C_\alpha^+$ is obtained from $C_0^+$ by a constant scaling!
Indeed,
\be
\label{scaling_trick}
z'\in C_\alpha^+ \,, ~z\in C_0^+  ~\to~  z' = (1+ \alpha)^{1/s} \, z \,. 
\ee
Since it appears frequently later,
we define the scaling function $I_{\alpha,s}$
as follows: 
\be
I_{\alpha, s} (z) \equiv  (1+ \alpha)^{1/s}  z\,.
\ee
Because of the reflection symmetry
about the imaginary axis, 
$C_\alpha^-$ is obtained from $C_0^-$ by the same constant scaling.
The identification for $\mathcal{P}_\alpha$ is
also properly transformed by the
scaling. Indeed, using
(\ref{indent_after_rep}) we have
\be
\label{indent_after_redikfp}
({z^+}')^s + (-{z^-}')^s 
=  1+ \alpha 
~~\hbox{for} ~\mathcal{P}_\alpha
~~~\hbox{and}~~~
(z^+)^s + (-z^-)^s 
=  1 
~~\hbox{for} ~\mathcal{P}_0 \,, 
\ee
and the scaling ${z^\pm}' = (1+ \alpha)^{1/s} \, z^\pm$ relates the 
identifications. 
We thus have a full mapping of the surfaces:
\be
\label{map_surfaces_scaled}
\boxed{\phantom{\biggl(}
\mathcal{P}_\alpha =  I_{\alpha, s} ( \,\mathcal{P}_0\,) 
~~\hbox{for special projectors} \,. ~~}
\ee
For a general projector,
this map is 
difficult to obtain and 
does not follow directly from the knowledge of $R(z)$ and $f(\xi)$.

We now claim that the map from $\mathcal{P}_\alpha$
to $\mathbb{H}$ is given by
the following function $h_\alpha$:
\begin{equation}
\label{f_alpha_from_f}
h_\alpha  = f_I \circ f^{-1} \circ
I_{\alpha,s}^{-1}  \,.
\end{equation}
The function $I_{\alpha,s}^{-1}$ scales $\mathcal{P}_\alpha$ 
down to $\mathcal{P}_0$,
with the identification applied to the boundary of $\mathcal{P}_0$.  
The function $f^{-1}$ then maps $\mathcal{P}_0$
to the upper-half disk with the inherited identification.
Finally, the function $f_I$ is defined by
\begin{equation}
f_I (\xi) = \frac{\xi}{1-\xi^2} \,.
\end{equation}
This is the function that defines the identity state: it maps
the upper-half disk of  $\xi$,
with the left and right parts of the semicircle 
boundary identified via $\xi\sim -1/\xi$, to $\mathbb{H}$.
It is then clear
that $h_\alpha$ maps $\mathcal{P}_\alpha$
to $\mathbb{H}$. 

The  surface state $P_\alpha$
corresponding to the surface $\mathcal{P}_\alpha$ 
is defined by 
\be
\label{set_up_for_p_a}
\langle \, \phi, P_\alpha \, \rangle \equiv 
\langle \, f\circ \phi (0) \, \rangle_{\mathcal{P}_\alpha}\,
 = \langle \, f_\alpha \circ \phi (0) \, \rangle_{\mathbb{H}}
\ee
for any state $\phi$ in the Fock space. 
The correlation function on $\mathcal{P}_\alpha$
in the projector frame has been mapped to that on the UHP
on the right-hand side, where $f_\alpha$ is given by
\be
f_\alpha =  h_\alpha \circ f =   f_I \circ f^{-1} \circ
I_{\alpha,s}^{-1}\circ f \,.
\ee
This is the expression 
obtained in~\cite{Rastelli:2006ap}. 
(See (3.35) of~\cite{Rastelli:2006ap}.)
In that work, however,
the presentation of
$\mathcal{P}_\alpha$ using the conformal frame
of the projector was not given, 
and a geometric proof of
the relation $P_\alpha \ast P_\beta = P_{\alpha+\beta}$
was not provided.
The above results will be useful later 
in our 
calculations
on the tachyon vacuum solutions.
For a general projector,
the calculation of $f_\alpha$ is complicated
because the map from $\mathcal{P}_\alpha$ to $\mathcal{P}_0$
is nontrivial.

\bigskip

We conclude this section with an example.  Aside from the sliver,
the simplest and 
most familiar projector is the
butterfly state. The butterfly is a special projector with $s=2$.
Recall
that the conformal frame of the butterfly is defined by 
\bigskip
\begin{equation}
z=f (\xi) = \frac{\xi}{\sqrt{1+\xi^2}} \,.
\end{equation}
Let us
see that the butterfly is related to the sliver 
through the reparameterization induced~by 
\be
R(z) = z^2\,.
\ee
The
full map (\ref{iothj}) between
the complements of the coordinate disks
is then given by  
\be
\label{iothj_butt}
z_{\cal S} = \widehat R (z) = 
\begin{cases} 
\phantom{-} z^2  \quad &\hbox{if} ~ z\in \mathcal{V}^+\,, \\[1.0ex]
-z^2   \quad &\hbox{if} ~ z\in \mathcal{V}^- \,.
\end{cases} 
\ee

Since the butterfly is a special projector with $s=2$,
the square of the coordinate curve must be 
a straight line 
or a set of straight lines~\cite{Rastelli:2006ap}. 
Points on the coordinate curve are $f(\xi)$ for $\xi = e^{i\theta}$, 
so we have 
\be
\label{butterfly_as_square}
z^2 = (f(e^{i\theta}))^2 =  {e^{2i\theta}\over 1 + e^{2i\theta}}
= {e^{i\theta}\over 2 \cos \theta} = {1\over 2}  + {i\over 2} \tan \theta \,.
\ee
The points here span a vertical line with real part equal to $1/2$. For 
$\theta\in [-{\pi\over 2} , {\pi\over 2}]$, 
we obtain 
the full vertical line
so we indeed find that $\widehat R$ maps $C_0^+ \to V_0^+$.
For $\theta\in [{\pi\over 2} , {3\pi\over 2}]$, 
(\ref{butterfly_as_square})
shows that $z^2$ also spans the full vertical line
with its 
real part equal to $1/2$.
With the minus sign
in the second case of (\ref{iothj_butt}), 
we find that $\widehat R$ maps $C_0^- \to V_0^-$.

 If we write $z = x + i\, y $, with $x$ and $y$
real, it follows from the real part of (\ref{butterfly_as_square}) that
the butterfly coordinate curve is part of the hyperbola
given by 
\be
\label{c_curve}
\Re\, (z^2) =  x^2 - y^2  = {1\over 2} \,.
\ee
In fact, the full coordinate curve is the part of the hyperbola that lies
on $\mathbb{H}$.

\begin{figure}
\centerline{\hbox{\epsfig{figure=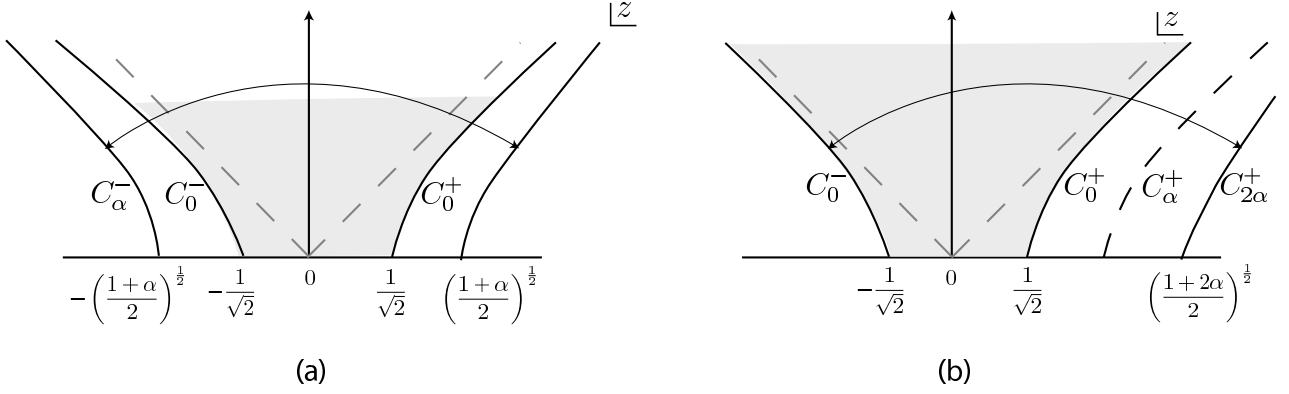, height=5.1cm}}}
\caption{(a) The surface $\mathcal{P}_\alpha$ in the butterfly
family. The curves $C_\alpha^-$ and $C_\alpha^+$ are identified.
The coordinate patch is that of the butterfly itself. (b) The same
surface, with the complement of the coordinate patch placed completely
to the right of the patch. The curves $C_0^-$ and $C_{2\alpha}^+$ are identified.}
\label{orzf3}
\end{figure}

Consider now the surface $\mathcal{P}_0$, namely, the region in $\mathbb{H}$
in between $C_0^-$ and $C_0^+$. Let $z_+ \in C_0^+$ and $z_-\in C_0^-$.
How do we write the identification of $C_0^-$ and $C_0^+$ as an analytic
relation between $z_-$ and $z_+$? From (\ref{indent_after_rep})   
we have 
\be
\label{tent_ident}
z_+^2 +  z_-^2 = 1 \,. 
\ee
This correctly identifies $z_-= -1/\sqrt{2}$ with $z_+ = 1/\sqrt{2}$.
We can confirm (\ref{tent_ident})  by recalling that
the identification is induced by that of $\xi$ and $-1/\xi$.
Therefore the point $z_- = f(\xi_-)$ 
is identified with $z_+= f(\xi_+)$ when $\xi_+ = -1/\xi_-$.  This gives
\begin{equation}
z_-^2 = \frac{\xi_-^2}{1+\xi_-^2} = \frac{1}{1+\xi_+^2}
= 1-z_+^2 
\end{equation}
in agreement with (\ref{tent_ident}).

The surface $\mathcal{P}_\alpha$
associated with the butterfly projector
is obtained by a dilation 
$z\to (1+\alpha)^{1/2} z$ of $\mathcal{P}_0$,
as we have seen in (\ref{map_surfaces_scaled}).
Under this dilation the bounding curves
$C_0^+$ and $C_0^-$ in 
(\ref{c_curve}) become the curves  $C^+_\alpha$ and $C^-_\alpha$ whose points
satisfy
\be
\label{c_curve_dil}
z\in C^\pm_\alpha \quad \to \quad
\Re (z^2)
= {1\over 2} (1+ \alpha) \,.
\ee
 Their identification is obtained from
(\ref{tent_ident}) by the dilation:
\be
\label{ident_alapha}
z_+^2 +  z_-^2 = 1+ \alpha \,.
\ee
The surface $\mathcal{P}_\alpha$ is the region
between $C_\alpha^-$ and $C_\alpha^+$. The coordinate disk 
can be viewed as $\mathcal{P}_0$, without identifications,
inside $\mathcal{P}_\alpha$.  The surface $\mathcal{P}_\alpha$
is shown in Figure~\ref{orzf3}(a).

We can use the identification (\ref{ident_alapha}) to move
the region $(C_\alpha^-, C_0^-)$
to the right of $C_\alpha^+$.
Since points $z_-\in C_0^-$ satisfy $\Re(z_-^2) = 1/2$, 
(\ref{ident_alapha}) shows that
under the identification they become 
\be
\Re (z_+^2)   = {1\over 2} (1+ 2\alpha) \quad\to \quad
z_+ \in C^+_{2\alpha}\,,
\ee
where we have used (\ref{c_curve_dil}).
 The surface $\mathcal{P}_\alpha$ can therefore
be described
as the region between $C_0^-$ and $C_{2 \alpha}^+$, with these two curves
identified via (\ref{ident_alapha}).  This presentation is shown
in Figure~\ref{orzf3}(b).

\sectiono{Solutions from reparameterizations}

In this section
we construct the tachyon vacuum solution
associated with a general twist-invariant projector.
We begin 
\S 4.1 with a review
of the algebraic structure of Schnabl's solution.
We then give a formal construction
of the solution associated with a general projector
using reparameterizations.
In \S 4.2 we present
the CFT description of the states $\psi_\alpha$ and $\psi'_\alpha$
for a general projector. 
In the last subsection
we analyze the various operator insertions in more detail
and geometrically confirm
that they obey the expected algebraic properties.

\subsection{Review of the algebraic construction}

Schnabl's solution $\Psi$ consists of two pieces
and is defined by a limit:
\begin{equation}
\Psi \label{sliversolution}
= \lim_{N \to \infty} \Bigl[ \, - \psi_N+
\sum_{n=0}^N \psi'_n \, \Bigr] \,.
\end{equation}
The ``phantom piece'' $\psi_N$ does not contribute
to inner products
with states in the Fock space
in the limit. 
Namely,
\begin{equation}
\lim_{N \to \infty} \langle \, \phi , \psi_N \, \rangle = 0
\end{equation}
for any state $\phi$ in the Fock space.
On the other hand, the piece involving the sum of $\psi'_n$
is the limit $\lambda \to 1$ of a state $\Psi_\lambda$,
\begin{equation}
\Psi_\lambda \equiv \sum_{n=0}^\infty \lambda^{n+1} \, \psi'_n \, ,
\label{lambda}
\end{equation}
which formally satisfies the equation of motion for all $\lambda$,
\begin{equation}
Q_B \Psi_\lambda + \Psi_\lambda \ast \Psi_\lambda = 0 \, .
\label{Psi_lambda-equation}
\end{equation}
The state $\Psi_\lambda$ can be formally written
as a pure-gauge configuration \cite{Okawa:2006vm}
and is considered to be gauge-equivalent
to $\Psi=0$ for $|\lambda |< 1$.
The equation (\ref{Psi_lambda-equation}) for any $\lambda$
is equivalent to
the following relations for $\psi'_n$ with integer $n$:
\begin{eqnarray}
Q_B \psi'_0 &=& 0 \,,
\label{psi'_0-relation}
\\
Q_B \psi'_n
&=&- \sum_{m=0}^{n-1} \psi'_{m} \ast \psi'_{n-m-1} \,, \qquad  n>0 \,.  
\label{psi'_n-relation}
\end{eqnarray}

\medskip
 
There is a simple algebraic construction
of the states  $\psi'_n$, which we now review.
It helps to use the abstract notation of \cite{Rastelli:2006ap}
even though for the time being
all the operators are meant to be
those associated with the sliver.
The left and right parts of the operator $L^+ = L + L^\star$
are denoted by $L^+_L$ and $L^+_R$, respectively,
and $L^+ = L^+_L+ L^+_R$.
The operator $K \equiv \widetilde{L}^+$ is defined by
$K = \widetilde{L}^+ = L^+_R - L^+_L$.
For the sliver, its explicit form derived
in \cite{Schnabl:2005gv} is
\begin{equation}
K = \widetilde{L}^+
= \frac{\pi}{2} \, K_1
= \frac{\pi}{2} \, \left( \, L_1 + L_{-1} \, \right) \,.
\end{equation}
The antighost operators
$B$, $B^\star$, $B^+ = B + B^\star$,
$\widetilde B^+ = B^+_R - B^+_L$ are
similarly defined by replacing
$T(z) \to b(z)$ or $L_n \to b_n$.
Thus for the sliver,
\begin{equation} \label{B1}
\widetilde{B}^+ 
= \frac{\pi}{2} \, \left( \, b_1 + b_{-1} \, \right) \,.
\end{equation}
In this language, we can write 
\begin{eqnarray} \label{defpsi0}
\psi_0 & = & C |P_1 \rangle \, ,
\\
\label{defpsin}
\psi_n & = & {}- C | P_1 \rangle \ast | P_{n-1} \rangle
\ast B^+_L C | P_1 \rangle \, , \quad n > 0\,,
\end{eqnarray}
as well as
\begin{eqnarray}
\psi'_0 &=& - Q_B B^+_L C | P_1 \rangle
\label{psip0alg} \\
\psi'_n &=& C | P_1 \rangle \ast | P_{n-1} \rangle
\ast B^+_L L^+_L C | P_1 \rangle \, , \quad n > 0 \, ,
\label{algebraic-definition}
\end{eqnarray}
where the 
 operator $C$ is
\begin{equation} \label{C1}
C \equiv \frac{2}{\pi} \, c_1 \,.
\end{equation}
Again, at this stage  all objects are defined in 
the sliver frame.
In particular,
$| P_\alpha \rangle$ is the wedge state $|W _\alpha \rangle$
and $|P_1 \rangle$ is just
the $SL(2,R)$-invariant vacuum $\ket{0}$. 

\medskip

It was algebraically shown in \cite{Okawa:2006vm} that
the string fields $\psi'_n$
defined by (\ref{psip0alg}) and (\ref{algebraic-definition})
satisfy (\ref{psi'_0-relation}) and (\ref{psi'_n-relation}).
In the proof, one uses
the abelian algebra $P_\alpha * P_\beta = P_{\alpha + \beta}$,
standard properties of the BRST operator
($Q_B$ is a nilpotent derivation
of the star algebra and annihilates the vacuum state),
as well as the following identities:
\begin{eqnarray}
&& \widetilde{B}^+ | P_1 \rangle
= ( B^+_R - B^+_L ) \, | P_1 \rangle = 0 \,, \
\phantom{\Big|} \label{iBp}
\\
&& \widetilde{B}^+ C | P_1 \rangle
= ( B^+_R - B^+_L ) \, C | P_1 \rangle = | P_1 \rangle \, ,
\phantom{\Big|}
\label{C-condition} \\
&& ( B^+_R \phi_1 ) \ast \phi_2
= (-1)^{\phi_1} \phi_1 \ast ( B^+_L \phi_2 ) \,.
\phantom{\Big|} \label{iRL}
\end{eqnarray}
The first two equations (\ref{iBp}) and (\ref{C-condition}) are
immediately checked using $|P_1 \rangle= \ket{0}$ and the
expansions (\ref{B1}) and (\ref{C1}).
The identity (\ref{iBp})
can also be understood as a special case of the
familiar conservation laws obeyed by wedge states,
\begin{equation}
\label{familiarcl}
\begin{split}
\widetilde{L}^+ | P_\alpha \rangle
& = ( L^+_R - L^+_L) | P_\alpha \rangle = 0 \,, \\
\widetilde{B}^+ | P_\alpha \rangle
& = ( B^+_R - B^+_L) | P_\alpha \rangle = 0 \,.
\end{split} 
\end{equation}
The last identity (\ref{iRL})
is obtained by observing that for any
 derivation $D = D_L + D_R$ one has
\be
( D_R \phi_1 ) \ast \phi_2
= {}- (-1)^{\phi_1\cdot D} \phi_1 \ast ( D_L \phi_2 ) \,.
\ee
For $D=\widetilde{B}^+$
we find (\ref{iRL}),
while for $D= K$ we obtain  
\be \label{lastid}
( L^+_R \phi_1 ) \ast \phi_2
=   \phi_1 \ast ( L^+_L \phi_2 ) \,.
\ee
Let us confirm  that $\psi'_n$ as defined
in (\ref{algebraic-definition}) is indeed the derivative with respect to $n$
of the state $\psi_n$ in (\ref{defpsin}).  
Since $| P_\alpha \rangle
= e^{-\frac{\alpha}{2} L^+} | {\cal I} \rangle $, 
we have
\begin{equation}
 \frac{d}{d \alpha} | P_\alpha \rangle
= -\frac{1}{2} \, L^+ | P_\alpha \rangle
= {}- L^+_R \, | P_\alpha \rangle \,,
\end{equation}
where we have used (\ref{familiarcl}).
With the help of (\ref{lastid}) we find that
\be
\frac{d}{dn}\psi_n = C |P_1 \rangle * L^+_R |P_{n-1} \rangle * B^+_L C |P_1 \rangle = 
C |P_1 \rangle *  |P_{n-1} \rangle * L^+_L B^+_L C |P_1 \rangle \, ,
\ee 
as claimed.
Note that $L^+_L$ and $B^+_L$ commute
because $L^+_L = \{ Q_B, B^+_L \}$ and $(B^+_L)^2 = 0$.

\smallskip

One can also show that the solution satisfies
the gauge condition $B \Psi = 0$.
The algebraic properties that guarantee this fact are
\begin{eqnarray}
\label{reqs_on_C}
&& \{ B \,, C\}= \{ B^\star \,, C\} = ~0 \,,\phantom{\Big|} \\
&& L \, C \, |P_1\rangle
= {}- C \, |P_1 \rangle \phantom{\Big|} , \nonumber
\end{eqnarray}
which follow immediately from 
the mode expansions on $B$,  $L$, and $C$ in the sliver frame.   
To show that (\ref{reqs_on_C})  imply $B \psi_n = B \psi'_n = 0$,
the following identities are useful.
Writing $B = {1\over 2} ( B^- + B^+_L + B^+_R)$,  one can prove that
\be
\label{prod_gf}
B (\psi_1 * \psi_2) =B \psi_1 * \psi_2 + (-1)^{\psi_1}
 \psi_1 * (B-B^+_L)\psi_2 \,.
\ee
For a larger number of factors we have
\be
\label{prod_gf_full}
B (\psi_1 * \psi_2 * \ldots \psi_n) =(B \psi_1) *\ldots *\psi_n
+\sum_{m=2}^n  (-)^{\sum_{k=1}^{m-1}\psi_k}
 ~\psi_1 * \ldots  *(B-B^+_L) \psi_m  *\ldots * \psi_n \,.
\ee
One can actually make
manifest the fact that $\psi'_n$ is annihilated by $B$
in the following way:
\be
\psi'_n = {1\over n}  B \Bigl( 
C \ket{P_1} \ast \ket{P_{n-1}} \ast ( L^+_L + {\textstyle{1\over n}}) C \ket{P_1}\Bigr) \, .
\ee

\medskip

We have seen in the previous section that
a generic single-split projector $P_\infty$ can be related
to the sliver $W_\infty$
by a reparameterization $\varphi$
as $P_\infty = U_\varphi^{-1} W_\infty$.
This allowed us to construct the abelian family $P_\alpha$
 from the wedge states by the same transformation
$P_\alpha \equiv U_\varphi^{-1} W_\alpha$.
We now proceed to define operators associated with $P_\infty$
by similarity transformations of the corresponding operators
associated with the sliver. From now on
we use the subscript ${\cal S}$ to denote 
objects in the sliver frame, and objects without the subscript
are those in the frame of $P_\infty$.
We have
\begin{eqnarray}
C & \equiv & U_\varphi^{-1} C_{\cal S} \, U_\varphi \, ,
\label{Csim}\\
L  & \equiv & U_\varphi^{-1} L_{\cal S} \, U_\varphi \, ,
\label{Lsim}\\ 
L^\star  & \equiv & U_\varphi^{-1} L_{\cal S}^\star \, U_\varphi \, ,
\label{Lssim}\\ 
L^\pm  & \equiv & U_\varphi^{-1} L^\pm_{\cal S} \, U_\varphi
= L \pm L^\star \, ,
\label{Lpmsim} \\ 
L^+_L & \equiv & U_\varphi^{-1} (L^+_L)_{\cal S} \, U_\varphi \, ,
\label{Lplsim} \\
L^+_R  & \equiv & U_\varphi^{-1} (L^+_R)_{\cal S} \, U_\varphi \,
\label{Lprsim} ,
\end{eqnarray}
and  analogous expressions for the antighost operators $B$, $B^\star$, $B^\pm$, $B^+_R$, $B^+_L$.
Because of the formal property (\ref{U1}), $L^\star$ in (\ref{Lssim}) {\it is} 
the BPZ conjugate of $L$ in (\ref{Lsim}), so our notation is consistent.
It is also consistent to use
$L^+_L$ and $L^+_R$ in (\ref{Lplsim}) and (\ref{Lprsim})
since reparameterizations preserve
the left/right decomposition of operators.
As we will see explicitly in \S 4.3,
the operators $L^+_L$ and $L^+_R$
are, respectively, the left and right parts
of the operator $L^+$ defined in (\ref{Lpmsim}).
It is also obvious that
all the algebraic properties
(\ref{iBp}), (\ref{C-condition}), (\ref{iRL}),
(\ref{familiarcl}), and (\ref{reqs_on_C})
are obeyed by the operators in the frame of $P_\infty$.

The states $\psi_n$ associated with $P_\infty$ are given by
\be
\psi_n \equiv U_\varphi^{-1} \, \psi_{n \; \cal S}= - C |P_1\rangle  * |P_{n-1}  \rangle * B^+_L   C |P_{1} \rangle \, ,
\ee
and $\psi'_n$ associated with $P_\infty$ are similarly obtained.
Finally, the solution $\Psi$ associated with $P_\infty$
is obtained from the sliver's solution $\Psi_{\cal S}$ as
\be
 \Psi = U_\varphi^{-1} \Psi_{\cal S} \, .
 \ee
Clearly, it takes the same form  (\ref{sliversolution}), 
with the understanding that the states $\psi'_n$ and $\psi_N$ 
are now those in the frame of $P_\infty$. 

\subsection{Solutions in the CFT formulation}

We now translate the above formal construction
into a geometric description.
In the CFT formulation,
the state $\psi_{n  \; \cal S}$ in the sliver frame 
 is defined by
\begin{equation} \label{psinScft}
\langle \, \phi, \psi_{n \, \cal S}\, \rangle
= \biggl\langle \, f_{\cal S} \circ \phi (0) \,\, c (1)
\int_{-V_\alpha^+} \frac{dz}{2 \pi i} \, b(z) \,\, c (n+1) \,
\biggr\rangle_{\mathcal{W}_{n+1}}\,,
\end{equation}
for any state $\phi$ in the Fock space,
where $1 < \alpha < 2 \, n+1$. A pictorial representation
of the correlator is given in Figure~\ref{orzf6}.  
The contour $V_\alpha^+$ is oriented in the direction of increasing imaginary
$z_{\cal S}$, and by $-V_\alpha^+$ we denote the same contour
with opposite orientation.
The expression (\ref{psinScft}) is the direct geometric translation
of the algebraic expression (\ref{defpsin}),
as explained in detail in~\cite{Okawa:2006vm}.
Recall the change in the normalization of $f_{\cal S}$.

\begin{figure}
\centerline{\hbox{\epsfig{figure=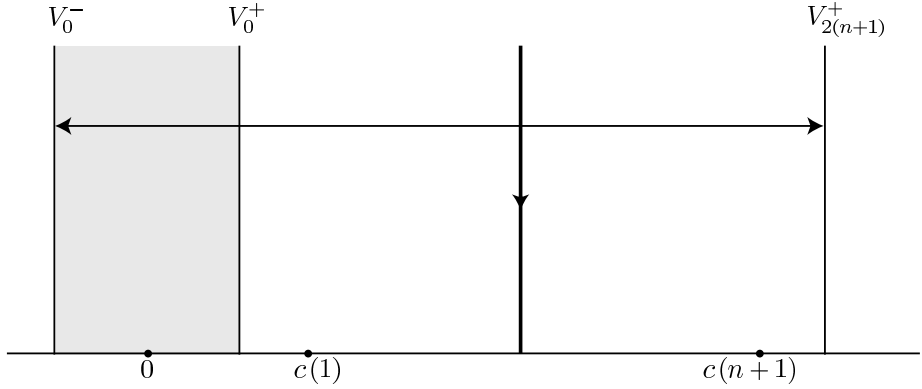, height=5.1cm}}}
\caption{A diagram of the correlator on $\mathcal{W}_{n+1}$ used in
 (\ref{psinScft}) to describe the solution in the sliver frame. Shown are
 ghost insertions at $z_S=1$ and $z_S = n+1$. The vertical
 line in between these insertions represents the antighost line integral.}
\label{orzf6}
\end{figure}

Let us apply the reparameterization $U_\varphi^{-1}$ to the state
$\psi_{n \; \cal S}$.
Geometrically, this amounts to mapping
the region $(V_0^+, V^+_{2(n+1)})$,
including the operator insertions,
by the conformal transformation $R^{-1}$
used to construct the state $| P_{n+1} \rangle$ from
the wedge state $| W_{n+1} \rangle$.
It is straightforward to calculate
the  transformations of the operator insertions in (\ref{psinScft}).
We find that the state $\psi_n$ associated
with a general projector is given by
\begin{equation} 
\langle \, \phi, \psi_n \, \rangle
= \langle \, f \circ \phi (0) \,\,
{\cal C} (1) \, {\cal B} \, {\cal C} (2 \, n+1) \,
\rangle_{\mathcal{P}_{n+1}}
\label{psi_n}
\end{equation}
for any state $\phi$ in the Fock space, where
\begin{equation}
{\cal C} (\alpha) \equiv
R' \left( R^{-1} \left( \frac{1+\alpha}{2} \right) \right) \,
c \left( R^{-1} \left( \frac{1+\alpha}{2} \right) \right) \,,
\qquad
{\cal B} \equiv \int \frac{dz}{2 \pi i} \,
\frac{b(z)}{R'(z)} \, .
\end{equation}
The contour of the integral for ${\cal B}$
can be taken to be $-C_\alpha^+$
with $1 < \alpha < 2 \, n+1$. 
(The orientation of the contour $C_\alpha^+$,
inherited from the orientation of $V_\alpha^+$,  is 
 directed towards increasing imaginary $z$).
In general, when ${\cal B}$ is located between two operators,
the contour of the integral must run between the two operators.
Note that ${\cal C} (\alpha)$ is nothing but 
the operator $c(z_{\cal S})$,
with $z_{\cal S} = {1\over 2}(1+ \alpha)$,
expressed in the  frame $z = R^{-1} (z_{\cal S})$. The argument  
$\alpha$ of $\cal{C}$ denotes the label of the line $C_\alpha^+$ that contains
the insertion.  The surface and insertions for the correlator indicated in (\ref{psi_n}) 
are shown in Figure~\ref{orzf7}.

\begin{figure} 
\centerline{\hbox{\epsfig{figure=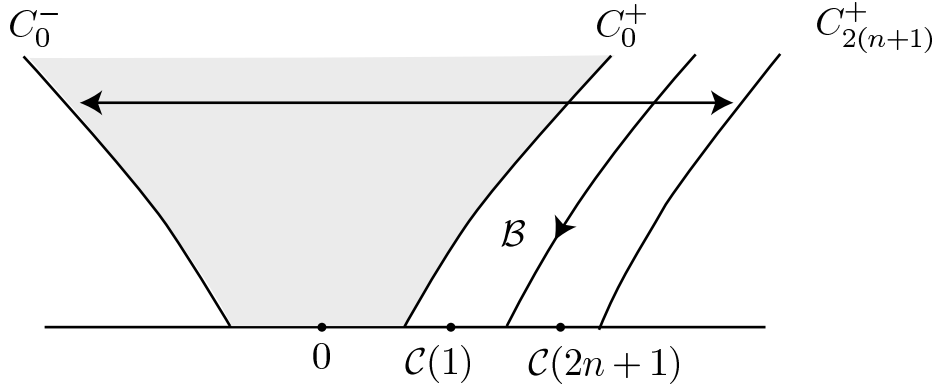, height=5.1cm}}}
\caption{The surface and insertions relevant to the correlator (\ref{psi_n})
used to define $\psi_n$. The surface $\mathcal{P}_{n+1}$ includes two
ghost insertions $\cal C$ and an antighost line integral $\cal B$.}
\label{orzf7}
\end{figure}

This definition of $\psi_n$ is valid for $n > 0$,
and $\psi_0$ can be defined by the limit $n \to 0$:
\begin{equation}
\psi_0 \equiv \lim_{n \to 0} \psi_n \,.
\end{equation}
Let us calculate $\psi_0$ explicitly.
The anticommutation relation of ${\cal B}$ and ${\cal C}$
is given by
\begin{equation}
\{ \, {\cal B} \,,\, {\cal C} (\alpha) \, \}
= {\cal B} \,{\cal C} (\alpha) + {\cal C} (\alpha) \, {\cal B}
= 1 \,.
\label{anticommutation}
\end{equation}
Note that the contour for ${\cal B}$
in the term ${\cal B} \,{\cal C} (\alpha)$
should be $- C_\beta^+$ with $\beta < \alpha$,
and the contour for ${\cal B}$
in the term ${\cal C} (\alpha) \, {\cal B}$
should be $- C_\beta^+$ with $\beta > \alpha$.
Using this anticommutation relation,
the inner product $\langle \, \phi, \psi_n \, \rangle$
in the limit $n \to 0$ is given by
\begin{equation}
\lim_{n \to 0} \langle \, \phi, \psi_n \, \rangle
= \langle \, f \circ \phi (0) \, {\cal C} (1) \,
\rangle_{\mathcal{P}_1} \,.
\end{equation}
This gives the CFT description of the state
$\psi_0 = C | P_1 \rangle $ in (\ref{defpsi0})
for a general projector.
It coincides with the state obtained by reparameterization
from  the sliver's $\psi_0$.

Another useful expression for the inner product $\langle \, \phi, \psi_n \, \rangle$
is
\begin{equation}
\label{useful_form}
\langle \, \phi, \psi_n \, \rangle
= {}- R'( R^{-1} (1) )^2 \,
\biggl\langle \, c( -R^{-1} (1) ) \,\,
f \circ \phi (0) \,\,
c( R^{-1} (1) )
\int_{C_\alpha^+} \frac{dz}{2 \pi i} \,
\frac{b(z)}{\widehat{R}' (z)} \,
\biggr\rangle_{\mathcal{P}_{n+1}} ,
\end{equation}
where $\alpha > 1$,
and we have mapped the operator ${\cal C} (2 \, n+1)$ to
$\widehat{R}'( \widehat{R}^{-1} (-1) ) \,
c( \widehat{R}^{-1} (-1) )
= R'( R^{-1} (1) ) \, c( -R^{-1} (1) )$
using the identification (\ref{indent_after_rep})
for the surface $\mathcal{P}_{n+1}$.
Note that $\phi$ must be Grassmann even
in order for the inner product to be nonvanishing.
We will use (\ref{useful_form}) in the next section.

Let us now consider $\psi'_n$.
Taking a derivative of $\psi_{n  \; \cal S}$ with respect to $n$
is equivalent to an insertion of the operator
\begin{equation}
\int_{-V_\alpha^+} \frac{dz}{2 \pi i} \, T(z)
\end{equation}
in (\ref{psinScft}), with $1 < \alpha < 2 \, n+1$.
See \cite{Okawa:2006vm} for more details.
Since the operator is transformed by  $R^{-1}$ to
\begin{equation}
{\cal L} \equiv \int \frac{dz}{2 \pi i} \,
\frac{T(z)}{R'(z)} \,,
\end{equation}
the geometric translation of (\ref{algebraic-definition}) 
for a general projector is
\begin{equation} \label{psipnfirst}
\langle \, \phi, \psi'_n \, \rangle
= \langle \, f \circ \phi (0) \,\,
{\cal C} (1) \, {\cal L} \, {\cal B} \, {\cal C} (2 \, n+1) \,
\rangle_{\mathcal{P}_{n+1}} \,,
\end{equation}
where the contour of the integral for ${\cal L}$
can be taken to be $-C_\alpha^+$ with $1 < \alpha < 2 \, n+1$.
The surface and insertions for this correlator are shown in
Figure~\ref{orzf8}.  

\begin{figure}  
\centerline{\hbox{\epsfig{figure=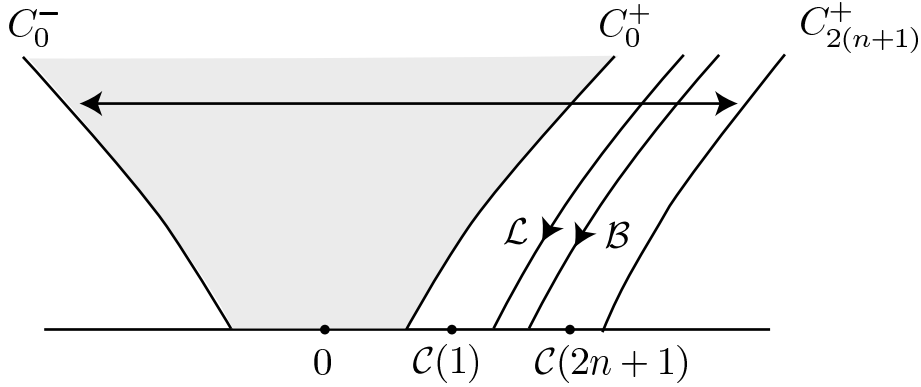, height=5.1cm}}}
\caption{The surface and insertions relevant to the correlator (\ref{psipnfirst})
used to define $\psi'_n$. The surface $\mathcal{P}_{n+1}$ includes two
ghost insertions $\cal C$, an antighost line integral $\cal B$, and 
a stress-tensor line integral $\cal L$.}
\label{orzf8}
\end{figure}

Note that ${\cal B}$ and ${\cal L}$ commute.
In general, when ${\cal L}$ is located between two operators,
the contour of the integral must run between the two operators.
The definition  (\ref{psipnfirst}) is valid for $n$ in the range $n > 0$.
As in the case of $\psi_0$, the state $\psi'_0$ can be
defined by the limit $n \to 0$:
\begin{equation}
\psi'_0 = \lim_{n \to 0} \psi'_n \,.
\end{equation}
Using the anticommutation relation (\ref{anticommutation}),
the inner product $\langle \, \phi, \psi'_n \, \rangle$
can be written as
\begin{equation} 
\begin{split} 
\langle \, \phi, \psi'_n \, \rangle
& = \langle \, f \circ \phi (0) \,\,
{\cal C} (1) \, {\cal B} \, {\cal L} \, {\cal C} (2 \, n+1) \,
\rangle_{\mathcal{P}_{n+1}}
\\
& = \langle \, f \circ \phi (0) \,\,
{\cal L} \, {\cal C} (2 \, n+1) \,
\rangle_{\mathcal{P}_{n+1}}
- \langle \, f \circ \phi (0) \,\,
{\cal B} \, {\cal C} (1) \, {\cal L} \, {\cal C} (2 \, n+1) \,
\rangle_{\mathcal{P}_{n+1}} \,.
\end{split}
\end{equation}
It is trivial to take the limit $n \to 0$ for the first term.
The limit of the second term can be calculated
using the formula
\begin{equation}
\lim_{\epsilon \to 0} {\cal C} (\alpha) \,
{\cal L} \, {\cal C} (\alpha+\epsilon)
= \lim_{\epsilon \to 0} {\cal C} (\alpha) \,
[ \, {\cal L}, {\cal C} (\alpha+\epsilon) \, ]
= Q_B \cdot {\cal C} (\alpha) \,,
\label{CLC}
\end{equation}
where $Q_B \cdot {\cal O}$
is the BRST transformation of ${\cal O}$.
The inner product $\langle \, \phi, \psi'_0 \, \rangle$ is thus
\begin{equation}
\langle \, \phi, \psi'_0 \, \rangle
= \langle \, f \circ \phi (0) \,\,
{\cal L} \, {\cal C} (1) \,
\rangle_{\mathcal{P}_{1}}
- \langle \, f \circ \phi (0) \,\,
{\cal B} \,\, Q_B \cdot {\cal C} (1) \,
\rangle_{\mathcal{P}_{1}} \,.
\end{equation}
This gives the geometric translation of the state
$\psi'_0 = - L^+_L C | P_1 \rangle + B^+_L Q_B C | P_1 \rangle
= {}- Q_B B^+_L C | P_1 \rangle$ in (\ref{psip0alg})
for a general projector,
as we will explain further in the next subsection.
The state coincides with the state obtained
by reparameterization from the sliver's  $\psi'_0$.

\subsection{Operator insertions in the geometric language}

The expressions of $\psi_n$ and $\psi'_n$
in (\ref{psi_n}), (\ref{useful_form}), and (\ref{psipnfirst})
are the central results of this section.
While the solution constructed from these states
are guaranteed to satisfy the equation of motion
because it is related to Schnabl's solution
by a reparameterization,
it is also possible to confirm this directly
without referring to the reparameterization.
In this subsection we offer a more detailed analysis
of how various operator insertions are presented
in the CFT formulation.
It is then straightforward to confirm
that the equation of motion is satisfied
using the formulas in this subsection.
The techniques developed in this subsection
will be useful in handling operator insertions
in the conformal frame of a general projector.

Let us begin with the operator $L$.
It is, by definition, obtained from $L_{\cal S}$ 
by the reparameterization $\varphi$,
where $\varphi$ is implicitly defined
by the relation $\widehat R ( f(t) )
= f_{\cal S} (  \varphi(t) )$
in (\ref{Rf}).
The function $f_{\cal S} (t)$
becomes 
$f_{\cal S} (  \varphi(t) ) = \widehat R ( f(t) )$,  
and thus $L$ in the general projector frame $z = f(\xi)$
is given by $L_{\cal S}$ in the sliver frame
$z_{\cal S} = f_{\cal S} (\xi_{\cal S})$
by the conformal transformation
$z = \widehat{R}^{-1} (z_{\cal S})$:
\begin{equation} \label{Lrepa}  
\begin{split}
L  & \equiv  U_{\varphi}^{-1} L_{\cal S} \, U_{\varphi}
= U_{\varphi}^{-1} \left( \int_{V_0^+-V_0^-}
\frac{d z_{\cal S}}{2 \pi i} \, z_{\cal S} \, T(z_{\cal S})
\right) U_{\varphi}
= \int_{C_0^+-C_0^-} \frac{d z}{2 \pi i} \,
\frac{\widehat{R} (z)}{\widehat{R}' (z)} \, T(z)
\\
&=   \int_{C_0^+} \frac{dz}{2 \pi i} \,
\frac{R(z)}{R'(z)} \, T(z)
+ \int_{C_0^-} \frac{dz}{2 \pi i} \,
\frac{R(-z)}{R'(-z)} \, T(z) \,. 
\end{split}
\end{equation}
In obtaining the second line we made use of (\ref{iothj}).  
For special projectors, $R(z) = z^s$
and the expression for $L$ simplifies to
\be \label{LL0}
L = \frac{1}{s} \oint \frac{dz}{2 \pi i} \, z \, T(z)
= \frac{ {\cal L}_0 }{s} \, .
\ee
The operator ${\cal L}_0$ is the Virasoro zero mode in the frame
of the projector. This is the definition of $L$ given
in \cite{Rastelli:2006ap}.
If the projector is not special,
(\ref{LL0}) does not hold. Generically the expansion
of $L$ in ordinary Virasoro operators $L_n$ contains
terms with negative $n$.

The inner product
$\langle \, L \, \phi, P_\alpha \, \rangle$
for any state $\phi$ in the Fock space is given by
\be
\label{lo_as_surf}
\begin{split}
\langle \, L \, \phi, P_\alpha \, \rangle
&= \biggl\langle \, \int_{C_0^+-C_0^-} \frac{d z}{2 \pi i} \,
\frac{\widehat{R} (z)}{\widehat{R}' (z)} \, T(z) \,
f \circ \phi (0) \, \biggr\rangle_{\mathcal{P}_\alpha}
\\
&= \biggl\langle \, \int_{C_0^+} \frac{d z}{2 \pi i} \,
\frac{R (z)}{R' (z)} \, T(z) \,
f \circ \phi (0) \, \biggr\rangle_{\mathcal{P}_\alpha}
+ \biggl\langle \, \int_{C_0^-} \frac{d z}{2 \pi i} \,
\frac{R (-z)}{R' (-z)} \, T(z) \,
f \circ \phi (0) \, \biggr\rangle_{\mathcal{P}_\alpha} .
\end{split}
\ee
This provides the CFT representation of the state
$L^\star \, | P_\alpha \rangle$
because
$\langle \, L \, \phi, P_\alpha \, \rangle
= \langle \, \phi, L^\star \, P_\alpha \, \rangle$.

Next, we wish to derive a representation of
$L \, | P_\alpha \rangle$.
To this end, we
 need an expression for
$\langle \, L^\star \, \phi, P_\alpha \, \rangle$.
While it is possible to construct
$L^\star$ from $L^\star_{\cal S}$ by the reparameterization $\varphi$
as in (\ref{Lrepa}),
it is instructive to understand BPZ conjugation
directly on the surface $\mathcal{P}_\alpha$.
BPZ conjugation is, by definition, performed  
by the map $I(\xi) = -1/\xi$ in the $\xi$ coordinate.
For an operator in the $z$-plane, 
BPZ conjugation requires mapping
the operator to the $\xi$ coordinate, performing the conjugation,
and mapping the resulting operator back to the $z$ coordinate.
The full conformal transformation is then
\be
z' = I_f (z) = f \circ I \circ f^{-1} (z)\,, \quad  I(\xi) = -1/\xi\,.
\ee
This relation between $z'$ and $z$ is nothing but
the identification between $z_+$ and $z_-$ for $\mathcal{P}_0$,
namely, 
\be
\label{gnkur}
R(z_+) + R(-z_-) = 1\,.
\ee
Let us apply this geometric understanding of
BPZ conjugation to the operator $L$.
The map $I_f$ transforms the 
two integrals in (\ref{Lrepa}) as follows:   
\begin{equation}
\label{983fd}
\begin{split}
\int_{C_0^+} \frac{dz_+}{2 \pi i} \,
\frac{R(z_+)}{R'(z_+)} \, T(z_+)
~&~\to {}- \int_{C_0^-} \frac{dz_-}{2 \pi i} \,
\frac{R(-z_-)}{R'(-z_-)} \, T(z_-)
+ \int_{C_0^-} \frac{dz_-}{2 \pi i} \,
\frac{T(z_-)}{R'(-z_-)} \,,
\\[1.0ex]
\int_{C_0^-} \frac{dz_-}{2 \pi i} \,
\frac{R(-z_-)}{R'(-z_-)} \, T(z_-)
~&~\to {}- \int_{C_0^+} \frac{dz_+}{2 \pi i} \,
\frac{R(z_+)}{R'(z_+)} \, T(z_+)
+ \int_{C_0^+} \frac{dz_+}{2 \pi i} \,
\frac{T(z_+)}{R'(z_+)} \,.
\end{split}
\end{equation}
Thus the inner product
$\langle \, L^\star \, \phi, P_\alpha \, \rangle$
is given by 
\begin{equation}
\label{lostar_cft}
\langle \, L^\star \, \phi, P_\alpha \, \rangle
= {}- \biggl\langle \, \int_{C_0^+- C_0^-} \frac{dz}{2 \pi i} \,
\frac{\widehat{R} (z)}{\widehat{R}' (z)} \, T(z) \,
f \circ \phi (0) \, \biggr\rangle_{\mathcal{P}_\alpha}
+ \biggl\langle \, \int_{C_0^++C_0^-} \frac{dz}{2 \pi i} \,
\frac{T(z)}{\widehat{R}' (z)} \,
f \circ \phi (0) \, \biggr\rangle_{\mathcal{P}_\alpha} \, .
\end{equation}
Recalling (\ref{lo_as_surf}), we can write
\be
\langle \, L^\star \, \phi, P_\alpha \, \rangle=
 {}-\langle \, L \, \phi, P_\alpha \, \rangle
+ \biggl\langle \, \int_{C_0^++C_0^-} \frac{dz}{2 \pi i} \,
\frac{T(z)}{\widehat{R}' (z)} \,
f \circ \phi (0) \, \biggr\rangle_{\mathcal{P}_\alpha} .
\end{equation}
It immediately follows that
\be \label{Lpint}
\langle \, L^+ \phi, P_\alpha \, \rangle
= \langle \,
( \, L+L^\star \, ) \, \phi,
P_\alpha \, \rangle
= \biggl\langle \,
\int_{C_0^++C_0^-} \frac{dz}{2 \pi i} \,
\frac{T(z)}{\widehat{R}' (z)} \,
f \circ \phi (0) \, \biggr\rangle_{\mathcal{P}_\alpha} ,
\ee
and thus the operator $L^+$ is
\begin{equation}
L^+ = \int_{C_0^+} \frac{dz}{2 \pi i} \,
\frac{T(z)}{R'(z)} \,
+ \int_{C_0^-} \frac{dz}{2 \pi i} \,
\frac{T(z)}{R'(-z)} \,.
\end{equation}
From these expressions, we easily confirm the algebra
$[ \, L, L^\star \, ] = L + L^\star$, 
\begin{equation}
\begin{split}
[ \, L, L^\star \, ]
= [ \, L, L + L^\star \, ]
&= \int_{C_0^++C_0^-} \frac{dw}{2 \pi i} \,
\frac{1}{\widehat{R}'(w)}
\oint \frac{dz}{2 \pi i} \,
\frac{\widehat{R} (z)}{\widehat{R}' (z)} \, T(z) \, T(w)
\\
&= \int_{C_0^++C_0^-} \frac{dw}{2 \pi i} \,
\frac{T(w)}{\widehat{R}' (w)}
= L + L^\star \,,
\end{split}
\end{equation}
where the contour of the integral of $z$
encircles $w$ counterclockwise,
and we have neglected surface terms
of the form $\widehat{R} (w) \, T(w) / \widehat{R}' (w)^2$
for integration by parts with respect to $w$.
Whether or not the surface terms vanish
should be checked for a given $\widehat{R} (z)$
by evaluating them in a coordinate
where the midpoint of the open string is
located at a finite point.

We now consider the operators $L^+_L$ and $L^+_R$.
Since  $C_0^-$ and $C_0^+$ are respectively the left and right
parts of the coordinate curve,
the expression in (\ref{Lpint}) splits as follows:
\be
\label{lrdecLplus}
\begin{split}
\langle \, L^+_R \, \phi, P_\alpha \, \rangle
&= \biggl\langle \, \int_{C_0^+} \frac{dz}{2 \pi i} \,
\frac{T(z)}{R'(z)} \,
f \circ \phi (0) \, \biggr\rangle_{\mathcal{P}_\alpha} ,
\\[1.0ex]
\langle \, L^+_L \, \phi, P_\alpha \, \rangle
&= \biggl\langle \, \int_{C_0^-} \frac{dz}{2 \pi i} \,
\frac{T(z)}{R'(-z)} \,
f \circ \phi (0) \, \biggr\rangle_{\mathcal{P}_\alpha} .
\end{split}
\ee
The BPZ conjugation map $I_f$ acts as
\begin{equation}
\label{if_l-s}
I_f : \quad \int_{C_0^+} \frac{dz_+}{2 \pi i} \,
\frac{T(z_+)}{R'(z_+)}
\to \int_{C_0^-} \frac{dz_-}{2 \pi i} \,
\frac{T(z_-)}{R'(-z_-)} \, ,
\end{equation}
so we see
\be
\label{bpz_on_L_plus_R}
(L^+_R)^\star = L^+_L\,.
\ee
Since BPZ conjugation is an involution,   
we also have $(L^+_L)^\star = L^+_R$. 

Using  the presentation of
$\mathcal{P}_\alpha$
as the region between $C_0^-$ and $C_{2\alpha}^+$ and recalling that these curves
are identified by (\ref{indent_after_rep}),
we can rewrite $\langle \, L^+_L \, \phi, P_\alpha \, \rangle$
in (\ref{lrdecLplus}) as
\begin{equation}
\label{gvnjeorh}
\langle \, L^+_L \, \phi, P_\alpha \, \rangle
= \biggl\langle \, \int_{C_{2 \alpha}^+} \frac{dz}{2 \pi i} \,
\frac{T(z)}{R'(z)} \,
f \circ \phi (0) \, \biggr\rangle_{\mathcal{P}_\alpha} .
\end{equation}
Since $(L^+_R)^\star = L^+_L$ and $(L^+_L)^\star = L^+_R$,
the inner products
$\langle \, \phi, L^+_R P_\alpha \, \rangle$
and  $\langle \, \phi, L^+_L P_\alpha \, \rangle$ are given by
\be
\label{L^+_RL-P_alpha}
\begin{split}
\langle \, \phi, L^+_R P_\alpha \, \rangle
&= \biggl\langle \, f \circ \phi (0) \,
\int_{C_{2 \alpha}^+} \frac{dz}{2 \pi i} \, \frac{T(z)}{R'(z)} \,
\biggr\rangle_{\mathcal{P}_\alpha} ,
\\[1.0ex]
\langle \, \phi, L^+_L P_\alpha \, \rangle
&= \biggl\langle \, f \circ \phi (0) \,
\int_{C_0^+} \frac{dz}{2 \pi i} \, \frac{T(z)}{R'(z)} \,
\biggr\rangle_{\mathcal{P}_\alpha} .
\end{split}
\ee
We see that the states $L^+_R | P_\alpha \rangle$
and $L^+_L | P_\alpha \rangle$ are both
represented as the region between $C_0^+$ and $C_{2 \alpha}^+$
with the  same operator  
 inserted on different locations:
it is on the right edge for $L^+_R | P_\alpha \rangle$
and on the left edge for $L^+_L | P_\alpha \rangle$.
Since there are no operator insertions in the region
between $C_0^+$ and $C_{2 \alpha}^+$,
the contour $C_{2 \alpha}^+$
can be deformed to $C_0^+$,
and we confirm that the states are the same:
\be
\label{slfdkjb}
L^+_R| P_\alpha \rangle=   L^+_L  \, | P_\alpha \rangle\,.
\ee

\medskip
Let us next consider the star multiplication
of states with insertions of $L^+_R$ or $L^+_L$.
We take $P_\alpha \ast (L^+_L P_\beta)$ as an example,
but the generalization to other cases is straightforward.
The operator $L_L^+$ of $L_L^+ P_\beta$ is represented
by an integral over $C_0^+$ on ${\cal P}_\beta$
in (\ref{L^+_RL-P_alpha}).
For the gluing of the star product we need the identification of curves
in two different coordinate systems. A curve  $C_q^+$
in the $z_<$ coordinate
is mapped to $C_{q+\gamma}^+$
in the $z_>$ coordinate
when $z_<$ and $z_>$ are related~by
\begin{equation}
R(z_>) = R(z_<) + \frac{\gamma}{2} \,.
\end{equation}
Under this identification
the operator insertion in (\ref{L^+_RL-P_alpha})
takes the same form in the two coordinates:
\begin{equation}
\int_{C_q^+} \frac{dz_<}{2 \pi i} \, \frac{T(z_<)}{R'(z_<)} \, 
= \int_{C_{q+\gamma}^+} \frac{dz_>}{2 \pi i} \,
\frac{T(z_>)}{R'(z_>)} \,.
\end{equation}
The operator integrated over $C_0^+$ on ${\cal P}_\beta$
is thus mapped to the same operator
integrated over $C_{2 \alpha}^+$ on the surface
${\cal P}_{\alpha+\beta} = (C_0^-, C_{2 \alpha+ 2 \beta}^+)$
for the star product $P_\alpha \ast (L^+_L P_\beta)$.
It follows from the first equation in (\ref{L^+_RL-P_alpha})
that the star product can also be interpreted
as $(L^+_R  P_\alpha ) \ast P_\beta$.
We have thus shown that 
\begin{equation}
\label{Ilv691}
(L^+_R  P_\alpha ) \ast P_\beta = P_\alpha \ast (L^+_L P_\beta) \,.
\end{equation}

\bigskip

The antighost field $b(z)$  transforms in the same way
as the energy-momentum tensor $T(z)$.
Therefore the formulas we have derived
for the energy-momentum tensor
based on its transformation
properties also apply to the antighost.
The equations in (\ref{lrdecLplus}), for example, become   
\be
\label{bbrdecLplus}
\begin{split}
\langle \, B^+_R \, \phi, P_\alpha \, \rangle
&= \biggl\langle \, \int_{C_0^+} \frac{dz}{2 \pi i} \,
\frac{b(z)}{R'(z)} \,
f \circ \phi (0) \, \biggr\rangle_{\mathcal{P}_\alpha} ,
\\[1.0ex]
\langle \, B^+_L \, \phi, P_\alpha \, \rangle
&= \biggl\langle \, \int_{C_0^-} \frac{dz}{2 \pi i} \,
\frac{b(z)}{R'(-z)} \,
f \circ \phi (0) \, \biggr\rangle_{\mathcal{P}_\alpha}
= \biggl\langle \, \int_{C_{2 \alpha}^+} \frac{dz}{2 \pi i} \,
\frac{b(z)}{R'(z)} \,
f \circ \phi (0) \, \biggr\rangle_{\mathcal{P}_\alpha} ,
\end{split}
\ee
and the equations in (\ref{L^+_RL-P_alpha}) become
\be
\begin{split}
\langle \, \phi, B^+_R P_\alpha \, \rangle
&= \biggl\langle \, f \circ \phi (0) \,
\int_{C_{2 \alpha}^+} \frac{dz}{2 \pi i} \, \frac{b(z)}{R'(z)} \,
\biggr\rangle_{\mathcal{P}_\alpha} ,
\\[1.0ex]
\langle \, \phi, B^+_L P_\alpha \, \rangle
&= \biggl\langle \, f \circ \phi (0) \,
\int_{C_0^+} \frac{dz}{2 \pi i} \, \frac{b(z)}{R'(z)} \,
\biggr\rangle_{\mathcal{P}_\alpha} .
\end{split}
\ee
We also have the analogs of (\ref{bpz_on_L_plus_R}), (\ref{slfdkjb}),
and (\ref{Ilv691})
\begin{eqnarray}
&& (B^+_R)^\star = B^+_L \,, \qquad
(B^+_L)^\star = B^+_R \,,
\\
&& \widetilde{B}^+ | P_\alpha \rangle
= (B^+_R - B^+_L ) \, | P_\alpha \rangle = 0 \,,
\\
&& (B^+_R  P_\alpha ) \ast P_\beta
= P_\alpha \ast (B^+_L P_\beta) \,.
\end{eqnarray}

Finally let us examine 
the operator $C$.
As we have discussed
in the calculation of $\psi_0$ in \S 4.2, 
the state $C | P_1 \, \rangle$
for a general projector is given by
\begin{equation}
\langle \, \phi, C P_1 \, \rangle
=\langle \, f \circ \phi (0) \, {\cal C} (1) \,
\rangle_{\mathcal{P}_1}
= R' ( R^{-1} (1) ) \, \langle \, f \circ \phi \,\,
c(R^{-1} (1)) \,
\rangle_{\mathcal{P}_1}
\label{C-definition}
\end{equation}
for any state $\phi$ in the Fock space,
where $\mathcal{P}_1$ is represented by the region
between $C_0^-$ and $C_2^+$.
Let us confirm that $C |P_1\rangle$ satisfies
the relation (\ref{C-condition}).
We need to show that
$\langle \, \phi, \widetilde{B}^+ C P_1 \, \rangle
= \langle \, \phi, P_1 \, \rangle$
for any state $\phi$ in the Fock space.
Since $\phi$ must be Grassmann odd
in order to have a nonvanishing inner product,
there is an extra minus sign
in taking the BPZ conjugate of $\widetilde{B}^+$,
and we have
\be
\langle \, \phi, \widetilde{B}^+ C P_1\, \rangle
= \langle \, \phi, (B^+_R-B^+_L) \, C P_1  \, \rangle
=-\langle \, (B^+_R-B^+_L)^\star \, \phi, C P_1 \, \rangle\, 
= \langle \, (B^+_R-B^+_L) \, \phi, C P_1 \, \rangle\,. 
\ee
The relevant correlation function can be written
using (\ref{bbrdecLplus}),
and it can be evaluated as follows:
\be
\begin{split}
\langle \, (B^+_R-B^+_L) \, \phi, C P_1 \, \rangle
&= R' ( R^{-1} (1) ) \, \biggl\langle \,
\int_{C_0^+-C_2^+} \frac{dz}{2 \pi i} \, \frac{b(z)}{R'(z)} \,
f \circ \phi (0) \, c(R^{-1} (1)) \,
\biggr\rangle_{\mathcal{P}_1}
\\[1.0ex]
&= R' ( R^{-1} (1) ) \, \left\langle \, f \circ \phi (0) \,
\left[ \, \oint \frac{dz}{2 \pi i} \, \frac{b(z)}{R'(z)} \,
c(R^{-1} (1)) \, \right] \, \right\rangle_{\mathcal{P}_1}
= \left\langle \, f \circ \phi (0) \,
\right\rangle_{\mathcal{P}_1} ,
\end{split}
\ee
where the contour of the integral in the last line
encircles $z=R^{-1} (1)$ counterclockwise.
This concludes the confirmation that
$\widetilde{B}^+ C | P_1 \rangle = | P_1 \rangle$.

\sectiono{Operator construction of the solution}\label{ope_cons_of_the_sol}

In this section we
give an explicit operator
construction of the  solution $\Psi$ for 
the most general single-split special 
projector
for arbitrary value of the reparameterization parameter $\beta$ introduced
in 
(\ref{deform_R}).  
We begin in \S\ref{hyperg-fam} with a discussion
of  single-split special projectors. They form
 a  ``hypergeometric collection,''  
indexed by a parameter $s \geq 1$.
Then in \S\ref{sol_ope_form} we derive 
an operator expression for the state $\psi_n$,
the key ingredient of $\Psi$ in (\ref{solPsi}).
The result, given in  
(\ref{psin_finfdfdal_new_more}), 
takes the form of normal-ordered operators acting
on the $SL(2,R)$-invariant 
vacuum.   It holds for any projector in 
the hypergeometric collection.

\subsection{The hypergeometric collection}\label{hyperg-fam}

In a previous paper~\cite{Rastelli:2006ap},
a family of special projectors with a parameter $s \geq 1$
was introduced.
It was demanded that the vector field $v_{\mathcal{L}_{-s}}$
associated with the Virasoro operator $\mathcal{L}_{-s}$
in the frame $z = \tilde f (\xi)$
take the form:\footnote{
We reserve the use of $f$ for the map
with a different normalization.
The map $\tilde f (\xi)$ here corresponds to $f (\xi)$
of~\cite{Rastelli:2006ap}.}
 \be \label{Lsansatz}
 v_{{\cal L}_{-s}}(\tilde f) \equiv \frac{s}{(\tilde f^s)'}= 
\frac{(1 + \xi^2)^s}{\xi^{s-1}} \,,
 \ee 
or, equivalently,
\be
\label{diff_eqn_hyper}
{d\tilde f^s\over d\xi} =    {s\xi^{s-1}\over (1+ \xi^2)^s}\,.
\ee
By integrating
this differential equation, 
$\tilde{f} (\xi)$ was found to be
 \be \label{fLs}
\tilde f(\xi) = \xi  \left(  {}_2F_1 \left[ \frac{s}{2}, s\,; 1 +\frac{s}{2}\,;
  -  \xi^2\right]\right)^{1/s} \,.
 \ee
  It turns out that for even $s$ the operator ${\cal L}_{-s}$  is proportional
to $L^+$ while for each odd $s$ it is proportional to $K= L^+_R - L^+_L$. More precisely,
we found that
\be
q(s) \, {\cal L}_{-s}\,=
\begin{cases}
L^+\quad {\rm for}\; s\; {\rm even} \, ,\\
K\phantom{L}\quad {\rm for}\; s\; {\rm odd}\,,  
\end{cases}
\quad \hbox{with} \quad q(s) =  \frac{\Gamma(s/2+1) \Gamma(s/2)}{\Gamma(s+1)}\,.
 \ee
It will be convenient to fix the normalization of $\tilde f(\xi)$ by introducing
a rescaled $f (\xi)$ with
$
f(\xi= 1) = 2^{-{1/ s}}$.
To implement this, we simply take
 \be \label{fLs}
 f(\xi) = 2^{-{1\over s}} \, {\tilde f (\xi)\over \tilde f(1)} = 2^{-{1\over s}}\, 
\xi  \left( { {}_2F_1 \left[ \frac{s}{2}, s\,; 1 +\frac{s}{2}\,;
  -  \xi^2\right]\over  {}_2F_1 \left[ \frac{s}{2}, s\,; 1 +\frac{s}{2}\,;
  - 1\right]   }\right)^{1/s} \,.
 \ee
Noting that
\be
{}_2F_1 \left[ \frac{s}{2}, s\,; 1 +\frac{s}{2}\,;
  - 1\right] \equiv \alpha (s)  ={  2^{-s} \sqrt{\pi} \, \Gamma [1 + {s\over 2}]
\over \Gamma [{1\over 2} + {s\over 2}] }\,,
\ee
a short computation shows that
 with the new normalization
\be
\label{LsvsL+}
{1\over s}\, {\cal L}_{-s}\,=
\begin{cases}
L^+\quad {\rm for}\; s\; {\rm even} \, ,\\
K\phantom{L}\quad {\rm for}\; s\; {\rm odd}\,.
\end{cases}
 \ee
This means that in the $z$-coordinate of the projector we have
\be
{1\over s}\, {1\over z^{s-1}} \,=
\begin{cases}
~v^+(z) \quad~~~~~ ~~{\rm for}\; s\; {\rm even} \, ,\\
\epsilon(t(z))\, v^+(z)\quad {\rm for}\; s\; {\rm odd}\,, 
\end{cases}
\quad  \hbox{with} 
~~ z =  f (t) \,,  ~ t = e^{i\theta}\,.
 \ee
In here we have used the step function $\epsilon (t)$ defined in~\cite{Rastelli:2006ap},
eqn. (2.37).   
 By definition, the vector $v$
corresponding to $L = \mathcal{L}_0/s$ is
\be
v =  {1\over s} \, z \,.
\ee 
It now follows from $v+ v^\star = v^+$ that 
\be
\label{putting_it_together}
v^\star (z) = 
\begin{cases}
 {1\over s} \, \cdot{1-z^s\over z^{s-1}}\quad ~~{\rm for}\; s\; {\rm even} \, ,\\
 {1\over s} \, \cdot{\epsilon-z^s\over z^{s-1}}  \quad {\rm for}\; s\; {\rm odd}\,.  
\end{cases}
 \ee

 The hypergeometric conformal frames are  projectors for all real $s \geq 1$:
 $ f( i) = \infty$. Moreover the midpoint $\xi = i$ is the only singular
 point, so the projectors are single-split.
 These properties and 
 the precise shape of the coordinate
 curve can be deduced from the differential
 equation (\ref{diff_eqn_hyper}).    A little algebra gives
 \be \label{diffF}
\frac{d F (\theta)}{d \theta} = \frac{i\, s}{2^{s+1}  \tilde f(1)^s} \frac{1}{(\cos \theta)^s}   \,, \quad F(\theta)    \equiv (f(e^{i \theta}))^s\,.
\ee
 By twist symmetry it is sufficient to consider the part of the curve with 
 $0 \leq \theta \leq \pi/2$.
 The differential equation (\ref{diffF}) must be supplemented with the initial condition $F(0) =f(1)^s = 1/2$.
 Since the right-hand side of (\ref{diffF}) is purely imaginary we see at once that $\Re (F(\theta)) = 1/2$ for $0 \leq \theta < \pi/2$.
 It  follows also 
 that for $s \geq 1$,  $\Im (F(\theta))$ is a monotonically increasing function in the interval $0 \leq \theta < \pi/2$ with $\lim_{\theta \to \pi/2^-} \Im (F(\theta)) = +\infty$. 
 We recognize  $\mathcal{F}_0 =
 \{ F(\theta) \, | 0 \leq \theta < \pi/2 \}$ as the vertical line  $V^+_0 = \{ z_{\cal S}\,  | \, \Re (z_{\cal S}) = 1/2) \}$, the
 positive part of the sliver's coordinate curve. We conclude that the reparameterization
 mapping the hypergeometric projector with $s >1$ to the sliver is simply
 \be
 z \to z_{\cal S} = R(z) = z^s \, , \quad \Re z > 0 \, ,
 \ee
 a fundamental fact that we had so far claimed without proof.

It seems to us plausible that   
the hypergeometric collection contains all the single-split special projectors. 
It was shown in~\cite{Rastelli:2006ap} (section 7.2) that for a conformal frame to be special
the function $z_{\cal S} = F(\theta)$, $0 \leq \theta \leq \pi/2$,
 needs to be piece-wise linear
in the $z_{\cal S}$-plane. 
On the other hand we also saw in~\cite{Rastelli:2006ap} (section 7.3)
that corners in  $\mathcal{F}_0$ seem to lead to operators $K$ that fail to kill
the identity, thus violating one of the conditions required to have a special projector.
If corners are not allowed anywhere, the intersection of $\mathcal{F}_0$ with the real
line must be orthogonal and then $\mathcal{F}_0=  V^+_0$, up to a real scaling constant.  This would imply
that all single-split projectors are in the hypergeometric collection.

For integer $s$ the hypergeometric function can be expressed in terms
of elementary functions. 
For the first few integer values  one finds  
\be
\begin{split}
s=1\,: \quad   f(\xi) &=  {2\over \pi} \arctan \xi \,,\\
s=2\,: \quad   f(\xi) &= {\xi\over \sqrt{1+ \xi^2} } \,,\\
s=3\,: \quad   f(\xi) & = \Bigl({2\over \pi}\Bigr)^{1\over 3} \Bigl(\arctan \xi
      - {\xi( 1- \xi^2)\over (1+ \xi^2)^2}\Bigr)^{1/3}\,,\\
s=4\,: \quad   f(\xi) &= x \Bigl(\,{ 3+x^2 \over (1+ x^2)^3}\,\Bigr)^{1/4}\,,\\
s=5\,: \quad   f(\xi) & = \Bigl({2\over \pi}\Bigr)^{1\over 5} \Bigl(\arctan \xi
       - {\xi( 1- \xi^2)(3+ 14\xi^2 + 3\xi^4)\over 3(1+ \xi^2)^4}\Bigr)^{1/5}\,,\\
s=6\,: \quad   f(\xi) &= x \Bigl(\,{ 10+ 5x^2+ x^4 \over (1+ x^2)^5}\,\Bigr)^{1/6}\,.
\end{split}
\ee
For $s=1$ we recover the sliver frame with a scaling. For $s=2$ we recover
the butterfly.  For $s=3$ we recover the projector in (7.56)
of \cite{Rastelli:2006ap}. For $s=4$ we have the projector with $a=4/3$
in (6.3) of \cite{Rastelli:2006ap}.  

For arbitrary $s$, a series expansion gives  $L = \mathcal{L}_0/s$ with
a simple analytic form:
\be
\label{Lfamily}
\begin{split}
\mathcal{L}_0 & =  \, L_0
+ 2\sum_{k=1}^\infty    { s!!\over (s-2k)!! } {s!!\over (s+2k)!!}  L_{2k} 
 \\[1.0ex]
&=  \, L_0 + {2s\over 2+s} \, L_2
+ {2 s(s-2)\over (2+s) (4+s)} L_4  +  
{2 s(s-2)(s-4) \over (2+s) (4+s)(6 +s)} L_6 + \ldots
\end{split}
\ee
For even $s$ the operator $L$ contains a finite number of terms
and therefore so does $L^+$. This is consistent with (\ref{LsvsL+}), since
according to (\ref{Lsansatz}) $\mathcal{L}_{-s}$ involves a finite 
number of operators for any integer $s$.

\subsection{The solution in operator form}\label{sol_ope_form}

To obtain the operator representation of the solution we will
begin with equation (\ref{useful_form}).  For notational
clarity  it is useful to 
introduce the definition
\be
r\equiv R^{-1} (1)\,, ~~\hbox{or} ~~  R(r) = 1\,.
\ee
Moreover, letting $n\to n-2$, we have that  (\ref{useful_form}) gives
\begin{equation}
\label{n_starting_point}
\langle \, \phi, \psi_{n-2} \, \rangle
= {}- (R'(r))^2 \,
\biggl\langle \, c( -r) \,\,
f \circ \phi (0) \,\,
c( r )
\int_{C_\gamma^+} \frac{dz}{2 \pi i} \,
\frac{b(z)}{{R}' (z)} \,
\biggr\rangle_{\mathcal{P}_{n-1}} ,
\end{equation}
with $1<\gamma \leq n-1$.  
The surface $\mathcal{P}_{n-1}$ in this correlator is defined by the reparameterization
function $R$.  Our goal is to obtain a formula for the state $\psi_{n-2}$ as a string of 
operators acting on the vacuum.  The operators must be normal ordered so that 
evaluation in the level expansion is possible.

In order to incorporate the reparameterizations that act within the family 
of surface states associated with a projector we take $R$ to be $\beta$-dependent as
in (\ref{deform_R}),  
\be
\label{R_from_R0}
R_\beta(z) = e^{-2 \beta} \Bigl( R_0(z) - {1\over 2} \Bigr) + {1\over 2}\,,
\ee
where  $R_0$ is the ``original'' function and $R_\beta$ the function
obtained by reparameterization. 
For generic projectors,
 the state  $\psi_{n-2}$ can be evaluated explicitly
only if certain conformal maps are known.  For the case of special
projectors in the hypergeometric collection, full and explicit evaluation is
possible.   Our result is an operator formula for $\psi_{n-2}$ that
depends on the parameter $s$ of the special projector and the
parameter $\beta$ in (\ref{R_from_R0}).

\subsubsection{Reparameterizations within a family}

Let us begin with some preparatory results concerning
the relations between operators and surfaces defined by $R_\beta$ and 
those defined by $R_0$. 
Using (\ref{R_from_R0}) one can readily verify that
\be
{R_\beta (z) \over R_\beta'(z)} = {R_0 (z) \over R_0'(z)} + {1\over 2} 
\Bigl( e^{2\beta} -1\Bigr) {1\over R_0'(z)}\,.
\ee
Letting $L, L^*$ denote  operators defined by $R$ and $\bar L, \bar L^\star$ 
denote operators defined by $R_0$, equation (\ref{Lrepa})
gives
\be
\begin{split}
L &= \int_{C_0^+} \frac{dz}{2 \pi i} \,
\frac{R(z)}{R'(z)} \, T(z)
+ \int_{C_0^-} \frac{dz}{2 \pi i} \,
\frac{R(-z)}{R'(-z)} \, T(z) \,,\\[1.0ex]
&= \bar L + {1\over 2} 
\Bigl( e^{2\beta} -1\Bigr)  \biggl[ ~\int_{C_0^+} \frac{dz}{2 \pi i} \,
\frac{T(z)}{R_0'(z)} 
+ \int_{C_0^-} \frac{dz}{2 \pi i} \,
\frac{T(z)}{R_0'(-z)} \, \biggr] \,.
\end{split}
\ee
We have therefore obtained
\be
L =  \bar L + {1\over 2} 
\bigl( e^{2\beta} -1\bigr) ( \bar L + \bar L^\star) \,.
\ee
Analogous relations hold for the operators 
associated with 
the antighost field $b(z)$.

It is interesting to examine $L$ for some special values of $\beta$.
As $\beta=0$, we get $L= \bar L$.  As $\beta$ becomes 
arbitrarily large and positive $L$ becomes proportional to $\bar{L}^+$:
\be
L \, \to \, {1\over 2} e^{2\beta} ( \bar L + \bar L^\star) \, , \quad \hbox{as} 
\quad \beta \to \infty\,.
\ee 
As $\beta$ becomes arbitrarily large and negative $L$ approaches $\bar{L}^-$:
\be
L \, \to \, {1\over 2} ( \bar L - \bar L^\star) \, , \quad \hbox{as} 
\quad \beta \to - \infty\,.
\ee 
The transition from $R_0$ to $R$ can be viewed as a reparameterization,
as discussed around equation (\ref{deform_R}).  Indeed, a short calculation gives
\be
L =  e^{\beta (\bar L-\bar L^\star)} \, \bar L\, e^{-\beta (\bar L-\bar L^\star)}\,,
\ee
showing that $\bar L-\bar L^\star$ generates the reparameterization
that maps the $R_0$-based operators to the $ R$-based operators.
 
\medskip
Let us compare surfaces defined by $R_\beta$  and surfaces defined by $R_0$.
Since $R$ maps $C_\alpha^+$ to
$V_\alpha^+$, we find
\be
z\in C_\alpha^+  \quad \to \quad
\Re ( R_\beta(z))
= {1\over 2} (1+ \alpha)\,.
\ee
For such $z$ we also have
\be
\Re ( R_0(z))
= e^{2\beta}\Bigl(  {1\over 2} (1+ \alpha) -{1\over 2} \Bigr) 
+ {1\over 2}
= {1\over 2} (1+ e^{2\beta}\alpha)\,.
\ee
Since we are focusing on a single curve in the projector we conclude that 
\be
C_\alpha^+  =  \bar C_{e^{2\beta}\alpha}^{+}\,, 
\ee
where the bar indicates a curve defined by $R_0$.
We thus have the identification of surfaces
\be
\mathcal{P}_\alpha =  \overline{\mathcal{P}}_{e^{2\beta}\alpha}\,,
\ee
where the overline indicates a surface defined by $R_0$.
Note that the surface $\mathcal{P}_0$ coincides with $\overline{\mathcal{P}}_0$.
This means that the function $z= f(\xi)$ that defines
the projector does not depend on  $\beta$.
\medskip
\noindent

The last ingredient we consider is the antighost
insertion in (\ref{n_starting_point}).  We wish to rewrite it in
terms of a closed contour integral that involves $R_0$.
  We begin by noting  the equality
\be
\label{f_step_one}
\int_{C_\gamma^+} {dz\over 2 \pi i}  {b(z)\over R'(z)}
  = e^{2 \beta} \int_{C_{n-1}^+} {dz\over 2 \pi i} {b(z)\over R_0'(z)}\,,
\ee
which follows from (\ref{R_from_R0}) and contour deformation.
To rewrite the right-hand side in terms of an
integral over a closed contour we recall that on the surface $\mathcal{P}_{n-1}$
the identification of points on $C_{n-1}^+$ and $C_{n-1}^-$
is given by  (\ref{indent_after_rep}):
\be
R_\beta(z^+) + R_\beta(-z^-) = n\,,
\ee
In terms of $R_0$ the identification reads
\be
\label{aux-eqn}
R_0(z^+) + R_0(-z^-) = 1 + (n-1) e^{2 \beta}.
\ee
We now consider the integral 
\be
 \int_{C_{n-1}^+-C_{n-1}^-} {dz\over 2 \pi i}\,\, {\widehat R_0 (z)
\over \widehat{R}_0'(z)}\,b(z) = 
\int_{C_{n-1}^+} {dz^+\over 2 \pi i}\,\, { R_0 (z^+)
\over {R}_0'(z^+)}\,b(z^+)  + \int_{C_{n-1}^-} {dz^-\over 2 \pi i}\,\, { R_0 (-z^-)
\over {R}_0'(-z^-)}\,b(z^-)\,.
\ee
Using (\ref{aux-eqn}) and its differential form $ R_0'(z^+) dz^+ - R_0' (-z^-) dz^{-}  = 0$,
we can write the second integral above as an integral over $C_{n-1}^+$.  We then
find a cancellation and we are left with
\be
 \int_{C_{n-1}^+-C_{n-1}^-} {dz\over 2 \pi i}\,\, {\widehat R_0 (z)
\over \widehat{R}_0'(z)}\,b(z) =  \bigl( 1 + (n-1) e^{2 \beta}
\bigr) \int_{C_{n-1}^+} {dz \over 2 \pi i}\,\, {b(z)
\over {R}_0'(z)}\,,
\ee
or, equivalently,
\be
\int_{C_{n-1}^+} {dz\over 2 \pi i} {b(z)\over R_0'(z)}
  ={ 1\over  1 + (n-1) e^{2\beta} }  
 \int_{C_{n-1}^+-C_{n-1}^-} {dz\over 2 \pi i}\,\, {\widehat R_0 (z)
\over \widehat{R}_0'(z)}\,b(z)\,.
\ee
Back in (\ref{f_step_one}) and using again contour deformation,
we find
\be
\label{ope_b_ins}
\int_{C_\gamma^+} {dz\over 2 \pi i}  {b(z)\over R_\beta'(z)}
  ={ e^{2 \beta}\over  1 + (n-1) e^{2\beta} }  
 \int_{C_\gamma^+-C_\gamma^-} {dz\over 2 \pi i}\, {\widehat R_0 (z)
\over \widehat{R}_0'(z)}\,b(z)  \,, \quad \hbox{on}\quad
\mathcal{P}_{n-1} \,.
\ee
This is our desired result.

\subsubsection{Operator formula}

We are now in a position to derive an operator result
beginning with (\ref{n_starting_point}).  As a first step
we use (\ref{ope_b_ins}) to obtain
\begin{equation}
\langle \, \phi, \psi_{n-2} \, \rangle
= {}-  \,{ R_0'(r )^2\,e^{-2 \beta}\over  1 + (n-1) e^{2\beta} } 
\biggl\langle \,
\int_{C_\gamma^+-C_\gamma^-} \frac{dz}{2 \pi i} \,
\frac{\widehat{R}_0 (z)}{\widehat{R}_0' (z)} \, b(z) \,\,
c( -r ) \,\,
f \circ \phi (0) \,\,
c(r) \,
\biggr\rangle_{\overline{\mathcal{P}}_{e^{2\beta} (n-1)}} .
\label{psi_n-for-ex}
\end{equation}
Note that we have expressed the surface in terms of the
function $R_0$.
Moving the antighost insertion contours inwards we pick up contributions
from each of the ghost insertions and we remain with an antighost insertion
that effectively surrounds the insertion of the test state $\phi$:
\be
\label{simp_most_gen_proj}
\begin{split}
\langle \, \phi, \psi_{n-2} \, \rangle
&= \,{  R_0'(r ) R_0(r)e^{-2 \beta}\over  1 + (n-1) e^{2\beta} }  \,\Bigl(\,
\bigl\langle ~c(-r) \,f\circ \phi~
\bigr\rangle_{\overline{\mathcal{P}}_{e^{2\beta} (n-1)}}
 + \bigl\langle\, f\circ \phi \,\, c(r)\,
\bigr\rangle_{\overline{\mathcal{P}}_{e^{2\beta} (n-1)}} \Bigr)  \\[1.0ex]
&~~{}+ \,{R_0'(r )^2\, e^{-2 \beta}\over  1 + (n-1) e^{2\beta} }  \, 
\biggl\langle 
c(-r)  \biggl[ \, \int_{C_\gamma^+-C_\gamma^-} \frac{dz}{2 \pi i} \,
\frac{\widehat{R}_0 (z)}{\widehat{R}_0' (z)} \, b(z) \,\,
f \circ \phi (0)  \biggr] \, c(r)  
\biggr\rangle_{\overline{\mathcal{P}}_{e^{2\beta} (n-1)}} .
\end{split}
\ee
Here $0\leq \gamma < 1$. 
This is
the most simplified expression we have obtained
for $\psi_{n-2}$ when the projector is completely general.

\medskip
Let us now assume that we have a special projector with  parameter $s$.
We thus take
\be
R_0 (z) = z^s \quad \to \quad {R_0 (z)\over R_0'(z)} = {1\over s} \, z \,,
\ee
which implies that 
\be
 \int_{C_\gamma^+-C_\gamma^-} \frac{dz}{2 \pi i} \,
\frac{\widehat{R}_0 (z)}{\widehat{R}_0' (z)} \, b(z)  = 
 {1\over s} \int_{C_\gamma^+ -C_\gamma^-} \frac{dz}{2 \pi i} \,
z\, b(z)  = {1\over s} \oint \frac{dz}{2 \pi i} \,
z\, b(z) \,.
\ee
Notice the great simplification: all that is left of the antighost 
insertion is a holomorphic integral encircling the origin.
We also define
\be
\label{define_a_n}
a_n \equiv  \Bigl( 1+ (n-1)e^{2\beta} \Bigr)^{-1/s} \,,
\ee
and confirm that 
\be
R_0(r) =r^s=  {1\over 2} \bigl( 1+ e^{2\beta} \bigr)  \,.\ee
Using the above relations (\ref{simp_most_gen_proj}) can be written as
\be
\begin{split}
\langle \, \phi, \psi_{n-2} \, \rangle
&= sr^{2s-1}(a_n)^s e^{-2 \beta} \,\Bigl(\,
\bigl\langle ~c(-r) \,f\circ \phi~
\bigr\rangle_{\overline{\mathcal{P}}_{e^{2\beta} (n-1)}}
 + \bigl\langle\, f\circ \phi \,\, c(r)\,
\bigr\rangle_{\overline{\mathcal{P}}_{e^{2\beta} (n-1)}} \Bigr)  \\[1.0ex]
&~~{}+ \,s\,r^{2s-2}(a_n)^s\, e^{-2 \beta} \, 
\biggl\langle 
c(-r)  \biggl[ \, \oint \frac{dz}{2 \pi i} \,
z\, b(z) \,\,
f \circ \phi (0)  \biggr] \, c(r)  
\biggr\rangle_{\overline{\mathcal{P}}_{e^{2\beta} (n-1)}} .
\label{psi_n-forex}
\end{split}
\ee
To map the correlators to the upper-half plane we first
 scale  $\overline{\mathcal{P}}_{e^{2\beta} (n-1)}$
down to $\mathcal{P}_0$.  This requires the scaling map
\be
z'= a_n z\,,
\ee
with $a_n$ defined in (\ref{define_a_n}). 
We let $\tilde f \equiv a_n \circ f$ and perform the scaling, finding 
\be
\begin{split}
\langle \, \phi, \psi_{n-2} \, \rangle
&= sr^{2s-1}(a_n)^{s-1} e^{-2 \beta} \,\Bigl(\,
\bigl\langle ~c(-a_nr) \,\tilde f\circ \phi~
\bigr\rangle_{\mathcal{P}_0} 
 + \bigl\langle\, \tilde f\circ \phi \,\, c(a_nr)\,
\bigr\rangle_{\mathcal{P}_0}  \Bigr)  \\[1.0ex]
&~~{}+ \,s\,r^{2s-2}(a_n)^{s-2}\, e^{-2 \beta} \, 
\biggl\langle 
c(-a_nr)  \biggl[ \, \oint \frac{dz}{2 \pi i} \,
z\, b(z) \,\,
\tilde f \circ \phi (0)  \biggr] \, c(a_nr)  
\biggr\rangle_{\mathcal{P}_0}  .
\label{psi_n-forex}
\end{split}
\ee
The map 
\be
g \equiv  f_I \circ f^{-1}
\ee
 takes $\mathcal{P}_0$ to the
upper half plane $\mathbb{H}$. Letting  
\be
\label{fn_minus_one_defined}
f_{n-1} \equiv g\circ \tilde f= f_I \circ f^{-1} \circ a_n \circ f\,,
\ee
we map the correlators by $g$ and find, noting that $g$ is 
an odd function, 
\be
\begin{split}
\langle \, \phi, \psi_{n-2} \, \rangle
&= sr^{s}(a_nr)^{s-1} {e^{-2 \beta}\over g'(a_nr)} \,\Bigl(\,
\bigl\langle ~c(-g(a_nr)) \, f_{n-1}\circ \phi~
\bigr\rangle_{\mathbb{H}}
 + \bigl\langle\,  f_{n-1}\circ \phi \,\, c(g(a_nr))\,
\bigr\rangle_{\mathbb{H}}  \Bigr)  \\[1.0ex]
&~~{}+ \,sr^{s}(a_nr)^{s-2}\,  {e^{-2 \beta}\over (g'(a_nr))^2} \, 
\biggl\langle 
c(-g(a_nr))  \bigl[ \,\widehat{B} \,\,
 f_{n-1} \circ \phi (0)  \bigr] \, c(g(a_nr))  
\biggr\rangle_{\mathbb{H}}  .
\label{psi_n-forex_99}
\end{split}
\ee
Here all correlators are now on the upper half plane $\mathbb{H}$
and  
\begin{equation}
\label{b_star_fineal}
\widehat{B} \equiv  \oint \frac{dz}{2 \pi i}
\frac{g^{-1}(z)}{(g^{-1})'(z)} \, b(z) \,.
\end{equation}
Note that the $\widehat B$ insertion is $\beta$ independent and $n$ independent.

Since the operator $I \circ f_{n-1} \circ \phi (0)$
corresponds to $\langle \phi | \, U_{f_{n-1}}^\star$
in the state-operator correspondence,
it is convenient to perform a final map by  $I(z) = -1/z$.  Noting that
the test state $\phi$ must be Grassman even, the result is  
\be\begin{split}
\langle \, \phi, \psi_{n-2} \, \rangle
&=\gamma \,s (a_n r)^{s-1} {g(a_n r)^2\over g'(a_nr)}~\Biggl[ ~
 \Bigl\langle I\circ f_{n-1}\circ \phi \,\, c \Bigl( \frac{1}{g(a_nr)} \Bigr)
\Bigr\rangle_{\mathbb{H}} +\Bigl\langle I\circ f_{n-1}\circ \phi ~c \Bigl( -\frac{1}{g(a_nr)} \Bigr) 
\Bigr\rangle_{\mathbb{H}}  \\[1.0ex]
&~~~~~~~~~~~~~~~{}\qquad + 
 {g(a_n r)^2\over a_nrg'(a_nr)}
\biggl\langle I\circ f_{n-1} \circ \phi (0)  \, \widehat{B}^\star~
c \Bigl( -\frac{1}{g(a_nr)} \Bigr) 
 \, c \Bigl( \frac{1}{g(a_nr)} \Bigr)
\biggr\rangle_{\mathbb{H}} \Biggr] \,,
\label{psi_n-forex434}
\end{split}
\ee
where we defined
\be
\gamma \equiv e^{-2\beta} r^s =  {1\over 2}  (1+e^{-2 \beta}) \,.
\ee 
We can now read out the operator expression for $\psi_{n-2}$:
\be
\label{psin_finfdfdal_new_more}
\boxed{
\begin{split}
 \psi_{n-2} &=~ \gamma\, s(a_n r)^{s-1}\,
   {g(a_n r)^2\over g'(a_nr)} \,
U_{f_{n-1}}^\star \biggl[ \, c \Bigl( -\frac{1}{g(a_nr)} \Bigr)
+ c \Bigl( \frac{1}{g(a_nr)} \Bigr) \phantom{(\Biggl()^X}\\[1.5ex]
& \qquad\qquad\qquad\qquad\qquad+ 
{g(a_n r)^2\over a_nrg'(a_nr)} \, \widehat{B}^\star \,
c \Bigl( -\frac{1}{g(a_nr)} \Bigr) \,
c \Bigl( \frac{1}{g(a_nr)} \Bigr) \, \biggr] | 0 \rangle \,.\phantom{\Biggr)} \\
\end{split}
}
\ee
Equation (\ref{psin_finfdfdal_new_more}) is the expected result:  
a formula for the state $\psi_{n-2}$ in which
operators act on the $SL(2,R)$-invariant 
vacuum.   The state depends on both
$s$ and $\beta$.  Moreover, as we will see in the following section, 
we can readily find a level expansion of the solution.  We recall that
the ``phantom" piece $\psi_N$ of the solution in (\ref{solPsi}) does not
contribute in the level expansion, so we have
\be
\label{sol_lev_exp}
\langle \, \phi, \Psi \, \rangle
= \sum_{n=2}^\infty \langle \, \phi, \psi'_{n-2} \, \rangle
\ee
for any state $\phi$ in the Fock space.

\sectiono{Level and other expansions}\label{sec_lev_and_oth_exp}

In this section we will expand and analyze the operator
form (\ref{psin_finfdfdal_new_more}) of the solution.
We  set up the level expansion computation for arbitrary 
$s$ and $\beta$ in \S\ref{lev_exp_pre}.
We proceed up to level four, but give the ingredients necessary
to carry the computations to arbitrary order, if so desired.

In \S\ref{sec-exp} we consider the special case $\beta=0$ 
and compute the vacuum expectation values of fields up to level four for
arbitrary values of $s$.  This allows us to compute the 
level zero, two, and four vacuum energies as functions
of $s$.  For $s\geq 1$  we find numerical evidence
consistent with convergence of the vacuum energy
to the expected value of minus 
the D-brane tension.  

Recall that for  $s<1$  the
special frames are not projectors.
The string field $\Psi$ 
which provides a solution for $s\geq 1$ is therefore not expected to 
provide
a solution for $s<1$. Indeed, for $s<1$ 
we find numerical evidence consistent with the energy
failing to converge to the expected value.

In \S\ref{no-siegel} we show that the tachyon
vacuum solution in the Siegel gauge cannot be obtained in
the present framework. The framework imposes  constraints
on expectation values that we show are not satisfied in the
most accurate version of the Siegel gauge solution known to date.

Finally, in \S\ref{pro_exp} we consider the limit $\beta\to \infty$ of
the solution.  This limit is of  some interest because the surface states used
to build the solution approach the surface state of the projector. 
For large $\beta$ the solution provides an analytic expression
closely related to  the  alternative level expansion scheme introduced in~\cite{Okawa:2003zc} and explored further in~\cite{Yang:2004xz}. 
 In this scheme,  the  string field solution is written in terms
of operators of increasing level inserted at the midpoint of a
regulated projector.  Our solution is given in terms
of exponentials of $\beta$ and has a leading divergent term as
well as terms that vanish as $\beta \to \infty$.

\subsection{Level expansion preliminaries}\label{lev_exp_pre}

We now set up the level expansion of the solution~(\ref{psin_finfdfdal_new_more}).
We begin by level expanding the operators $U_{f_{n-1}}^\star$
and  $\widehat{B}^*$.  We then write out the 
level four string field and compute the expectation values of the
various components.  The results are given in terms of infinite
sums that we
 evaluate numerically.

\medskip
The 
operator $U_{f_{n-1}}^\star$
 is  defined by the function $f_{n-1}(\xi)$
introduced in (\ref{fn_minus_one_defined}):
\be
\label{f000sd}
f_{n-1} =  f_I \circ f^{-1} \circ a_n \circ f \,.
\ee
It is most convenient to obtain a  factorized 
form in which
\be
U_{f_{n-1}} =  e^{\bar t_0 L_0}  \, e^{\bar t_2 L_2} \, 
e^{\bar t_4 L_4} \, e^{\bar t_6  
L_6}  \cdots  \,,
\ee
with calculable coefficients $\bar t_n$. The bpz dual is immediately written 
\be
\label{xhjk}
U_{f_{n-1}}^\star =  
\cdots  e^{\bar t_6   L_{-6}} \,e^{\bar t_4 L_{-4}}  \, e^{\bar t_2 L_{-2}} \,  e^{\bar t_0 L_0} \,.
\ee
Given 
an arbitrary function $f(\xi)$ that defines a surface state and has 
an expansion 
\be
f(\xi) =  \xi + f_2 \xi^3   + f_4 \xi^5 +  f_6 \xi^7 + f_8 \xi^9 + \cdots\,,
\ee
the first few $\bar t_n$ coefficients are obtained following the
steps indicated in appendix~A of~\cite{Schnabl:2005gv}.  We find
that they are given by
\be
\label{split_virasoro}
\begin{split}
\bar t_2 &=  f_2\,, \\[1.5ex]
\bar t_4 &= f_4 - {3\over 2} f_2^2\,, \\[1.5ex]
\bar t_6 &= f_6 - 3 f_2f_4 + 2 f_2^3 \,, \\[1.5ex]
\bar t_8 &= f_8 - 3 f_2f_6- {5\over 2} f_4^2 + 9 f_2^2 f_4 - {19\over 4} f_2^4\,.
\end{split}
\ee
Using this result and  the power series expansion of $f_{n-1}$ 
we can readily calculate the coefficients $\bar t_n$ needed
to obtain $U_{f_{n-1}}$ to level four:
\be
\label{9909}
e^{\bar t_0} = a_n \,, \quad  \bar t_2 = {-s + 2a_n^2 (1+s) \over 2+s}\,,
\quad  \bar t_4 = - {(s-2)s + 8a_n^4 (1+s)   \over  2 (2+s) (4+s)}\,.
\ee
With these we get
\be
\label{xh899}
U_{f_{n-1}}^\star =  
\cdots  \,e^{\bar t_4 L_{-4}}  \, e^{\bar t_2 L_{-2}} \,  (a_n)^{L_0} \,.
\ee
The expansion of $\widehat{B}^\star$ is easier to obtain. 
Recalling  (\ref{b_star_fineal}) and the
relation  $g = f_I \circ f^{-1}$  we find 
\be
\label{Bhat_exp_485}
\widehat{B} = \sum_{n=0}^\infty \beta_n \, b_n =
  b_0 + {4(1+s)\over 2+s} \, b_2
- {16 (1+s)\over (2+s) (4+s)} b_4   +
\cdots~\,.
\ee
Note that both the Virasoro operators and the antighost 
operators in the above expansions are even moded.
\medskip

The level expansion of the string 
is obtained by  the action on the vacuum of arbitrary ghost
oscillators, even moded Virasoro operators,  and even
moded antighost oscillators. 
The  string field up to level four is thus given~by 
\be
\label{level4-field}
\begin{split}
\Psi_4 &= {}- \Bigl( \, t\, c_1\ket{0} \\[1.0ex]
&\qquad \quad ~
+ u\, c_{-1} \ket{0}  + v\, L_{-2} c_1 \ket{0}
+ w \, b_{-2} c_0 c_1 \ket{0} \\[1.0ex]
&\qquad \quad ~
+ A\, L_{-4} c_1\ket{0} + B L_{-2} L_{-2} c_1 \ket{0} 
+ C \, c_{-3} \ket{0}  + E \, b_{-2} c_{-2} c_1 \ket{0} 
+ F \, L_{-2} c_{-1} \ket{0} \\[1.0ex]
&\qquad \quad ~
+ w_2 \,b_{-2} c_{-1} c_0 \ket{0} + w_3 \,b_{-4} c_{0} c_1 \ket{0} + 
w_4 \, L_{-2} b_{-2} c_0 c_1 \ket{0} \, \Bigr) \,.
\end{split}
\ee
The first line contains the level-zero tachyon, the 
second line contains the three level-two fields, and the last two lines
contain the eight level-four fields.  In this expansion the Virasoro
operators include matter and ghost contributions and
have zero central charge.

To describe the solution, assume a general expansion 
in a basis of Fock space states
\be
\label{def_exp_coeff_78}
\Psi = \sum_i  \phi^{(i)} \ket{\mathcal{O}_i}\,.
\ee
Up to level four, the states $\ket{\mathcal{O}_i}$ and the expansion
coefficients $\phi^{(i)}$ are those in (\ref{level4-field}).  Our goal is to
compute those expansion coefficients, since they are the expectation
values of the component fields.
 Assume now that $\psi_{n-2}$, given in (\ref{psin_finfdfdal_new_more}), is also expanded
in the same basis:
\be
\label{def_coef_exP_0}
\psi_{n-2} = \sum_i  \phi^{(i)}_n \ket{\mathcal{O}_i}\,.
\ee
Using (\ref{sol_lev_exp}) we have
\be 
\Psi = \sum_{n=2}^\infty \psi'_{n-2} = \sum_i 
 \sum_{n=2}^\infty (\partial_n \phi^{(i)}_n) \ket{\mathcal{O}_i}\,.
\ee  
Comparing with (\ref{def_exp_coeff_78}) we find that the vevs are given by
\be
\label{comp_sol}
\phi^{(i)} = \sum_{n=2}^\infty  \partial_n \, \phi^{(i)}_n \,.
\ee

\medskip
We can now expand the solution (\ref{psin_finfdfdal_new_more})
to level four. Since the combination $a_nr$ appears repeatedly 
both by itself and as the argument of $g$ we introduce the notation
\be
  \tilde a \equiv  a_n r\,,\quad
g \equiv  g (\tilde a)\,.
\ee
Using the expansion  (\ref{xh899}) of $U_{f_{n-1}}^\star$ and 
the expansion (\ref{Bhat_exp_485}) of $\widehat{B}$, 
together with 
(\ref{def_coef_exP_0}), we find that the expansion of 
(\ref{psin_finfdfdal_new_more}) yields
\be
\label{lkop}
\begin{split}
& t_n = 2\gamma r s \, \tilde{a}^{s-2} \,\, {g^2\over g'} \Bigl( 1 - {g \over 
\tilde{a}g'} \Bigr)\,,\quad
u_n ={\gamma\over r} {\tilde{a}^2 \over g^2} \,\,t_n \,, \quad 
v_n =\gamma  r\bar t_2\, t_n  \,, \quad
w_n = -2{\gamma\over r} s \beta_2 \, \tilde{a}^{s-1}  {g^3\over g'^2}\,, \\[1.0ex]
& A_n = \gamma r\bar t_4 \, t_n \,, \quad  B_n = {1\over 2}\,\gamma  r\bar{t}_2^2 \, t_n\,, 
\quad C_n = {\gamma\over r^3} {\tilde{a}^4\over g^4} \, t_n \,, 
\quad E_n = - 2{\gamma\over r^3} s \, \tilde{a}^{s+1} \, \beta_2 \,{g\over g'^2}\,,
\\[1.0ex]
& F_n ={\gamma\over r} {\tilde{a}^2\over g^2} \, \bar t_2 \, t_n \,, \quad
(w_2)_n =  -{ \gamma\over r^3} E_n\,, \quad  (w_3)_n = -2{\gamma\over r^3} s \, \tilde{a}^{s+1} \, \beta_4  {g^3\over g'^2} \,,\\[1.0ex]
&(w_4)_n = - 2{\gamma\over r} s \, \tilde{a}^{s-1} \, \bar t_2 \beta_2 \, {g^3\over g'^2} \,.
\end{split}
\ee
The powers of $r$ here arise from the factor $(a_n)^{L_0} = (\tilde{a}/r)^{L_0}$ in
$U_{f_{n-1}}^\star$ --- see (\ref{xh899}).  In the above formulae all appearances of $a_n$
are in the combination $\tilde a$.  Note, however, that the coefficients $\bar t_2$ and 
$\bar t_4$ have $a_n$ dependence.  Following (\ref{comp_sol}), the expectation value of $A$,
for example, would be given by
\be
A  = \sum_{n=2}^\infty  \partial_n A_n \,.
\ee
For arbitrary $\beta$ and $s$, the derivatives with respect to $n$ 
of the component fields in (\ref{lkop}) give long and complicated expressions.
Therefore,  we do not attempt any further simplification of the string field.

\subsection{Level Expansion for $\beta=0$}\label{sec-exp}

In this subsection we set $\beta=0$ and explore the
solution for various values of $s$.  
We calculate explicitly the expectation values of
level four fields and use them evaluate the approximate energy
of the solution.  We find numerical evidence consistent with the
energy converging to the expected value of $-1$ (in units of the
D-brane tension) for $s\geq 1$.  
For $s<1$ we can still use (\ref{psin_finfdfdal_new_more}) to calculate a string field
but given that the $s<1$ surface states are not projectors,
we have no reason to believe that the constructed field is a solution.
Indeed, a level computation of the energy in those cases suggests that
it does not converge to minus one.

For $\beta=0$ we have $r=1$  and
 the solution in  (\ref{psin_finfdfdal_new_more}) reduces to
\be
\label{psin_finfdfdal_new_more_xx}
\begin{split}
 \psi_{n-2} &=~ s(a_n)^{s-1}   {g(a_n )^2\over g'(a_n)} \,
U_{f_{n-1}}^\star \biggl[ \, c \Bigl( -\frac{1}{g(a_n)} \Bigr)
+ c \Bigl( \frac{1}{g(a_n)} \Bigr) \\[1.5ex]
& \qquad\qquad\qquad\qquad\qquad+ 
{g(a_n )^2\over a_ng'(a_n)} \, \widehat{B}^\star \,
c \Bigl( -\frac{1}{g(a_n)} \Bigr) \,
c \Bigl( \frac{1}{g(a_n)} \Bigr) \, \biggr] | 0 \rangle \,. \\
\end{split}
\ee
This time we write
\be
a\equiv a_n =  n^{-1/s}\,, \quad  g \equiv  g(a) \,, \quad g' \equiv g'(a)\,,
\ee
and the results in (\ref{lkop}) simplify to 
\be
\begin{split}
& t_n = 2s \, a^{s-2} \,\, {g^2\over g'} \Bigl( 1 - {g \over 
ag'} \Bigr)\,,\quad
u_n = {a^2 \over g^2} \,\,t_n \,, \quad 
v_n = \bar t_2\, t_n  \,, \quad
w_n = -2s \beta_2 \, a^{s-1}  {g^3\over g'^2}\,, \\[1.0ex]
& A_n = \bar t_4 \, t_n \,, \quad  B_n = {1\over 2}\, \bar{t}_2^2 \, t_n\,, 
\quad C_n =  {a^4\over g^4} \, t_n \,, \quad E_n = 
- 2s \, a^{s+1} \, \beta_2 \,{g\over g'^2}\,,
\\[1.0ex]
& F_n = {a^2\over g^2} \, \bar t_2 \, t_n \,, \quad
(w_2)_n =  - E_n\,, \quad  (w_3)_n = -2s \, a^{s+1} \, \beta_4  {g^3\over g'^2} \,, \quad
(w_4)_n = - 2s \, a^{s-1} \, \bar t_2 \beta_2 \, {g^3\over g'^2} \,.
\end{split}
\ee
These formulae, together with (\ref{comp_sol}) allow the
evaluation of the level four expectation values.  
As in~\cite{Schnabl:2005gv}, no simple closed form seems possible
and the computation must be done numerically.

\medskip

The level four string field in (\ref{level4-field}) can be rewritten 
using matter Virasoro operators. Expanding the Virasoro operators in (\ref{level4-field})  into matter and 
ghost parts one obtains the string field 
\be
\label{n_s_f_matt}
\begin{split}
\Psi_4 &= {}- \Bigl( \, t'\, c_1\ket{0}  \\[1.0ex]
&\qquad \quad ~
+ u'\, c_{-1} \ket{0}  + v'\, L_{-2}^m c_1 \ket{0}  
+ w' \, b_{-2} c_0 c_1 \ket{0} \\[1.0ex]
&\qquad \quad ~
+A'\, L^m_{-4} c_1\ket{0} + B' L^m_{-2} L^m_{-2} c_1 \ket{0} 
+ C' \, c_{-3} \ket{0}+ D' \,b_{-3} c_{-1} c_1 \ket{0}  + E' \, b_{-2} c_{-2} c_1 \ket{0} 
\\[1.0ex]
&\qquad \quad ~
+ F' \, L^m_{-2} c_{-1} \ket{0} + w'_2 \,b_{-2} c_{-1} c_0 \ket{0} + w'_3 \,b_{-4} c_{0} c_1 \ket{0} + 
w'_4 \, L_{-2}^m b_{-2} c_0 c_1 \ket{0} \, \Bigr) \,,
\end{split}
\ee
where the primed fields are given by
\begin{align}
\label{level4_matter_vir}
t'&=t\qquad  &u'&= u+ 3v\nonumber\\
v'&= v\qquad  &w' &= w- 2v\nonumber\\[0.5ex]
A' &= A\qquad  &B' &= B\nonumber\\
C' &= C + 7A + 15B + 5F\qquad
&D' &= -5A + 3B + F
\\
E' &= E - 6A - 8B + 4w_4\qquad
&F'& = F + 6B\nonumber\\
w_2' &= w_2 + 12B + 2F - 3w_4\qquad
&w_3' &= w_3 - 4A\nonumber\\
w_4' &= w_4 - 4B  &  &\nonumber
\end{align}

Note that in (\ref{n_s_f_matt}) we had to introduce a  field $D'$ to 
multiply the state $b_{-3} c_{-1} c_1 \ket{0}$.  Since $\widehat B$ only
has even-moded oscillators, that state  arises from (\ref{psin_finfdfdal_new_more})  
only after expanding the total Virasoro operators in $U_{f_{n-1}}^\star$ 
into matter and ghost parts.  Note also that the state $L_{-3}^m c_0 |0\rangle$ 
does not arise in the expansion.  The expansion in ghost and matter
parts cannot generate odd-moded Virasoro operators, only odd-moded
antighost operators.

We can now consider some numerical work.
For $s=1$ we find the expectation value
$t= 0.553466$, $u=  0.0436719$, 
$v=0.137646$, and  $w=   0.131082$.  These imply
$t'=0.553466$, $u' = 0.45661$, $v' = 0.137646$, 
and $w' =-0.14421$
in complete agreement with~\cite{Schnabl:2005gv}.
We have also checked that the  expectation values of the level
four fields for $s=1$ agree with those in~\cite{Schnabl:2005gv}.
For $s=2$ we find
\be
t= 0.65779\,,  \quad    u=  0.04634 \,,\quad
v=     0.32231\,, \quad   w=   0.14857\,.
\ee
Vacuum expectation values for these and other values of 
$s$ are listed in Table~1.

\begin{table}
\begin{center}
\renewcommand{\arraystretch}{1.65}
\begin{tabular}{|c|c|c|c|c|c|c|c|}
\hline
  & $s=0.6$ & $s=0.8$ & $s=1$ & $s=1.2 $ & $s=1.4$ & $s=2.0$ & $s=3.0$\\ \hline
$t(s)$  & $ 0.52860 $ &$0.53755$ & $0.55347$ & $0.57278$ &  $0.59361$ & 0.65779 & $0.75882$
\\ \hline
$u(s)$  & $0.02935$ & $0.03881$& $0.04367$ & $0.04600$ & $0.04694$ & $0.04634$ & $0.04268$
 \\  \hline
$v(s)$ & $0.04541$ &  $0.09289$ & $0.13765$ & $0.17939$ & $0.21840$ & $0.32231$ & $0.46548$
\\  \hline
$w(s)$ & $0.09945$ &  $0.11908$ & $0.13108$ & $0.13860$ & $0.14330$ & $0.14857$ & $0.14617$
\\  \hline
\end{tabular}
\caption{\small The expectation values of all fields up to level two calculated
using the exact analytic expressions as a function of the parameter $s$.}\label{tab2:sec}
\end{center}
\end{table}
\bigskip

The energy, normalized to minus one, can be computed
using the vevs of the fields and the kinetic terms in
the string field theory. To level zero, two, and four we get 
\be
\begin{split}
E_0 &= {2\pi^2\over 3} \Bigl( -{1\over 2} t^2\Bigr)\,, \\[1.0ex]
E_2 &= E_0 + {2\pi^2\over 3} \Bigl( -{1\over 2}  u^2 + 3 u (v-w)
+ 2 (v-w)^2\Bigr)\,, \\[1.0ex]
E_4 &= E_2 +  {2\pi^2\over 3} \Bigl(~ 4A^2 + 24 AB + 5AC - 6AE + 
18AF - 8Aw_3 - 24Aw_4 \\
&\qquad\quad  ~~~~~~ 
- 3BC + 8BE -  24 Bw_2 - 24 Bw_3 + CF - Cw_2 - 5Cw_3 + 3Cw_4 \\
&\qquad \quad ~~~~~~
- {3\over 2} E^2 + 6 EF +   3Ew_2 + 6Ew_3 - 8 Ew_4 
  - {13\over 2} F^2 - 5 Fw_2 - 18 F w_3 \\
&\qquad \quad ~~~~~~ - 2 w_2^2 + 
  24 w_2 w_4 + 4 w_3^2 + 24 w_3 w_4\Bigr)\,.
\end{split}
\ee

\begin{table}
\begin{center}
\renewcommand{\arraystretch}{1.65}
\begin{tabular}{|c|c|c|c|c|c|c|c|}
\hline
  & $s=0.6$ & $s=0.8$ & $s=1$ & $s=1.2 $ & $s=1.4$ & $s=2.0$ & $s=3.0$\\ \hline
$E_0(s)$  & $ -0.91925 $ &$- 0.95064$ & $-1.00777$ & $-1.07934$ &  $-1.15927$ & -1.42348 & $-1.8943$
\\ \hline
$E_2(s)$  & $-0.91495$   & $- 0.96663$  & $-1.00782$ & $-1.02736$ & $-1.02271$ & $-0.87438$ & $-0.2896$
 \\  \hline
$E_4(s)$ & $-0.91389$ &  $- 0.97221$ & $-1.0045$ & $-1.00843$ & $-0.99591$ & $-0.98916$ & $-1.4827$
\\  \hline
\end{tabular}
\caption{\small The energy calculated at levels zero, two, and four, for several
values of the parameter $s$.}\label{tab2:ch17}
\end{center}
\end{table}
\bigskip

In Figure~\ref{orz1fig} we plot  energies as a function of 
$s\in [0.6, 2.0]$.  There are three curves: 
the level-zero energy $E_0(s)$, the level-two energy $E_2(s)$, and 
the level-four energy $E_4(s)$. At each level the energy was computed
using the exact numerical values for all the fields. 
For $s\geq 1$ the
various curves are consistent with an energy that approaches the correct
value.
For $s<1$ the plot suggests that the energy will
not approach the correct
value.
Some particular values
are also tabulated in Table~2.  Note how efficient the convergence is for
$s=2$, while for $s=0.6$ it appears that the energy will not move much beyond
 the value $-0.91$.   

\begin{figure}
\centerline{\hbox{\epsfig{figure=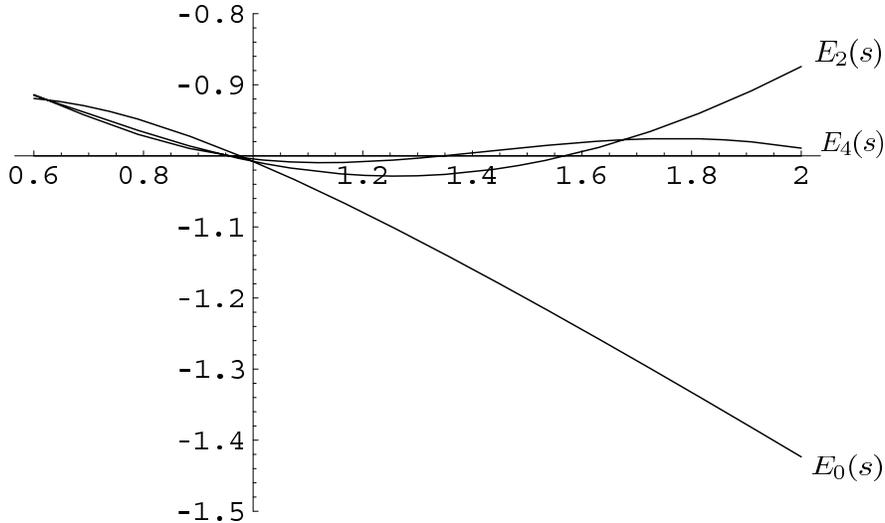, height=6.9cm}}}
\caption{Plot of the energies $E_0(s), E_2(s),$ and $E_4(s)$ computed
at levels zero, two, and four, respectively. The exact value is $-1$.}
\label{orz1fig}
\end{figure}

\subsection{No Siegel gauge in the family}\label{no-siegel}

The solution for the tachyon vacuum in the Siegel gauge
is a state in the universal subspace of the total CFT: the
ghost number one subspace spanned by all states built on the vacuum
by acting with finite numbers
of ghost and antighost oscillators as well as finite number
of matter Virasoro operators.
Apart from an  $SU(1,1)$ symmetry that relates certain expectation values
no additional relations are known.

It is clear from the form of $\psi_n$
that the solution $\Psi$ 
belongs to
a {\em constrained} universal space where states are
built acting on the vacuum with
arbitrary ghost oscillators, {\em even}-moded antighost oscillators,
and {\em even}-moded total Virasoro operators. Before imposing any
gauge condition, the level four universal subspace contains 10 states,
while the level four constrained space
has only 8 states. 

As we show now, at level four the Siegel gauge
expectation values must satisfy an additional relation if it 
is to lie on the  constrained universal space.  This condition
is not satisfied.

In the Siegel gauge we can use the expansion (\ref{n_s_f_matt})
of the string field.  The question is whether the values of the primed
fields in the Siegel gauge are consistent with expectation values for
the unprimed fields.  Can we solve for the unprimed fields using
(\ref{level4_matter_vir})?  There is a constraint, however. 
We readily find that 
\be
\label{d-pred}
D' = -5A' - 3B' + F'\,.
\ee
This is a constraint that must be satisfied by the Siegel gauge solution, if it is to have
the structural form required by the general $s$ solution.  From ~\cite{Gaiotto:2002wy} we have 
\be
\begin{split}
A' &= -0.005049\,,\\
B' &= -0.000681\,,\\
F'& =\phantom{-} 0.001234\,.
\end{split}
\ee
This together with (\ref{d-pred}) predicts  $D'=0.028522$. The value from 
\cite{Gaiotto:2002wy}, however, is $D'= 0.01976$, in clear disagreement.
We conclude that we cannot reach the Siegel gauge solution for 
any value of the parameter $s$.

\subsection{Projector expansion}\label{pro_exp}

In~\cite{Okawa:2003zc}  a variant of level
expansion was proposed in which the  string field solution is written in terms
of operators of increasing level inserted at the midpoint of a
regulated projector surface state.  The original discussion used the 
butterfly state but this was extended to large classes of 
projectors 
in \cite{Yang:2004xz}.  In this section we show how
to obtain a possibly related expansion  using the~$\beta$ parameter
in the limit of large $\beta$.

In the solution (\ref{psin_finfdfdal_new_more}) and in its level
expansion we noted the repeated appearance of 
 $a_nr= \tilde{a}$, which is 
given by
\be
\tilde{a} = a_n r  = \Bigl[ {1\over 2}\cdot {1 + e^{2\beta}\over 1+ (n-1) e^{2\beta} } \Bigr]^{1/s}\,. 
\ee
For $\beta \to \infty$ we get a
finite limit
\be
\label{abardef}
\lim_{\beta\to \infty}  \tilde{a} =  \Bigl(\, {1\over 2n-2}\, \Bigr)^{1/s} \equiv  \bar a\,.
\ee
We also note that for large  $\beta$ 
\be
r\simeq  2^{-1/s} e^{2\beta/s} \,,  \quad   a_n \simeq  2^{1/s} \bar{a}\, e^{-2\beta/s}\,.
\ee

Let us separate the factor $U_f^\star$ from $U_{f_{n-1}}^\star$.
We recall (\ref{f000sd}), which implies that 
\be
U_{f_{n-1}} = U_{f_I \circ f^{-1}} \, (a_n)^{L_0} \, U_f \,.
\ee
It follows that
\be
U_{f_{n-1}}^\star = \, U_f^\star\, \,\Bigl[ (a_n)^{L_0} \, U_{f_I \circ f^{-1}}^\star\,
(a_n)^{-L_0} \Bigr]\,\,
(a_n)^{L_0}\,.
\ee
Since $f_{n-1}$  is independent of the overall scale
of $f$, we can assume that $f(z) \sim z + \ldots$ in evaluating
$U_{f_I \circ f^{-1}}$. We can then write an expansion without an
$L_0$ term:  
\be
U_{f_I \circ f^{-1}}^\star =  
\cdots  e^{\bar d_6   L_{-6}} \,e^{\bar d_4 L_{-4}}  \, e^{\bar d_2 L_{-2}} \,.
\ee
Here the  $\bar d_n$ are calculable coefficiencts that are independent of $\beta$.
We then have
\be
U_{f_{n-1}}^\star = \, U_f^\star\, \Bigl(\,\cdots  e^{{\bar d_6 a_n^6}   L_{-6}} 
\,e^{{\bar d_4 a_n^4} L_{-4}}  \, 
e^{ {\bar d_2 a_n^2} L_{-2}}\,\Bigr)\, (a_n )^{L_0}\,.
\ee
The string field
will be an expansion in powers of $e^{2\beta/s}$.  The leading term in the
expansion of the string field will occur when $U_{f_{n-1}}^\star$ acts on
the tachyon state, the state with $L_0 = -1$.  In this case, to leading order
in $e^{2\beta/s}$, the above factor in parenthesis is equal to one, and we have
\be
U_{f_{n-1}}^\star c_1 |0\rangle  \simeq  \, U_f^\star c_1 |0\rangle \cdot
\,  {1 \over \bar a} \,2^{-1/s}\, e^{2\beta/s}\,.
\ee
It now follows from (\ref{psin_finfdfdal_new_more}) that
\be
\label{psin_ore}
 \psi_{n-2} \simeq  {1\over 2} \, s\, \bar{a}^{s-1}\,
   {g(\bar{a})^2\over g'(\bar{a})} \, {1 \over \bar a} \,2^{-1/s}\, e^{2\beta/s}
\,U_{f}^\star  c_1 |0\rangle \cdot  2 \Bigl( 1 - {g(\bar a) \over 
\bar{a}g'(\bar a)} \Bigr)\,,
\ee
or, equivalently, 
\be
\label{psin_oren}
 \psi_{n-2} \simeq \,U_{f}^\star  c_1 |0\rangle \cdot 
  \,\,2^{-1-{1\over s}} s\, e^{2\beta/s} \,2\bar{a}^{s-2}\,
   {g(\bar{a})^2\over g'(\bar{a})} \,
 \Bigl( 1 - {g(\bar a) \over 
\bar{a}g'(\bar a)} \Bigr)\,.
\ee
This means that to leading order in the expansion the string field
is given by
\be
|\Psi\rangle  = U_f^\star \, c_1 |0\rangle \,\cdot
\,2^{-1-{1\over s}} \, e^{2\beta/s} \,2 \sum_{n=2}^\infty \partial_n \Bigl(\bar{a}^{s-2}\,
   {g(\bar{a})^2\over g'(\bar{a})} \,
 \Bigl( 1 - {g(\bar a) \over 
\bar{a}g'(\bar a)} \Bigr)
 \Bigr) \,.  
\ee
This is the general result, valid for all arbitrary $s \geq 1$.  Note that this term
diverges parameterically with $\beta$. 
For the case of the sliver, the string field becomes 
\be
|\Psi\rangle  =  U_f^\star \, c_1 |0\rangle \,\cdot
{1\over 4} e^{2\beta} \, \cdot 2 \sum_{n=2}^\infty \partial_n \Bigl( \, {g^2(\bar a)\over 
\bar{a}g'(\bar a)} \Bigl( 1 - {g(\bar a) \over 
\bar{a}g'(\bar a)} \Bigr)
 \Bigr) \,, \qquad  s=1\,, 
\ee
with $g(z) = {1\over 2} \tan (\pi z)$.  Recalling the definition of $\bar a$ in
(\ref{abardef}) one can easily evaluate the above expression numerically.
The result is 
\be
|\Psi\rangle  =  U_f^\star \, c_1 |0\rangle \,\cdot
{1\over 4} e^{2\beta}\, \cdot (0.39545107 ) \,.  
\ee
We will not attempt the calculation of the subleading terms 
in the solution.  In the work of~\cite{Okawa:2003zc} the leading
term of the solution is a divergent coefficient that multiplies a 
ghost insertion on a regulated
projector.  The regulation parameter and the divergent coefficient
are related, and this helps produce finite energy.  
While the expansion of the solution around the sliver
in this subsection is well defined
in calculating coefficients in front of states in the Fock space,
it is not well defined
in calculating the energy of the solution.
It would be interesting to find a more systematic way
to expand the solution for large $\beta$,
in particular, in the context of VSFT.

\sectiono{Concluding Remarks}

We find it tantalizing  that projectors play a significant role
in the construction of solutions of OSFT.  Projectors
are essentially the solutions of vacuum string field theory (VSFT),
so this fact should help relate OSFT to VSFT and, with some luck,
to obtain a regular form of VSFT. In addition to finding new solutions
of OSFT, the development of VSFT may pave the way for further
progress in this field. 

The role of projectors was somewhat hidden in the tachyon vacuum
solution of Schnabl~\cite{Schnabl:2005gv}.  The 
$\mathcal{L}_0$, $\mathcal{L}_0^\star$ structure
associated with the geometry of the 
wedge states seemed to be the central and necessary ingredient.  
In~\cite{Rastelli:2006ap} it was found that the $\mathcal{L}_0$, $\mathcal{L}_0^\star$  structure is not unique to the wedge states. Including other conditions required by solvability, one is led to special projectors.

In this work we have used reparameterizations
to show that {\it any}
twist-invariant,
single-split projector furnishes a solution.  It is not
required to have a special projector, but the form of
the solution simplifies considerably for that case.  This is
a satisfying conclusion: each single-split projector furnishes a solution
in a different gauge, and all single-split projectors are allowed. 

Our methods
using reparameterizations 
do not immediately apply
to multiple-split projectors, {\it i.e.},
conformal frames where the coordinate curve goes to infinity at other points besides the
string midpoint. These projectors are not related by regular reparameterizations
to the sliver.  
Examples  
of multiple-split special projectors were given in~\cite{Rastelli:2006ap}.
It is not difficult to construct
formal solutions for a certain class of multiple-split
special projectors
by inserting operators analogous to those in section 4,
but it is not obvious
if the calculation of their energies is well defined.

While the idea of using reparameterizations is certainly
not new, it was generally felt that concrete computations
would be difficult since the operators that perform
reparameterizations are extremely difficult to construct.
We found a way to implement the necessary reparameterizations without
constructing the operators.  

One particularly interesting by-product is the construction
of an abelian algebra of states for any projector. The surface
states interpolate between the identity and the projector. 
For the sliver this is the familiar algebra of wedge states.   We believe,
although we have not proven, 
that the wedge states are the unique states that interpolate
between the identity and the sliver and star-multiply among themselves.  
If this is the case, the possibility of reparameterizations implies that
the interpolating family must be a canonical unique object for
any projector.  In this sense there is no preferred projector and our
use of the sliver is recognized to be just a technical tool.

\vspace{1cm}

{\bf \large Acknowledgments}

 \medskip

We would like to thank Ian Ellwood and Wati Taylor
for helpful conversations. 
The work of LR is  supported in part
 by the National Science Foundation Grant No. PHY-0354776. 
  Any opinions, findings, and
conclusions or recommendations expressed in this material are
those of the authors and do not necessarily reflect the views of
the National Science Foundation.
The work of BZ is supported in part
by the U.S. DOE  grant DE-FC02-94ER40818.


\medskip

\begin{thebibliography}{10}


%\cite{Witten:1985cc}
\bibitem{Witten:1985cc}
  E.~Witten,
   ``Noncommutative Geometry And String Field Theory,''
  %
  Nucl.\ Phys.\ B {\bf 268}, 253 (1986).
  %%CITATION = NUPHA,B268,253;%%






%\cite{Schnabl:2005gv}
\bibitem{Schnabl:2005gv}
  M.~Schnabl,
  ``Analytic solution for tachyon condensation
  in open string field theory,''
  arXiv:hep-th/0511286.
  %%CITATION = HEP-TH 0511286;%%

%\cite{Okawa:2006vm}
\bibitem{Okawa:2006vm}
  Y.~Okawa,
   ``Comments on Schnabl's analytic solution
  for tachyon condensation in Witten's open string field theory,''
  JHEP {\bf 0604}, 055 (2006)
  [arXiv:hep-th/0603159].
  %%CITATION = HEP-TH 0603159;%%


\bibitem{Fuchs:2006hw}
  E.~Fuchs and M.~Kroyter,
  ``On the validity of the solution of string field theory,''
  JHEP {\bf 0605}, 006 (2006)
  [arXiv:hep-th/0603195].
  %%CITATION = HEP-TH 0603195;%%


%\cite{Rastelli:2006ap}
\bibitem{Rastelli:2006ap}
  L.~Rastelli and B.~Zwiebach,
  ``Solving open string field theory with special projectors,''
  arXiv:hep-th/0606131.
  %%CITATION = HEP-TH 0606131;%%
 



\bibitem{Ellwood:2006ba}
  I.~Ellwood and M.~Schnabl,
  ``Proof of vanishing cohomology at the tachyon vacuum,''
  arXiv:hep-th/0606142.
  %%CITATION = HEP-TH 0606142;%%
  
  \bibitem{Fuchs:2006an}
  E.~Fuchs and M.~Kroyter,
  ``Schnabl's ${\cal L}_0$ 
  operator in the continuous basis,''
  arXiv:hep-th/0605254.
  %%CITATION = HEP-TH 0605254;%%
  %\cite{Fuchs:2006gs}
%\bibitem{Fuchs:2006gs}
  E.~Fuchs and M.~Kroyter,
  ``Universal regularization for string field theory,''
  arXiv:hep-th/0610298.
  %%CITATION = HEP-TH 0610298;%%
  
  
  \bibitem{Fuji:2006me}
  H.~Fuji, S.~Nakayama and H.~Suzuki,
  ``Open string amplitudes in various gauges,''
  arXiv:hep-th/0609047.
  %%CITATION = HEP-TH 0609047;%%




%\cite{Rastelli:2000iu}
\bibitem{Rastelli:2000iu}
  L.~Rastelli and B.~Zwiebach,
  ``Tachyon potentials, star products and universality,''
  JHEP {\bf 0109}, 038 (2001)
  [arXiv:hep-th/0006240].
  %%CITATION = HEP-TH 0006240;%%
  
  %\cite{Kostelecky:2000hz}
\bibitem{Kostelecky:2000hz}
  V.~A.~Kostelecky and R.~Potting,
  ``Analytical construction of a nonperturbative vacuum for the open  bosonic
  string,''
  Phys.\ Rev.\ D {\bf 63}, 046007 (2001)
  [arXiv:hep-th/0008252].
  %%CITATION = HEP-TH 0008252;%%


%\cite{Rastelli:2001jb}
\bibitem{Rastelli:2001jb}
 L.~Rastelli, A.~Sen and B.~Zwiebach,
  ``Classical solutions in string field theory around the tachyon vacuum,''
  Adv.\ Theor.\ Math.\ Phys.\  {\bf 5}, 393 (2002)
  [arXiv:hep-th/0102112].




%\cite{Rastelli:2001vb}
\bibitem{Rastelli:2001vb}
  L.~Rastelli, A.~Sen and B.~Zwiebach,
  ``Boundary CFT construction of D-branes in vacuum string field theory,''
  JHEP {\bf 0111}, 045 (2001)
  [arXiv:hep-th/0105168].
  %%CITATION = HEP-TH 0105168;%%


\bibitem{LC}
%\cite{LeClair:1988sp}
  A.~LeClair, M.~E.~Peskin and C.~R.~Preitschopf,
   ``String Field Theory On The Conformal Plane. 1. Kinematical Principles,''
  %
  Nucl.\ Phys.\ B {\bf 317}, 411 (1989).
  %%CITATION = NUPHA,B317,411;%%
%\cite{LeClair:1988sj}
   ``String Field Theory On The Conformal Plane. 2. Generalized Gluing,''
  %
  Nucl.\ Phys.\ B {\bf 317}, 464 (1989).
  %%CITATION = NUPHA,B317,464;%%



%\cite{Gaiotto:2002kf}
\bibitem{Gaiotto:2002kf}
  D.~Gaiotto, L.~Rastelli, A.~Sen and B.~Zwiebach,
  ``Star algebra projectors,''
  JHEP {\bf 0204}, 060 (2002)
  [arXiv:hep-th/0202151].
  %%CITATION = HEP-TH 0202151;%%



%\cite{Rastelli:2001rj}
\bibitem{Rastelli:2001rj}
  L.~Rastelli, A.~Sen and B.~Zwiebach,
  ``Half strings, projectors, and multiple D-branes in vacuum string field
  theory,''
  JHEP {\bf 0111}, 035 (2001)
  [arXiv:hep-th/0105058].
  %%CITATION = HEP-TH 0105058;%%





%\cite{Gross:2001rk}
\bibitem{Gross:2001rk}
  D.~J.~Gross and W.~Taylor,
  ``Split string field theory. I,''
  JHEP {\bf 0108}, 009 (2001)
  [arXiv:hep-th/0105059].
  %%CITATION = HEP-TH 0105059;%%
  ``Split string field theory. II,''
  JHEP {\bf 0108}, 010 (2001)
  [arXiv:hep-th/0106036].
  %%CITATION = HEP-TH 0106036;%%





%\cite{Witten:1986qs}
\bibitem{Witten:1986qs}
  E.~Witten,
  ``Interacting Field Theory Of Open Superstrings,''
  Nucl.\ Phys.\ B {\bf 276}, 291 (1986).
  %%CITATION = NUPHA,B276,291;%%


%\cite{Qiu:1987dv}
\bibitem{Qiu:1987dv}
  Z.~Qiu and A.~Strominger,
  ``Gauge symmetries in (super)string field theory,''
  Phys.\ Rev.\ D {\bf 36}, 1794 (1987).
  %%CITATION = PHRVA,D36,1794;%%








\bibitem{vsft}
%\cite{Rastelli:2001uv}
%\bibitem{Rastelli:2001uv}
  L.~Rastelli, A.~Sen and B.~Zwiebach,
  ``Vacuum string field theory,'' arXiv:hep-th/0106010.
  %%CITATION = HEP-TH 0106010;%%
%\cite{Gaiotto:2001ji}
%\bibitem{Gaiotto:2001ji}

 
  %\cite{Gaiotto:2001ji}
\bibitem{Gaiotto:2001ji}
  D.~Gaiotto, L.~Rastelli, A.~Sen and B.~Zwiebach,
  ``Ghost structure and closed strings in vacuum string field theory,''
  Adv.\ Theor.\ Math.\ Phys.\  {\bf 6}, 403 (2003)
  [arXiv:hep-th/0111129].
  %%CITATION = HEP-TH 0111129;%%


%\cite{Okawa:2002pd}
\bibitem{Okawa:2002pd}
  Y.~Okawa,
  ``Open string states and D-brane tension from vacuum string field theory,''
  JHEP {\bf 0207}, 003 (2002)
  [arXiv:hep-th/0204012].
  %%CITATION = HEP-TH 0204012;%%

%\cite{Drukker:2005hr}
\bibitem{Drukker:2005hr}
  N.~Drukker and Y.~Okawa,
  ``Vacuum string field theory without matter-ghost factorization,''
  JHEP {\bf 0506}, 032 (2005)
  [arXiv:hep-th/0503068].
  %%CITATION = HEP-TH 0503068;%%


  %\cite{Okawa:2003zc}
\bibitem{Okawa:2003zc}
  Y.~Okawa,
  ``Solving Witten's string field theory using the butterfly state,''
  Phys.\ Rev.\ D {\bf 69}, 086001 (2004)
  [arXiv:hep-th/0311115].
  %%CITATION = HEP-TH 0311115;%%
  
  %\cite{Yang:2004xz}
\bibitem{Yang:2004xz}
  H.~Yang,
  ``Solving Witten's SFT by insertion of operators on projectors,''
  JHEP {\bf 0409}, 002 (2004)
  [arXiv:hep-th/0406023].
  %%CITATION = HEP-TH 0406023;%%
  
  %\cite{Gaiotto:2002wy}
\bibitem{Gaiotto:2002wy}
  D.~Gaiotto and L.~Rastelli,
  ``Experimental string field theory,''
  JHEP {\bf 0308}, 048 (2003)
  [arXiv:hep-th/0211012].
  %%CITATION = HEP-TH 0211012;%%

\end{thebibliography}
\end{document}